\documentclass[% 
 reprint,
showpacs,preprintnumbers,
 amsmath,amssymb,
 aps,
prb,
]{revtex4-1}

\usepackage{graphicx}% Include figure files
\usepackage{dcolumn}% Align table columns on decimal point
\usepackage{bm}% bold math
\usepackage{color}
\usepackage[usenames,dvipsnames]{xcolor}
\usepackage{ulem}
\usepackage{braket} 
\usepackage{longtable}

\allowdisplaybreaks

\begin{document}

\preprint{APS/123-QED}

\title{
Classification of atomic-scale multipoles under crystallographic point groups\\ and application to linear response tensors 
}

\author{Satoru Hayami$^{1}$, Megumi Yatsushiro$^{1}$, Yuki Yanagi$^{2}$, and Hiroaki Kusunose$^2$}
\affiliation{
$^1$Department of Physics, Hokkaido University, Sapporo 060-0810, Japan \\
$^2$Department of Physics, Meiji University, Kawasaki 214-8571, Japan 
}

\begin{abstract}
Four types of atomic-scale multipoles, electric, magnetic, magnetic toroidal, and electric toroidal multipoles, give a complete set to describe arbitrary degrees of freedom for coupled charge, spin, and orbital of electrons.   
We here present a systematic classification of these multipole degrees of freedom towards the application in condensed matter physics. 
Starting from the multipole description under the rotation group in real space, we generalize the concept of multipoles in momentum space with the spin degree of freedom. 
We show how multipoles affect the electronic band structures and linear responses, such as the magneto-electric effect, magneto-current (magneto-gyrotropic) effect, spin conductivity, Piezo-electric effect, and so on.  
Moreover, we exhibit a complete table to represent the active multipoles under 32 crystallographic point groups. 
Our comprehensive and systematic analyses will give a foundation to identify enigmatic electronic order parameters and a guide to evaluate peculiar cross-correlated phenomena in condensed matter physics from 
microscopic point of view. 
\end{abstract}
\maketitle

\section{Introduction}
\label{sec:Introduction}

\begin{figure}[htb!]
\begin{center}
\includegraphics[width=1.0 \hsize]{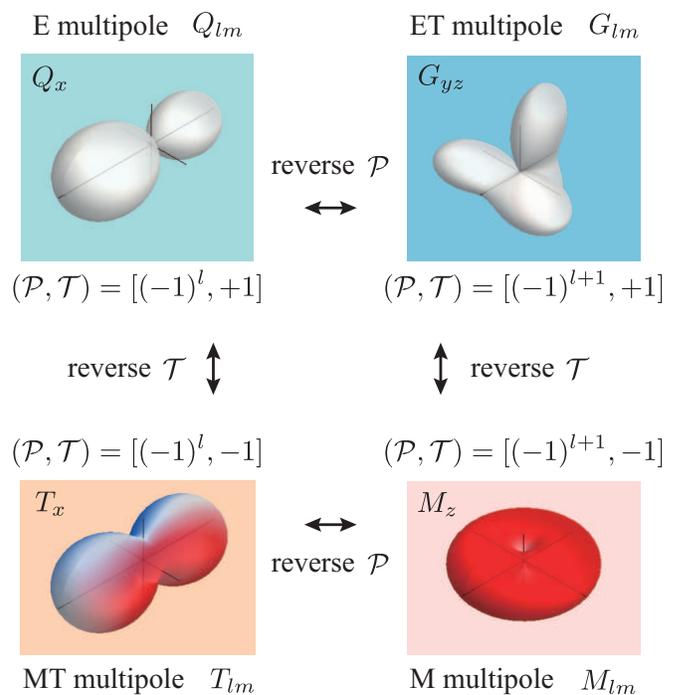} 
\caption{
\label{Fig:multipole}
Four types of multipoles under the space-time inversion group. 
The E (M) multipole transforms the MT (ET) multipole by reversing the time-reversal property, while the E (M) multipole transforms ET (MT) multipole by reversing the spatial-inversion property. 
Schematic pictures of the wave functions for the E dipole, M dipole, MT dipole, and ET quadrupole are shown in each panel. 
The shape and color map of the pictures represent the electric charge density and the $z$ component of the orbital angular-momentum density, respectively. 
}
\end{center}
\end{figure}

The multipole moments characterize electric charge and current distributions, whose concept has been widely developed in various fields of physics at different length scales, such as classical electromagnetism
~\cite{dubovik1975multipole,LandauLifshitz198001,Jackson3rd1999,nanz2016toroidal}, nuclear physics~\cite{Blin-Stoyle_RevModPhys.28.75,zel1958relation,Flambaum_PhysRevC.56.1641,Blatt1991}, solid-state physics~\cite{Kusunose_JPSJ.77.064710,kuramoto2008electronic,Santini_RevModPhys.81.807,kuramoto2009multipole,suzuki2018first}, and metamaterials~\cite{kaelberer2010toroidal,fedotov2013resonant,Fan_PhysRevB.87.115417,Savinov_PhysRevB.89.205112,papasimakis2016electromagnetic}. 
It was well-known that there are four-types of fundamental multipoles according to their spatial inversion and time-reversal properties~\cite{dubovik1986axial,dubovik1990toroid,nanz2016toroidal,hayami2018microscopic}; 
electric (E: polar/true tensor with time-reversal even), magnetic (M: axial/pseudo tensor with time-reversal odd),
magnetic toroidal (MT: polar/true tensor with time-reversal odd), and electric toroidal (ET: axial/pseudo tensor with time-reversal even) multipoles. 
Recent study has shown that the four fundamental multipoles in atomic scale constitute a complete set to span the Hilbert space under the space-time inversion group.
They can be applied to not only a classical but also a quantum-mechanical picture~\cite{hayami2018microscopic}. 
The mutual relationship between four multipoles and their schematic pictures in the quantum-mechanical representation are shown in Fig.~\ref{Fig:multipole}. 

In condensed matter physics, the multipoles have been recognized as important quantities to describe multiple degrees of freedom in electrons, e.g., charge, spin, and orbital, in a unified way.  
Especially, the atomic-scale multipoles have been extensively studied in $f$-electron systems, as the interplay between the Coulomb interaction and spin-orbit coupling due to strongly localized nature gives rise to anisotropic charge and spin distributions~\cite{Kusunose_JPSJ.77.064710,Santini_RevModPhys.81.807,kuramoto2009multipole}. 
In fact, higher-rank multipole orders beyond the conventional E and M dipole orders have been established by mutual interplay between experimental and theoretical investigations, for instance, E quadrupole in Pr$T_2X_{20}$($T=$Ir, Rh, V, Ti and $X=$Al, Zn)~\cite{Onimaru_PhysRevLett.106.177001,sakai2012superconductivity,Tsujimoto_PhysRevLett.113.267001,matsumoto2015effect,kubo2015absence,onimaru2016exotic,Onimaru_PhysRevB.94.075134} and M octupole in Ce$_{1-x}$La$_x$B$_6$~\cite{shiina1997magnetic,tanaka2005manipulating,Mannix_PhysRevLett.95.117206,Matsumura_PhysRevLett.103.017203}. 

On the other hand, such a multipole concept can be applied to a cluster consisting of several atomic sites, which is so-called a cluster multipole~\cite{Yanase_JPSJ.83.014703,Hayami_PhysRevB.90.081115,hitomi2014electric,Hayami_1742-6596-592-1-012131,hayami2016emergent,Suzuki_PhysRevB.95.094406,hitomi2016electric,Watanabe_PhysRevB.96.064432,matsumoto2017symmetry,yanagi2017optical,hayami2017mean,Hayami_PhysRevB.97.024414}.  
The complex electric and magnetic orderings in cluster are regarded as the ferroic arrangement of higher-order cluster multipoles from the symmetry aspect. 
For example, the all-in/all-out ordering of magnetic moments on the pyrochlore structure is regarded as a M octupole order (precisely speaking, M pseudoscalar in the point group)~\cite{Arima_doi:10.7566/JPSJ.82.013705}, while a staggered antiferromagnetic ordering on the zigzag chain is regarded as a ferroic order of MT dipole and M quadrupole~\cite{Yanase_JPSJ.83.014703,Hayami_doi:10.7566/JPSJ.84.064717,hayami2016emergent}. 
More recently, the concept of multipoles is extended to multiple hybrid orbitals~\cite{hayami2018microscopic}, and to the momentum space relevant with topologically nontrivial excitations through the Berry curvature~\cite{Fu_PhysRevLett.115.026401,benalcazar2017quantized,Benalcazar_PhysRevB.96.245115,shitade2018theory,gao2018theory}.

In this way, studies of the multipoles are useful to cover various unconventional order parameters in a systematic manner and understand/expect physical phenomena from the symmetry viewpoint. 
For example, the nematic order in iron-based superconductors~\cite{chuang2010nematic,Goto_JPSJ.80.073702,Yoshizawa_JPSJ.81.024604,fernandes2014drives,Johnston2015Iron,HOSONO2015399} and magnetic insulators~\cite{Chen_PhysRevLett.27.1383,andreev1984spin,lacroix2011introduction, Nic_PhysRevLett.96.027213}, Pomeranchuk instability~\cite{pomeranchuk1959sov,Yamase_JPSJ.69.332,Halboth_PhysRevLett.85.5162,Honerkamp_PhysRevB.72.115103,yamase2007van,Yamase_PhysRevB.75.155117}, anisotropic density wave including staggered flux phases~\cite{affleck1988large,Nayak_PhysRevB.62.4880,Chakravarty_PhysRevB.63.094503,Raghu_PhysRevLett.100.156401,Allais_PhysRevB.90.155114,Ikeda_PhysRevLett.81.3723,Fujimoto_PhysRevLett.106.196407}, excitonic insulators~\cite{jerome1967excitonic,Halperin_RevModPhys.40.755,Kunes_PhysRevB.90.235112,Kaneko_PhysRevB.94.125127,yamaguchi2017multipole}, and spin chirality accompanying Berry phase~\cite{Ohgushi_PhysRevB.62.R6065,taguchi2001spin,Nagaosa_RevModPhys.82.1539} are also described by multipole terminology. 
It is advantageous for the multipole description that once a type of the multipoles is identified, it is easy to understand/predict how the band structure is deformed and what types of cross-correlated couplings and transport phenomena occur. 
For example, the magneto-electric effects in Cr$_2$O$_3$~\cite{dzyaloshinskii1960magneto,astrov1960magnetoelectric,popov1999magnetic} and Co$_4$Nb$_2$O$_9$~\cite{fischer1972new,Khanh_PhysRevB.93.075117,Khanh_PhysRevB.96.094434,Yanagi2017,Yanagi_PhysRevB.97.020404} are understood by the M quadrupole, magneto-current effect in metallic UNi$_4$B by the MT dipole~\cite{Hayami_PhysRevB.90.024432,Hayami_1742-6596-592-1-012101,saito2018evidence}, valley splitting in the electronic band structure by the MT octupole~\cite{li2013coupling}, nonreciprocal magnon excitations in BiTeBr~\cite{ideue2017bulk} and $\alpha$-Cu$_2$V$_2$O$_7$~\cite{Gitgeatpong_PhysRevB.92.024423,Hayami_doi:10.7566/JPSJ.85.053705,Gitgeatpong_PhysRevB.95.245119,Gitgeatpong_PhysRevLett.119.047201,takashima2018nonreciprocal} by the E and MT dipoles, orbital Edelstein effect in Te by the ET monopole and quadrupole~\cite{furukawa2017observation,yoda2018orbital}, and large anomalous Hall effect, Nernst effect, and Kerr rotation in Mn$_3$Sn by the M octupole~\cite{nakatsuji2015large,Suzuki_PhysRevB.95.094406,ikhlas2017large,kuroda2017evidence,higo2018large}. 
Moreover, unconventional superconductivities are also discussed in the language of multipoles~\cite{Sumita_PhysRevB.93.224507,Nomoto_PhysRevB.94.174513,hattori2017local,Sumita_PhysRevLett.119.027001}.

In the present paper, we push forward these multipole studies in a more general framework. 
Our purpose of this paper is to give a comprehensive classification of multipoles in real and momentum spaces and to elucidate physical properties brought by the multipole degrees of freedom. 
Starting from the multipole description under the rotation group in real space, we generalize the concept of multipoles in momentum space with the spin degree of freedom. 
We show how multipoles affect the electronic band structures and linear responses, such as the magneto-electric effect and magneto-current (magneto-gyrotropic) effect.  
Moreover, we demonstrate a complete table to display the active multipoles under 32 crystallographic point groups. 
Our systematic study will encourage a direction for material design based on multipole (electronic) degrees of freedom.

The organization of this paper is as follows. 
In Sec.~\ref{sec:Multipole descriptions}, we give a definition of four multipoles in both real and momentum spaces. 
In Sec.~\ref{sec:Physical properties under multipoles}, we show how the emergent multipoles affect the electronic band structures and linear responses including magneto-electric and magneto-current(magneto-gyrotropic) phenomena.  
After presenting the crystalline-electric-field potential under all the point groups, we clarify what types of multipoles are active and become potential order parameters under point-group irreducible representation in Sec.~\ref{sec:Point-group irreducible representations}. 
Section~\ref{sec:Summary} is devoted to a summary of the present paper. 
In Appendix~\ref{sec:Atomic basis wave functions}, we show the atomic basis wave functions for $s$, $p$, $d$, and $f$ orbitals. 
We discuss a possible extension of the momentum-space multipoles to multi-orbital system in Appendix~\ref{sec:k-space multipoles in multi-orbital systems}. 
In Appendix~\ref{sec:Space-time inversion operations of linear response tensor}, we show transformation properties of linear response tensors under the spatial inversion and time-reversal operations. 
We present which multipoles are relevant with rank-3 linear response tensors, such as the spin conductivity tensor and Piezo-electric(current) tensor in Appendix~\ref{sec:Other linear response tensors}. 
In Sec.~\ref{sec:Oh and its subgroups}, we show the active multipoles for atomic basis functions for the cubic $O_{\rm h}$ and its subgroups. 
Appendix~\ref{sec:Oh and its subgroups} contains various tables for the hexagonal $D_{\rm 6h}$ and its subgroups to complete all the point groups.

\section{Definition of multipoles}
\label{sec:Multipole descriptions}

The concept of multipoles is useful to describe electronic degrees of freedom in a systematic way, and
it is capable to express arbitrary type of order parameters in phase transition and physical responses to external fields at microscopic level. 
As mentioned in the Introduction, there are four types of multipoles, E (electric: $Q_{lm}$), M (magnetic: $M_{lm}$), MT (magnetic toroidal: $T_{lm}$), and ET (electric toroidal: $G_{lm}$), according to the time-reversal and spatial inversion properties~\cite{dubovik1975multipole,dubovik1986axial,dubovik1990toroid}.  

In this section, we introduce microscopic expressions of four types of multipoles. 
First, we give the expressions in real space in Sec.~\ref{sec:Multipoles in real space}, and then we give them in momentum space in Sec.~\ref{sec:Multipoles in momentum space}. 
With these prescriptions, we summarize in Table~\ref{tab_activated_multipole} which type of multipoles are activated in the atomic basis functions with the spin degree of freedom.

\subsection{Multipoles in real space}
\label{sec:Multipoles in real space}

Based on the standard multipole expansion of the electromagnetic potentials $\phi(\bm{r})$ and $\bm{A}(\bm{r})$ in the presence of the source electric current $\bm{j}_{\rm e} (\bm{r})= c \bm{\nabla}\times \bm{M}(\bm{r})$ and the magnetic current $\bm{j}_{\rm m}(\bm{r})=c \bm{\nabla}\times \bm{P}(\bm{r})$ ($\bm{M}$ and $\bm{P}$ are the magnetization and electric polarization), and by inserting the quantum-mechanical expressions of $\bm{j}_{\rm e}$ and $\bm{j}_{\rm m}$ including spin contributions, two of the present authors have shown the quantum-mechanical expressions of multipoles as~\cite{hayami2018microscopic} 
\begin{align}
\label{eq:Emultipole_Q}
\hat{Q}_{lm} &= -e\sum_j O_{lm} (\bm{r}_j), \\
\label{eq:Mmultipole_Q}
\hat{M}_{lm} &= -\mu_{\rm B}\sum_j \bm{m}_{l}(\bm{r}_{j})\cdot\bm{\nabla} O_{lm}(\bm{r}_j),\\
\label{eq:MTmultipole_Q}
\hat{T}_{lm} &=  -\mu_{\rm B}\sum_j \bm{t}_{l}(\bm{r}_{j})\cdot \bm{\nabla} O_{lm}(\bm{r}_j), \\
\label{eq:ETmultipole_Q}
\hat{G}_{lm} &= -e\sum_{j}\sum_{\alpha\beta}^{x,y,z}g_{l}^{\alpha\beta}\nabla_{\alpha}\nabla_{\beta} O_{lm}(\bm{r}_{j}),
\end{align}
where $\bm{m}_l$, $\bm{t}_{l}$, and $g^{\alpha\beta}_l$ are defined in terms of the dimensionless orbital and spin angular-momentum operators $\bm{l}_{j}$ and $\bm{\sigma}_{j}/2$ of an electron at $\bm{r}_j$ as
\begin{align}
\label{eq:m}
\bm{m}_{l}(\bm{r}_{j})&=\frac{2 \bm{l}_j}{l+1} + \bm{\sigma}_j, \\
\label{eq:t}
\bm{t}_{l}(\bm{r}_{j})&=\frac{\bm{r}_j}{l+1} \times \left( \frac{2\bm{l}_j}{l+2} + \bm{\sigma}_j \right), \\
\label{eq:g}
g_{l}^{\alpha\beta}(\bm{r}_{j})&=m_{l}^{\alpha}(\bm{r}_{j})t_{l}^{\beta}(\bm{r}_{j}).
\end{align}
We have introduced 
\begin{align}
O_{lm}(\bm{r})=\sqrt{\frac{4\pi}{2l+1}}r^l Y^{*}_{lm}(\hat{\bm{r}}),
\end{align}
where $Y_{lm}(\hat{\bm{r}})$ is the spherical harmonics as a function of angles $\hat{\bm{r}}=\bm{r}/r$~\cite{Blatt1991,Kusunose_JPSJ.77.064710} with the azimuthal and magnetic quantum numbers, $l$ and $m$ ($-l \leq m \leq l$). 
We adopt the phase convention so as to satisfy $Y_{lm}(\hat{\bm{r}})=(-1)^{m}Y_{l-m}^{*}(\hat{\bm{r}})$. 
The index $l$ represents the rank of multipoles: $l=0$ (monopole), $l=1$ (dipole), $l=2$ (quadrupole), $l=3$ (octupole), $l=4$ (hexadecapole), and so on. 
Strictly speaking, the rank is meaningful as the irreducible representation only if rotational symmetry is preserved, otherwise the same irreducible representations appear in different ranks in general, and they mix with each other. 
Hereafter, we omit ``elementary charge'' for multipoles, i.e., $-e$ and $-\mu_{\rm B}$, for notational simplicity.

\begin{table*}[htb!]
\caption{
Even-parity cubic harmonics (E multipoles in unit of $-e$ in the cubic $O_{\rm h}$ group). 
The linear combinations of the tesseral harmonics are shown where $(lm)$ and $(lm)'$ correspond to $O_{lm}^{\rm (c)}(\bm{r})$ and $O_{lm}^{\rm (s)}(\bm{r})$, respectively.
In the irreducible representation (irrep.), the subscript represents the spatial parity (even:$g$, odd:$u$), while the superscript represents the time-reversal property (even:$+$, odd:$-$).
A$_{1}$, A$_{2}$, E, T$_{1}$, and T$_{2}$ correspond to $\Gamma_{1}$, $\Gamma_{2}$, $\Gamma_{3}$, $\Gamma_{4}$, and $\Gamma_{5}$, respectively, in the Bethe notation. 
}
\label{tab_Oh_CEFe}
\centering
\begingroup
\renewcommand{\arraystretch}{1.0}
 \begin{tabular}{ccccc}
 \hline \hline
rank & irrep. & symbol & definition & linear combination \\ \hline
$0$  & ${\rm A}_{1g}^{+}$ & $Q_0$ & $1$ & $(00)$ \\
\hline
$2$  & ${\rm E}_{g}^{+}$ & $Q_{u}$, $Q_{v}$ & $\displaystyle \frac{1}{2}(3z^2-r^2)$, $\displaystyle \frac{\sqrt{3}}{2}(x^2-y^2)$ & $(20)$, $(22)$ \\
     & ${\rm T}_{2g}^{+}$ & $Q_{yz}$, $Q_{zx}$, $Q_{xy}$ & $\displaystyle  \sqrt{3}y z$, $\sqrt{3}z x$, $\sqrt{3}x y$ 
     & $(21)'$, $(21)$, $(22)'$\\
\hline
$4$ & ${\rm A}_{1g}^{+}$ & $Q_{4}$ & $\displaystyle  \frac{5\sqrt{21}}{12}\left(x^4+y^4+z^4-\frac{3}{5} r^4 \right)$ & $\displaystyle  (4) \equiv \frac{1}{2\sqrt{3}} [\sqrt{5} (44)+\sqrt{7}(40)]$ \\
      & ${\rm E}_{g}^{+}$ & $Q_{4u}$& $\displaystyle \frac{7\sqrt{15}}{6}\left[z^4-\frac{x^4+y^4}{2}-\frac{3}{7} r^2(3z^2-r^2)\right]$ & $\displaystyle  -\frac{1}{2\sqrt{3}}[\sqrt{7}(44)-\sqrt{5}(40)]$ \\ 
     & & $Q_{4v}$ & $\displaystyle \frac{7\sqrt{5}}{4}\left[x^4-y^4-\frac{6}{7}r^2 (x^2-y^2)\right]$ & $-(42)$ \\
     & ${\rm T}_{1g}^{+}$ & $Q^{\alpha}_{4x}$ & $\displaystyle \frac{\sqrt{35}}{2}y z(y^2-z^2)$ & $\displaystyle -\frac{1}{2\sqrt{2}}[(43)'+\sqrt{7}(41)']$ \\
     &  & $Q^{\alpha}_{4y}$ & $\displaystyle \frac{\sqrt{35}}{2}z x(z^2-x^2)$ & $\displaystyle -\frac{1}{2\sqrt{2}}[(43)-\sqrt{7}(41)]$\\
     &  & $Q^{\alpha}_{4z}$ & $\displaystyle \frac{\sqrt{35}}{2}x y(x^2-y^2)$ & $(44)'$ \\
     & ${\rm T}_{2g}^{+}$ & $Q^{\beta}_{4x}$ & $\displaystyle \frac{\sqrt{5}}{2}yz(7x^2-r^2)$ & $\displaystyle \frac{1}{2\sqrt{2}}[\sqrt{7}(43)'-(41)']$\\
     & & $Q^{\beta}_{4y}$ & $\displaystyle \frac{\sqrt{5}}{2}zx(7y^2-r^2)$ &$\displaystyle -\frac{1}{2\sqrt{2}}[\sqrt{7}(43)+(41)]$ \\
     & & $Q^{\beta}_{4z}$ & $\displaystyle \frac{\sqrt{5}}{2}xy(7z^2-r^2)$ & $(42)'$ \\
\hline
$6$ & ${\rm A}_{1g}^{+}$  & $Q_{6}$ & $\displaystyle  \frac{231\sqrt{2}}{8}\biggl[x^{2}y^{2}z^{2}+\frac{r^{2}}{22}\left(x^{4}+y^{4}+z^{4}-\frac{3}{5}r^{4}\right)-\frac{r^{6}}{105}\biggr]$ & $\displaystyle  (6)\equiv \frac{1}{2\sqrt{2}}[-\sqrt{7}(64)+(60)]$
\\
& ${\rm A}_{2g}^{+}$  & $Q_{6t}$ 
 & $\displaystyle  \frac{\sqrt{2310}}{8}\bigl[x^{4}\left(y^{2}-z^{2}\right)+y^{4}\left(z^{2}-x^{2}\right)+z^{4}\left(x^{2}-y^{2}\right)\bigr]$ & $\displaystyle  (6{\rm t})\equiv \frac{1}{4}[-\sqrt{5}(66)+\sqrt{11}(62)]$
\\
& ${\rm E}_{g}^{+}$  & $Q_{6u}$ & $\displaystyle  \frac{11\sqrt{14}}{4}\biggl[z^{6}-\frac{x^{6}+y^{6}}{2}-\frac{15}{11}r^{2} (z^{4}-\frac{x^{4}+y^{4}}{2} )+\frac{5}{22}r^{4}\left(3z^{2}-r^{2}\right)\biggr]$ & $\displaystyle \frac{1}{2\sqrt{2}}[(64)+\sqrt{7}(60)]$
\\
&   & $Q_{6v}$ & $\displaystyle \frac{11\sqrt{42}}{8}\biggl[x^{6}-y^{6}-\frac{15}{11}r^{2}(x^{4}-y^{4})+\frac{5}{11}r^{4}\left(x^{2}-y^{2}\right)\biggr]$ & $\displaystyle \frac{1}{4}[\sqrt{11}(66)+\sqrt{5}(62)]$
\\
& ${\rm T}_{1g}^{+}$  & $Q^{\alpha}_{6x}$ & $\displaystyle  \frac{3\sqrt{7}}{4}yz\left(y^{2}-z^{2}\right)\left(11x^{2}-r^{2}\right)$ & $\displaystyle \frac{1}{8}[-\sqrt{22}(65)'-\sqrt{30}(63)'+2\sqrt{3}(61)']$
\\
&  & $Q^{\alpha}_{6y}$ & $ \displaystyle \frac{3\sqrt{7}}{4}zx\left(z^{2}-x^{2}\right)\left(11y^{2}-r^{2}\right)$ & $\displaystyle \frac{1}{8}[\sqrt{22}(65)-\sqrt{30}(63)-2\sqrt{3}(61)]$
\\
&  & $Q^{\alpha}_{6z}$ & $\displaystyle  \frac{3\sqrt{7}}{4}xy\left(x^{2}-y^{2}\right)\left(11z^{2}-r^{2}\right)$ & $(64)'$
\\
& ${\rm T}_{2g}^{+}$  & $Q^{\beta 1 }_{6x}$ & $\displaystyle  \frac{\sqrt{462}}{2}yz\left[y^{4}+z^{4}-\frac{5}{8}\left(y^{2}+z^{2}\right)^{2}\right]$ & $\displaystyle \frac{1}{16}[\sqrt{3}(65)'+\sqrt{55}(63)'+3\sqrt{22}(61)']$
\\
& & $Q^{\beta 1}_{6y}$ & $\displaystyle  \frac{\sqrt{462}}{2}zx\left[z^{4}+x^{4}-\frac{5}{8}\left(z^{2}+x^{2}\right)^{2}\right]$& $\displaystyle \frac{1}{16}[\sqrt{3}(65)-\sqrt{55}(63)+3\sqrt{22}(61)]$
\\
& & $Q^{\beta 1}_{6z}$ & $\displaystyle  \frac{\sqrt{462}}{2}xy\left[x^{4}+y^{4}-\frac{5}{8}\left(x^{2}+y^{2}\right)^{2}\right]$ & $(66)'$
\\
& ${\rm T}_{2g}^{+}$  & $Q^{\beta 2}_{6x}$ & $\displaystyle \frac{\sqrt{210}}{16}yz\left(33x^{4}-18x^{2}r^{2}+r^{4}\right)$ & $\displaystyle \frac{1}{16}[\sqrt{165}(65)'-9(63)'+\sqrt{10}(61)']$
\\
& & $Q^{\beta 2}_{6y}$ & $\displaystyle \frac{\sqrt{210}}{16}zx\left(33y^{4}-18y^{2}r^{2}+r^{4}\right)$ & $\displaystyle \frac{1}{16}[\sqrt{165}(65)+9(63)+\sqrt{10}(61)]$
\\
& & $Q^{\beta 2}_{6z}$ & $\displaystyle \frac{\sqrt{210}}{16}xy\left(33z^{4}-18z^{2}r^{2}+r^{4}\right)$ & $(62)'$
\\
\hline\hline
\end{tabular}
\endgroup
\end{table*}

In the rotation group, arbitrary linear combinations within the same rank can be taken as an irreducible basis set.
It is often useful to introduce the tesseral harmonics $O_{lm}^{\rm (c)}(\bm{r})$ and $O_{lm}^{\rm (s)}(\bm{r})$, which correspond to the real expressions for the spherical harmonics~\cite{hutchings1964point}: 
\begin{align}
&
O_{l0}^{\rm (c)}(\bm{r})\equiv O_{l0}(\bm{r}),
\\&
O_{lm}^{\rm (c)}(\bm{r})\equiv\frac{(-1)^{m}}{\sqrt{2}}[O_{lm}(\bm{r})+O_{lm}^{*}(\bm{r})],
\\&
O_{lm}^{\rm (s)}(\bm{r})\equiv\frac{(-1)^{m}}{\sqrt{2}i}[O_{lm}(\bm{r})-O_{lm}^{*}(\bm{r})],
\end{align}
for $m=1,2,\cdots,l$.
Since the multipole operator $\hat{X}_{lm}$ ($X=Q,M,T,G$) transforms like $O_{lm}(\bm{r})$ by the rotational operation, the similar linear combinations are applied to $\hat{X}_{lm}$ as well.
In what follows, the symbols $O_{lm}$ and $\hat{X}_{lm}$ are used to represent the set of harmonics in the real expression, either the tesseral harmonics, cubic harmonics, hexagonal harmonics, and so on with the rank $l$.
For later convenience, the even- and odd-parity cubic harmonics in $O_{\rm h}$ group are summarized in Tables~\ref{tab_Oh_CEFe} and \ref{tab_Oh_CEFo}, respectively.
The hexagonal harmonics in the $D_{\rm 6h}$ group are also summarized in Tables~\ref{tab_D6h_CEFe} and \ref{tab_D6h_CEFo} in Appendix~\ref{sec:Electric multipoles and CEF parameters}.

\begin{table*}[htb!]
\caption{
Odd-parity cubic harmonics (E multipoles in unit of $-e$ in the cubic $O_{\rm h}$ group). 
}
\label{tab_Oh_CEFo}
\centering
\begingroup
\renewcommand{\arraystretch}{1.0}
 \begin{tabular}{ccccc}
 \hline \hline
rank & irrep. & symbol & definition & linear combination \\ \hline
 $1$   & ${\rm T}_{1u}^{+}$ & $Q_x$, $Q_y$, $Q_z$ & $x$, $y$, $z$ & $(11)$, $(11)'$, $(10)$ \\ \hline
3  & ${\rm A}_{2u}^{+}$ & $Q_{xyz}$ & $\sqrt{15}x y z$ & $(32)'$ \\
 & ${\rm T}_{1u}^{+}$ & $Q_{x}^{\alpha}$ & $\displaystyle  \frac{1}{2}x(5x^{2}-3r^{2})$ & $\displaystyle  \frac{1}{2\sqrt{2}}[\sqrt{5}(33)-\sqrt{3}(31)]$\\
  &  & $Q_{y}^{\alpha}$ & $\displaystyle  \frac{1}{2}y(5y^{2}-3r^{2})$ & $\displaystyle  -\frac{1}{2\sqrt{2}}[\sqrt{5}(33)'+\sqrt{3}(31)']$\\
   &  &$Q_{z}^{\alpha}$ & $\displaystyle \frac{1}{2}z(5z^{2}-3r^{2})$ & $(30)$ \\
 & ${\rm T}_{2u}^{+}$ & $Q_{x}^{\beta}$ & $\displaystyle  \frac{\sqrt{15}}{2}x(y^{2}-z^{2})$ & $\displaystyle -\frac{1}{2\sqrt{2}}[\sqrt{3}(33)+\sqrt{5}(31)]$\\
 & & $Q_{y}^{\beta}$ & $\displaystyle \frac{\sqrt{15}}{2}y(z^{2}-x^{2})$ & $\displaystyle \frac{1}{2\sqrt{2}}[-\sqrt{3}(33)'+\sqrt{5}(31)']$\\
 & & $Q_{z}^{\beta}$ & $\displaystyle \frac{\sqrt{15}}{2}z(x^{2}-y^{2})$ & $(32)$ \\
\hline
$5$ & ${\rm E}_{u}^{+}$  & $Q_{5u}$, $Q_{5v}$ & $\displaystyle \frac{3\sqrt{35}}{2}xyz\left(x^{2}-y^{2}\right)$, $\displaystyle -\frac{\sqrt{105}}{2}xyz\left(3z^{2}-r^{2}\right)$ & $(54)'$, $-(52)'$
\\
& ${\rm T}_{1u}^{+}$ & $Q^{\alpha 1}_{5x}$ & $\displaystyle \frac{x}{8}\left[8x^{4}-40x^{2}\left(y^{2}+z^{2}\right)+15\left(y^{2}+z^{2}\right)^{2}\right]$ & $\displaystyle \frac{1}{8\sqrt{2}}[3\sqrt{7}(55)-\sqrt{35}(53)+\sqrt{30}(51)]$
\\
& & $Q^{\alpha 1}_{5y}$ & $\displaystyle \frac{y}{8}\left[8y^{4}-40y^{2}\left(z^{2}+x^{2}\right)+15\left(z^{2}+x^{2}\right)^{2}\right]$& $\displaystyle \frac{1}{8\sqrt{2}}[3\sqrt{7}(55)'+\sqrt{35}(53)'+\sqrt{30}(51)']$
\\
& & $Q^{\alpha 1}_{5z}$ & $\displaystyle \frac{z}{8}\left[8z^{4}-40z^{2}\left(x^{2}+y^{2}\right)+15\left(x^{2}+y^{2}\right)^{2}\right]$ & $(50)$
\\
& ${\rm T}_{1u}^{+}$ & $Q^{\alpha 2}_{5x}$ & $\displaystyle \frac{3\sqrt{35}}{2}x\left[y^{4}+z^{4}-\frac{3}{4}\left(y^{2}+z^{2}\right)^{2}\right]$ & $\displaystyle \frac{1}{16}[\sqrt{10}(55)+9\sqrt{2}(53)+2\sqrt{21}(51)]$
\\
& &$Q^{\alpha 2}_{5y}$ & $\displaystyle \frac{3\sqrt{35}}{2}y\left[z^{4}+x^{4}-\frac{3}{4}\left(z^{2}+x^{2}\right)^{2}\right]$& $\displaystyle \frac{1}{16}[\sqrt{10}(55)'-9\sqrt{2}(53)'+2\sqrt{21}(51)']$
\\
& &$Q^{\alpha 2}_{5z}$ & $\displaystyle \frac{3\sqrt{35}}{2}z\left[x^{4}+y^{4}-\frac{3}{4}\left(x^{2}+y^{2}\right)^{2}\right]$ & $(54)$
\\
& ${\rm T}_{2u}^{+}$ & $Q^{\beta}_{5x}$ & $\displaystyle \frac{\sqrt{105}}{4}x\left(y^{2}-z^{2}\right)\left(3x^{2}-r^{2}\right)$ & $\displaystyle \frac{1}{4\sqrt{2}}[-\sqrt{15}(55)-\sqrt{3}(53)+\sqrt{14}(51)]$
\\
& & $Q^{\beta}_{5y}$ & $\displaystyle \frac{\sqrt{105}}{4}y\left(z^{2}-x^{2}\right)\left(3y^{2}-r^{2}\right)$ & $\displaystyle \frac{1}{4\sqrt{2}}[\sqrt{15}(55)'-\sqrt{3}(53)'-\sqrt{14}(51)']$
\\ 
& & $Q^{\beta}_{5z}$ & $\displaystyle \frac{\sqrt{105}}{4}z\left(x^{2}-y^{2}\right)\left(3z^{2}-r^{2}\right)$ & $(52)$
\\
\hline\hline
\end{tabular}
\endgroup
\end{table*}

These four multipole operators are clearly independent with each other under the space-time inversion operations as shown in Fig.~\ref{Fig:multipole}. 
It is readily confirmed by using the facts that $O_{lm} (\bm{r})$ has parity $(-1)^l$ under the spatial inversion, and $\bm{l}$ and $\bm{\sigma}$ are odd under the time-reversal operation. 
$\hat{Q}_{lm}$ and $\hat{T}_{lm}$ in Eqs.~(\ref{eq:Emultipole_Q}) and (\ref{eq:MTmultipole_Q}) (see also the left panels in Fig.~\ref{Fig:multipole}) represent the polar E and MT multipole operators, while $\hat{M}_{lm}$ and $\hat{G}_{lm}$ in Eqs.~(\ref{eq:Mmultipole_Q}) and (\ref{eq:ETmultipole_Q}) (see also the right panels in Fig.~\ref{Fig:multipole}) represent the axial M and ET multipole operators. 
Note that M monopole $\hat{M}_{0}\equiv \hat{M}_{00}$, MT monopole $\hat{T}_{0}\equiv \hat{T}_{00}$, ET monopole $\hat{G}_{0}\equiv \hat{G}_{00}$, and ET dipole $\hat{G}_{1m}\equiv \hat{\bm{G}}=(\hat{G}_{x}, \hat{G}_{y}, \hat{G}_{z})$ vanish in Eqs.~(\ref{eq:Mmultipole_Q})-(\ref{eq:ETmultipole_Q}) owing to the identity $\bm{\nabla}O_{00}(\bm{r})=0$ and $\nabla_\alpha \nabla_\beta O_{1m}(\bm{r})=0$. 

In the spinless basis functions under the rotation group, $(L,M)$ [$L=0(s),1(p),2(d),3(f)$ and $M=-L,-L+1,\cdots,L$] shown in Appendix~\ref{sec:Atomic basis wave functions}, the multipoles given by Eqs.~(\ref{eq:Emultipole_Q})-(\ref{eq:ETmultipole_Q}) without spin contributions constitute a complete set to express an arbitrary degree of freedom\cite{hayami2018microscopic}.
In other words, the even-rank E and odd-rank M multipoles are active in non-hybrid (intra) orbitals, such as $d$-$d$ and $f$-$f$ orbitals.
Meanwhile, the odd(even)-rank E and MT and even(odd)-rank M and ET multipoles are active in odd(even)-parity hybrid (inter) orbitals, such as $p$-$d$ ($s$-$d$) and $d$-$f$ ($p$-$f$) orbitals~\cite{hayami2018microscopic}. 
The total number of these independent active multipoles equals to the number of matrix elements in the relevant Hilbert space.

Since $\hat{X}_{lm}$ transforms like $O_{lm}(\bm{r})$ by the rotational operation, the matrix elements of four multipoles are related with each other.
According to the Wigner-Eckart theorem, the matrix elements can be divided into the purely angular part and common part as
\begin{align}
\braket{L_{1}M_{1}|\hat{X}_{lm}|L_{2}M_{2}}=\braket{L_{1}||\hat{X}_{l}||L_{2}}\braket{L_{1}M_{1}|L_{2}M_{2};lm},
\end{align}
where $\braket{L_{1}||\hat{X}_{l}||L_{2}}$ and $\braket{L_{1}M_{1}|L_{2}M_{2};lm}$ are the reduced matrix element independent of $M_{1}$ and $M_{2}$, and the Clebsch-Gordan coefficient, respectively.
Note that the reduced matrix elements for $\hat{Q}_{lm}$ and $\hat{T}_{lm}$ vanish when those for $\hat{M}_{lm}$ and $\hat{G}_{lm}$ are nonzero, and vice versa (see Table~\ref{tab_activated_multipole} for example).
Therefore, the matrix elements for $\hat{T}_{lm}$ are proportional to those of $\hat{Q}_{lm}$ and the proportional coefficient, $R_{l}(L_{1},L_{2})$, is independent of $M_{1}$ and $M_{2}$, i.e.,
\begin{align}
\braket{L_{1}M_{1}|\hat{T}_{lm}|L_{2}M_{2}}=R_{l}(L_{1},L_{2})\braket{L_{1}M_{1}|\hat{Q}_{lm}|L_{2}M_{2}}.
\end{align}
Similar proportionality holds between $\hat{G}_{lm}$ and $\hat{M}_{lm}$ as
\begin{align}
\braket{L_{1}M_{1}|\hat{G}_{lm}|L_{2}M_{2}}=R_{l}(L_{1},L_{2})\braket{L_{1}M_{1}|\hat{M}_{lm}|L_{2}M_{2}}.
\end{align}
The proportional coefficients turn out to be common as
\begin{align}
R_{l}(L_{1},L_{2})
&=\frac{\braket{L_{1}||\hat{T}_{l}||L_{2}}}{\braket{L_{1}||\hat{Q}_{l}||L_{2}}}
=\frac{\braket{L_{1}||\hat{G}_{l}||L_{2}}}{\braket{L_{1}||\hat{M}_{l}||L_{2}}}
\cr&
=\frac{L_{1}(L_{1}+1)-L_{2}(L_{2}+1)}{(l+1)(l+2)}i.
\end{align}

The result is consistent with the fact that the toroidal multipoles, $\hat{T}_{lm}$ and $\hat{G}_{lm}$, vanish for non-hybrid orbitals, i.e., $L_1 = L_2$.
For the spinful basis as discussed below the similar proportionality also hold, in which the spinless basis $(L,M)$ is replaced by spinful one $(J,J_{z})$.
In other words, we can define arbitrary type of MT and ET multipoles in the total (orbital) angular-momentum basis through the proportionality, and it is sufficient to calculate the matrix elements of the E and M multipoles.

Next, let us extend to the situation where the total angular-momentum basis $(J, J_z)$ is appropriate rather than $(L, M)$ by taking into account the spin degree of freedom. 
In the one-electron state, the total angular momentum $J$ is represented by $J=L-1/2$ or $L+1/2$ and $J_{z}=-J,-J+1,\cdots,J$. 
As similar to the argument for the spinless basis, the active multipoles are uniquely identified for the spinful basis as well from the following conditions: (1) the even-rank E multipoles and the odd-rank M multipoles are active in non-hybrid orbitals, such as $s$-$s$ and $p$-$p$ orbitals with the same total angular momentum $J$. 
(2) the even(odd)-rank E and MT multipoles and odd(even)-rank M and ET multipoles are active in the even(odd)-parity hybrid orbitals with different $J$ or $L$. 
(3) The rank of active multipole $l$ is determined by $|J_1-J_2|\leq l \leq J_1+ J_2$ where $J_1$ and $J_2$ are total angular momenta in the bra- and ket-basis functions. 
For example, in $p$-$d$ orbitals with the same total angular momentum $J=3/2$, the number of the independent active multipoles is 32 as indicated by the 11th row in Table~\ref{tab_activated_multipole}, which consist of M/ET monopole, E/MT dipole, M/ET quadrupole, and E/MT octupole without excess or deficiency. 
We summarize all the active multipoles in the spinful non-hybrid and hybrid orbitals in Table~\ref{tab_activated_multipole}. 
In contrast to the cases for the spinless basis, $\hat{M}_{0}$, $\hat{G}_{0}$, and $\hat{\bm{G}}$ are active in the spinful basis.

\begin{table*}[hbt!]
\begin{center}
\caption{
Active multipoles in non-hybrid (intra) and hybrid (inter) orbitals with the spin degree of freedom. 
The number of independent multipoles within the relevant Hilbert space is indicated in parentheses. The columns of $\mathcal{P}$ and $l$ represent the spatial parity and the rank of the active multipoles, respectively. 
The column of ``orbital'' indicates the atomic orbital that consists of the basis. 
}
\label{tab_activated_multipole}
\begingroup
\renewcommand{\arraystretch}{1.4}
 \begin{tabular}{lccccccccccc}
 \hline \hline
 \multicolumn{2}{l}{basis} & orbital &$\mathcal{P}$ & $l=0$ 
 $(1)$ & $1$ $(3)$ & $2$ $(5)$ & $3$ $(7)$ & $4$ $(9)$ & $5$ $(11)$& $6$ $(13)$ & 7 $(15)$     \\
\hline
1/2-1/2 &(4) & $s$-$s$, $p$-$p$& $+$    &E& M& --&--&-- &--&--&--\\ 
3/2-3/2 &(16)& $p$-$p$, $d$-$d$&           &E&M&E&M&--&--&--&--\\ 
5/2-5/2 &(36)& $d$-$d$, $f$-$f$&         &E&M&E&M&E&M&-- &--     \\ 
7/2-7/2 &(64)& $f$-$f$&         &E&M&E&M&E&M&E &M     \\  \hline
1/2-3/2 &(16)& $s$-$d$, $p$-$p$& $+$   &--&M/ET&E/MT&--&--&--&--&-- \\ 
1/2-5/2 &(24)& $s$-$d$&         &--&---&E/MT&M/ET &--&--&--&--\\ 
3/2-5/2 &(48) &$p$-$f$, $d$-$d$&           &--&M/ET&E/MT&M/ET&E/MT&--&--&--\\ 
3/2-7/2 &(64)& $p$-$f$&           & --&---&E/MT&M/ET&E/MT&M/ET&-- &--  \\
5/2-7/2 &(96)& $f$-$f$&           & --&M/ET&E/MT&M/ET&E/MT&M/ET&E/MT &--  \\ \hline
1/2-1/2 &(8)& $s$-$p$ & $-$    &M/ET& E/MT& --&--&-- &--&--&--\\ 
3/2-3/2 &(32)& $p$-$d$&           &M/ET&E/MT&M/ET&E/MT&--&--&--&--\\ 
5/2-5/2 &(72)& $d$-$f$&         &M/ET&E/MT&M/ET&E/MT&M/ET&E/MT&-- &--     \\ \hline
1/2-3/2 &(16)& $s$-$p$, $p$-$d$& $-$   &--&E/MT&M/ET&--&--&--&--&-- \\ 
1/2-5/2 &(24)&  $s$-$f$, $p$-$d$&     &--&---&M/ET&E/MT &--&--&--&--\\ 
1/2-7/2 &(32)&  $s$-$f$&       &--&---&---&E/MT &M/ET&--&--&--\\ 
3/2-5/2 &(48)& $p$-$d$, $d$-$f$&           &--&E/MT&M/ET&E/MT&M/ET&--&--&--\\ 
3/2-7/2 &(64)&  $d$-$f$&      & --&---&M/ET&E/MT&M/ET&E/MT&-- &--  \\
5/2-7/2 &(96)&  $d$-$f$&    &--&E/MT&M/ET&E/MT&M/ET&E/MT&M/ET&--\\ \hline\hline
\end{tabular}
\endgroup
\end{center}
\end{table*}

The necessity of M and ET monopoles and ET dipole, which do not appear in the multipole expansion, implies that the corresponding quantum-mechanical operators $\hat{M}_{0}$, $\hat{G}_{0}$, and $\hat{\bm{G}}$ are definable in the spinful basis. 
To obtain explicit forms of these multipoles, we focus on the fact that the inner product between the M dipole $\hat{\bm{M}}=\bm{m}_{1}(\bm{r})$ and the position vector $\bm{r}$ has the same symmetry as the M monopole in the rotation group.  
Thus, we define the M monopole as 
\begin{align}
\label{eq:Mmonopole}
\hat{M}_0' &\equiv\sum_{j} \bm{m}_1 (\bm{r}_j) \cdot \bm{r}_j = \sum_{j}\bm{\sigma}_j\cdot \bm{r}_j, 
\end{align}
where we have used $\bm{l}\cdot \bm{r}=0$, and the prime represents that it does not appear in the multipole expansion. 
$\hat{M}_0'$ in Eq.~(\ref{eq:Mmonopole}) clearly exhibits a property of a time-reversal-odd pseudoscalar and is valid only for the spinful basis. 
Similarly, by using the fact that the MT dipole $\hat{\bm{T}}=\bm{t}_{1}(\bm{r})$ has the same spatial inversion property as the E dipole,
it is natural to define a time-reversal-even pseudoscalar $\hat{G}_0'$ as 
\begin{align}
\label{eq:ETmonopole}
\hat{G}_0' &\equiv 6 \sum_j \bm{m}_1 (\bm{r}_j)\cdot \bm{t}_1 (\bm{r}_j) =\sum_j  \bm{l}_j \cdot (\bm{r}_j\times \bm{\sigma}_j),   
\end{align}
where we have used $(\bm{r}\times \bm{l})\cdot \bm{l}=(\bm{r}\times\bm{\sigma})\cdot \bm{\sigma}=0$ and the prefactor $6$ for notational simplicity. 
Since $\hat{G}_{0}'$ acts as an elementary charge for the ET multipoles, the ET dipole $\hat{\bm{G}}'\equiv(\hat{G}_{x}',\hat{G}_{y}',\hat{G}_{z}')$ is obtained by multiplying the position vector $\bm{r}$ to $\hat{G}_0'$ as 
\begin{align}
\label{eq:ETdipole}
\hat{\bm{G}}' &\equiv \sum_{j} (\hat{G}_0')_j \bm{r}_j =\sum_{j} \bm{r}_j \left[ \bm{l}_j \cdot (\bm{r}_j\times \bm{\sigma}_j)\right]. 
\end{align}
It is noteworthy that $\hat{G}_0'$ and $\hat{\bm{G}}'$ are also valid only for the spinful basis. 

On the other hand, the MT monopole is identically zero even by taking an inner product between the MT dipole $\bm{t}_1(\bm{r})$ and the position vector $\bm{r}$, since $\bm{t}_1 (\bm{r})$ is perpendicular to $\bm{r}$. 
This is consistent with the fact that the MT monopole is unnecessary to span an arbitrary matrix in both non-hybrid and hybrid orbitals as shown in Table~\ref{tab_activated_multipole}. 
It is interesting to note that the MT monopole becomes active in hybrid orbitals with different additional quantum number within the same total angular momentum, \textit{e.g.}, $2p$-$4p$ and $3d$-$5d$ orbitals. 
From the correspondence between the E and MT multipoles, the expression of $\hat{T}_0'$ should be given by 
\begin{align}
\hat{T}_{0}'\equiv i,  
\end{align}
which means that $\hat{T}_0$ represents the imaginary hybridization between the different orbitals. 

Finally, let us remark on the further extension of multipoles in the spinful basis. 
In the spinful basis, the scalar product of $\bm{l}$ and $\bm{\sigma}$ can also be regarded as the E monopole, i.e., 
\begin{align}
\label{eq:Emonopole}
\hat{Q}_{0}'\equiv \sum_{j} \bm{l}_j \cdot \bm{\sigma}_j. 
\end{align}
Then, by combining this elementary charge $\hat{Q}_{0}'$ with the polynomial $O_{lm}(\bm{r})$, we obtain the higher-rank E multipoles including the spin degree of freedom. 
Similar extensions for $\hat{M}_{lm}'$, $\hat{T}_{lm}'$, and $\hat{G}_{lm}'$ are possible by combining the elementary charges, $\hat{M}_{0}'$, $\hat{T}_{0}'$, and $\hat{G}_{0}'$ with $O_{lm}(\bm{r})$. 

The proper expressions of the even- and odd-parity multipole operators up to $l=4$ in the cubic $O_{\rm h}$ group are summarized in Tables~\ref{tab_real_Oe} and \ref{tab_real_Oo}.
For the hexagonal $D_{\rm 6h}$ group, the proper expressions are also summarized in Tables~\ref{tab_real_D6e} and \ref{tab_real_D6o} in Appendix~\ref{sec:Electric multipoles and CEF parameters}.
These expressions are sufficient to consider symmetry-classified multipoles in any point groups as they are subgroup of $O_{\rm h}$ or $D_{\rm 6h}$ groups. 

\begin{table*}[htb!]
\begin{center}
\caption{
Operator expressions of even-parity multipoles up to $l=4$ in the cubic $O_{\rm h}$ group.
For a noncommute product, a symmetrized expression should be used, e.g., $AB\to (AB+B^{\dagger}A^{\dagger})/2$.
$\bm{m}_{l}$, $\bm{t}_{l}$, and $g^{\alpha\beta}_{l}$ are defined in Eqs.~(\ref{eq:m})-(\ref{eq:g}). 
The multipole with the prime requires the complementary definitions other than Eqs.~(\ref{eq:Emultipole_Q})-(\ref{eq:ETmultipole_Q}). 
}
\label{tab_real_Oe}
\begingroup
\renewcommand{\arraystretch}{1.25}
 \begin{tabular}{ccccc}
 \hline \hline
rank &type & irrep. & symbol & definition \\ \hline
$0$ & E  & ${\rm A}_{1g}^{+}$ & $Q_0$ & $1$ \\ 
  & MT  & ${\rm A}_{1g}^{-}$ & $T_0'$ & $i$ \\ 
   \hline
$1$ & M  & ${\rm T}_{1g}^{-}$ & $M_x$, $M_y$, $M_z$ & $m_{1}^{x}$, $m_{1}^{y}$, $m_{1}^{z}$ \\
    & ET & ${\rm T}_{1g}^{+}$ & $G_x'$, $G_y'$, $G_z'$ & $x\bm{l}\cdot (\bm{r}\times \bm{\sigma})$, $y\bm{l}\cdot (\bm{r}\times \bm{\sigma})$, $z\bm{l}\cdot (\bm{r}\times \bm{\sigma})$ \\ \hline
$2$ & E  & ${\rm E}_{g}^{+}$ & $Q_{u}$, $Q_{v}$ & $\displaystyle \frac{1}{2}(3z^2-r^2)$, $\displaystyle \frac{\sqrt{3}}{2}(x^2-y^2)$ \\
    &    & ${\rm T}_{2g}^{+}$ & $Q_{yz}$, $Q_{zx}$, $Q_{xy}$ & $\sqrt{3}yz$, $\sqrt{3}zx$, $\sqrt{3}xy$ \\
    & MT & ${\rm E}_{g}^{-}$ & $T_{u}$, $T_{v}$ & $3zt_{2}^{z}-\bm{r}\cdot \bm{t}_2$, $\sqrt{3} (xt_{2}^{x}-yt_{2}^{y})$ \\
    &    & ${\rm T}_{2g}^{-}$ & $T_{yz}$, $T_{zx}$, $T_{xy}$ & $\sqrt{3}(yt_{2}^{z}+zt_{2}^{y})$, $\sqrt{3}(zt_{2}^{x}+xt_{2}^{z})$, $\sqrt{3}(xt_{2}^{y}+yt_{2}^{x})$ \\ 
\hline
3 & M & ${\rm A}_{2g}^{-}$ & $M_{xyz}$ & $\sqrt{15}(yzm_{3}^{x}+zxm_{3}^{y}+xym_{3}^{z})$ \\
  &   &  ${\rm T}_{1g}^{-}$ & $M_{x}^{\alpha}$, $M_{y}^{\alpha}$, $M_{z}^{\alpha}$ & $\displaystyle 3\left[\frac{1}{2}(3x^{2}-r^{2})m_{3}^{x}-x(ym_{3}^{y}+zm_{3}^{z})\right]$, (cyclic) \\
  &   & ${\rm T}_{2g}^{-}$ & $M_{x}^{\beta}$, $M_{y}^{\beta}$, $M_{z}^{\beta}$ & $\displaystyle \sqrt{15}\left[\frac{1}{2}(y^{2}-z^{2})m_{3}^{x}+x(ym_{3}^{y}-zm_{3}^{z})\right]$, (cyclic) \\
  & ET & ${\rm A}_{2g}^{+}$ & $G_{xyz}$ & $2\sqrt{15}(xg_{3}^{yz}+yg_{3}^{zx}+zg_{3}^{xy})$ \\
  &   & ${\rm T}_{1g}^{+}$ & $G_{x}^{\alpha}$, $G_{y}^{\alpha}$, $G_{z}^{\alpha}$ & $\displaystyle 9xg_{3}^{xx}-6(yg_{3}^{xy}+zg_{3}^{zx})-3x\sum_{\alpha}g_{3}^{\alpha\alpha}$, (cyclic) \\
  &   & ${\rm T}_{2g}^{+}$ & $G_{x}^{\beta}$, $G_{y}^{\beta}$, $G_{z}^{\beta}$ & $\sqrt{15}[2(yg_{3}^{xy}-zg_{3}^{zx})+x(g_{3}^{yy}-g_{3}^{zz})]$, (cyclic) \\
\hline
$4$ & E & ${\rm A}_{1g}^{+}$ & $Q_{4}$ & $\displaystyle  \frac{5\sqrt{21}}{12}\left(x^4+y^4+z^4-\frac{3}{5} r^4\right)$ \\
   &  & ${\rm E}_{g}^{+}$ & $Q_{4u}$, $Q_{4v}$ & $\displaystyle  \frac{7\sqrt{15}}{6}\left[z^4-\frac{x^4+y^4}{2}-\frac{3}{7} r^2(3z^2-r^2)\right]$, $\displaystyle  \frac{7\sqrt{5}}{4}\left[x^4-y^4-\frac{6}{7}r^2 (x^2-y^2)\right]$ \\
  &  &  ${\rm T}_{1g}^{+}$ & $Q^{\alpha}_{4x}$, $Q^{\alpha}_{4y}$, $Q^{\alpha}_{4z}$ & $\displaystyle \frac{\sqrt{35}}{2}y z(y^2-z^2)$, (cyclic) \\
  &   &  ${\rm T}_{2g}^{+}$ & $Q^{\beta}_{4x}$, $Q^{\beta}_{4y}$, $Q^{\beta}_{4z}$ & $\displaystyle \frac{\sqrt{5}}{2}yz(7x^2-r^2)$, (cyclic) \\
   & MT & ${\rm A}_{1g}^{-}$ & $T_{4}$ & $\displaystyle  \frac{5\sqrt{21}}{3}\left[x^3 t^x_4+y^3 t^y_4+z^3 t^z_4-\frac{3}{5} r^2 (\bm{r}\cdot \bm{t}_4)\right]$ \\
   &  &  ${\rm E}_{g}^{-}$ & $T_{4u}$ & $\displaystyle \frac{\sqrt{15}}{3}\left[ \sum_\alpha \alpha (5 \alpha^2-3r^2)t^\alpha_4 -3 x (x^2-3y^2)t^x_4 + 3y (3x^2-y^2)t^y_4\right]$ \\
   &  &   & $T_{4v}$ &$\displaystyle \sqrt{5}[x(x^2-3z^2)t^x_4 -y (y^2-3z^2)t^y_4-3z (x^2-y^2)t^z_4]$ \\
  &     & ${\rm T}_{1g}^{-}$ & $T^{\alpha}_{4x}$, $T^{\alpha}_{4y}$, $T^{\alpha}_{4z}$ & $\displaystyle \frac{\sqrt{35}}{2} [z(3y^2-z^2)t^y_4 + y (y^2-3z^2)t^z_4]$, (cyclic) \\
  &    & ${\rm T}_{2g}^{-}$ & $T^{\beta}_{4x}$, $T^{\beta}_{4y}$, $T^{\beta}_{4z}$ & $\displaystyle 6\sqrt{5}xyz t_4^x +\frac{\sqrt{5}}{2}(7x^2-r^2)(z t^y_4 + y t^z_4)-\sqrt{5} yz (y t^y_4 + z t^z_4)$, (cyclic) \\
\hline\hline
\end{tabular}
\endgroup
\end{center}
\end{table*}

\begin{table*}[htb!]
\begin{center}
\caption{
Operator expressions of odd-parity multipoles up to $l=4$ in the cubic $O_{\rm h}$ group.
}
\label{tab_real_Oo}
\begingroup
\renewcommand{\arraystretch}{1.25}
 \begin{tabular}{ccccc}
 \hline \hline
rank & type & irrep. & symbol & definition \\ \hline
$0$ & M  & ${\rm A}_{1u}^{-}$ & $M_0'$ & $\bm{r}\cdot \bm{\sigma}$ \\ 
    & ET  & ${\rm A}_{1u}^{+}$ & $G_0'$ & $\bm{l}\cdot (\bm{r}\times \bm{\sigma})$ \\ 
   \hline
$1$ & E  & ${\rm T}_{1u}^{+}$ & $Q_x$, $Q_y$, $Q_z$ & $x$, $y$, $z$ \\
    & MT & ${\rm T}_{1u}^{-}$ & $T_x$, $T_y$, $T_z$ & $t_{1}^{x}$, $t_{1}^{y}$, $t_{1}^{z}$ \\ 
\hline
$2$ & M  & ${\rm E}_{u}^{-}$ & $M_{u}$, $M_{v}$ & $3zm_{2}^{z}-\bm{r}\cdot \bm{m}_2$, $\sqrt{3} (xm_{2}^{x}-ym_{2}^{y})$ \\
    &    & ${\rm T}_{2u}^{-}$ & $M_{yz}$, $M_{zx}$, $M_{xy}$ & $\sqrt{3}(ym_{2}^{z}+zm_{2}^{y})$, $\sqrt{3}(zm_{2}^{x}+xm_{2}^{z})$, $\sqrt{3}(xm_{2}^{y}+ym_{2}^{x})$ \\
    & ET & ${\rm E}_{u}^{+}$ & $G_{u}$, $G_{v}$ & $3g_{2}^{zz} -\sum_{\alpha}g_{2}^{\alpha\alpha}$, $\sqrt{3}(g_{2}^{xx}-g_{2}^{yy})$ \\
    &    & ${\rm T}_{2u}^{+}$ & $G_{yz}$, $G_{zx}$, $G_{xy}$ & $2\sqrt{3}g_{2}^{yz}$, $2\sqrt{3}g_{2}^{zx}$, $2\sqrt{3}g_{2}^{xy}$ \\
\hline
3 & E & ${\rm A}_{2u}^{+}$ & $Q_{xyz}$ & $\sqrt{15}xyz$ \\
  &   & ${\rm T}_{1u}^{+}$ & $Q_{x}^{\alpha}$, $Q_{y}^{\alpha}$, $Q_{z}^{\alpha}$ & $\displaystyle \frac{1}{2}x(5x^{2}-3r^{2})$, (cyclic) \\
  &   & ${\rm T}_{2u}^{+}$ & $Q_{x}^{\beta}$, $Q_{y}^{\beta}$, $Q_{z}^{\beta}$ & $\displaystyle \frac{\sqrt{15}}{2}x(y^{2}-z^{2})$, (cyclic) \\
  & MT & ${\rm A}_{2u}^{-}$ & $T_{xyz}$ & $\sqrt{15}(yzt_{3}^{x}+zxt_{3}^{y}+xyt_{3}^{z})$ \\
  &   & ${\rm T}_{1u}^{-}$ & $T_{x}^{\alpha}$, $T_{y}^{\alpha}$, $T_{z}^{\alpha}$ & $\displaystyle 3\left[\frac{1}{2}(3x^{2}-r^{2})t_{3}^{x}-x(yt_{3}^{y}+zt_{3}^{z})\right]$, (cyclic) \\
  &   & ${\rm T}_{2u}^{-}$ & $T_{x}^{\beta}$, $T_{y}^{\beta}$, $T_{z}^{\beta}$ & $\displaystyle \sqrt{15}\left[\frac{1}{2}(y^{2}-z^{2})t_{3}^{x}+x(yt_{3}^{y}-zt_{3}^{z})\right]$, (cyclic) \\
\hline
$4$ & M & ${\rm A}_{1u}^{-}$ & $M_{4}$ & $\displaystyle  \frac{5\sqrt{21}}{3}\left[x^3 m^x_4+y^3 m^y_4+z^3 m^z_4-\frac{3}{5} r^2 (\bm{r}\cdot \bm{m}_4)\right]$ \\
   &  & ${\rm E}_{u}^{-}$ & $M_{4u}$ & $\displaystyle \frac{\sqrt{15}}{3}\left[ \sum_\alpha \alpha (5 \alpha^2-3r^2)m^\alpha_4 -3 x (x^2-3y^2)m^x_4 + 3y (3x^2-y^2)m^y_4\right]$ \\
      &  & & $M_{4v}$ & $\displaystyle \sqrt{5}[x(x^2-3z^2)m^x_4 -y (y^2-3z^2)m^y_4-3z (x^2-y^2)m^z_4]$ \\
  &    & ${\rm T}_{1u}^{-}$ & $M^{\alpha}_{4x}$, $M^{\alpha}_{4y}$, $M^{\alpha}_{4z}$ & $\displaystyle \frac{\sqrt{35}}{2} [z(3y^2-z^2)m^y_4 + y (y^2-3z^2)m^z_4]$, (cyclic) \\
  &   & ${\rm T}_{2u}^{-}$ & $M^{\beta}_{4x}$, $M^{\beta}_{4y}$, $M^{\beta}_{4z}$ & $\displaystyle 6\sqrt{5}xyz m_4^x +\frac{\sqrt{5}}{2}(7x^2-r^2)(z m^y_4 + y m^z_4)-\sqrt{5} yz (y m^y_4 + z m^z_4)$, (cyclic) \\
   & ET & ${\rm A}_{1u}^{+}$ & $G_{4}$ & $\displaystyle -4 \sqrt{21}(xy g^{xy}_4+ yz g^{yz}_4+zx g^{zx}_4)+\sqrt{21} \sum_{\alpha} (3 \alpha^2 -r^2) g^{\alpha\alpha}_4$ \\
   &  & ${\rm E}_{u}^{+}$ & $G_{4u}$ & $\sqrt{15}[(3y^2-r^2)g^{xx}_4+(3x^2-r^2)g^{yy}_4+(3z^2-r^2)g^{zz}_4+4(2 x y g^{xy}_4- x z g^{xz}_4- y z g^{yz}_4)]$\\
   &  &  & $G_{4v}$ & $3\sqrt{5}[(x^2-z^2)g^{xx}_4+(z^2-y^2)g^{yy}_4+(y^2-x^2)g^{zz}_4-4(xz g^{xz}_4-yz g^{yz}_4)]$ \\
  &  &  ${\rm T}_{1u}^{+}$ & $G^{\alpha}_{4x}$, $G^{\alpha}_{4y}$, $G^{\alpha}_{4z}$ & $3\sqrt{35}[y z (g^{yy}_4 -g^{zz}_4)+ (y^2-z^2) g^{yz}_4]$, (cyclic) \\
  &    & ${\rm T}_{2u}^{+}$ & $G^{\beta}_{4x}$, $G^{\beta}_{4y}$, $G^{\beta}_{4z}$ & $3\sqrt{5}[4 x(y g^{xz}_4+ z g^{xy}_4 )+ yz (2 g^{xx}_4 -g^{yy}_4 - g^{zz}_4)+ (3x^2-r^2)g^{yz}_4]$, (cyclic) \\
\hline\hline
\end{tabular}
\endgroup
\end{center}
\end{table*}

\subsection{Multipoles in momentum space}
\label{sec:Multipoles in momentum space}

The classification of multipoles is also applicable in momentum space. 
We present the momentum-based multipoles in the single-band systems, which are expressed in terms of ($\bm{k}$, $\sigma_{0}$, $\bm{\sigma}$) where $\bm{k}$ is the wave vector of electrons, and $\sigma_{0}$ and $\bm{\sigma}$ are the identity and Pauli matrices in the spin degree of freedom. 
Extension to multi-band systems are not unique. 
We discuss a possible extension in Appendix~\ref{sec:k-space multipoles in multi-orbital systems}. 

To construct the multipoles in momentum space, we first introduce the harmonics $O_{lm}(\bm{k})$ by replacing the position vector $\bm{r}$ with the wave vector $\bm{k}$ in $O_{lm}(\bm{r})$.
Reflecting the polar vector of $\bm{k}$, the harmonics $\sigma_{0}O_{lm}(\bm{k})$ represent the polar-type E and MT multipole degrees of freedom. 
As $\bm{k}$ is odd under time-reversal operation, the even-rank $\sigma_{0}O_{lm} (\bm{k})$ represents the E multipoles, while the odd-rank $\sigma_{0}O_{lm}(\bm{k})$ represents the MT multipoles. 
By multiplying $O_{lm}(\bm{k})$ by the operator $(\bm{k}\times \bm{\sigma})\cdot \bm{\nabla}_{\bm{k}}$ that reverses its time-reversal property with keeping its parity and rank, the expressions of the odd-rank E multipole and the even-rank MT multipole are obtained. 
Thus, the expressions of E and MT multipole in momentum space are given by 
\begin{align}
&
Q_{lm}(\bm{k})\equiv\begin{cases}
\sigma_{0}O_{lm}(\bm{k}) & (l=0,2,4,6,\cdots) \\
(\bm{k}\times\bm{\sigma})\cdot\bm{\nabla}_{\bm{k}}O_{lm}(\bm{k}) & (l=1,3,5,\cdots)
\end{cases}
\\&
T_{lm}(\bm{k})\equiv\begin{cases}
0 & (l=0) \\
(\bm{k}\times\bm{\sigma})\cdot\bm{\nabla}_{\bm{k}}O_{lm}(\bm{k}) & (l=2,4,6,\cdots) \\
\sigma_{0}O_{lm}(\bm{k}) & (l=1,3,5,\cdots)
\end{cases}
\end{align}
Note that $T_{0}$ vanishes owing to the derivative of $O_{lm}(\bm{k})$. 

Similarly, the axial-type M and ET multipoles in momentum space is expressed by multiplying $\sigma_{0}O_{lm}(\bm{k})$ by the operator $\bm{\sigma}\cdot \bm{\nabla}_{\bm{k}}$ that reverses its parity with keeping its time-reversal property and rank. 
Therefore, the odd-rank M multipoles and the even-rank ET multipoles are given by $\bm{\sigma}\cdot \bm{\nabla}_{\bm{k}}O_{lm}(\bm{k})$. 
On the other hand, it is impossible to construct the even-rank M multipoles and the odd-rank ET multipoles within the single-band systems. 
They require the multi-orbital or sublattice degree of freedom, as discussed in Appendix~\ref{sec:k-space multipoles in multi-orbital systems}.
Thus, the M and ET multipoles are given by 
\begin{align}
&
M_{lm}(\bm{k})\equiv\begin{cases}
0 & (l=0,2,4,6,\cdots) \\ 
\bm{\sigma}\cdot\bm{\nabla}_{\bm{k}}O_{lm}(\bm{k}) & (l=1,3,5,\cdots)
\end{cases}
\label{eq:mmom}
\\&
G_{lm}(\bm{k})\equiv\begin{cases}
\bm{k}\cdot\bm{\sigma} & (l=0) \\
\bm{\sigma}\cdot\bm{\nabla}_{\bm{k}}O_{lm}(\bm{k}) & (l=2,4,6,\cdots) \\
0 & (l=1,3,5,\cdots) 
\end{cases}
\label{eq:etmom}
\end{align}
We complementarily define $G_0(\bm{k})$ as $\bm{k}\cdot \bm{\sigma}$ from the symmetry consideration. 
The even- and odd-parity multipole operators in momentum space in the $O_{\rm h}$ group up to $l=4$ are summarized in Tables~\ref{tab_k-multipole-e} and \ref{tab_k-multipole-o}.
The hexagonal version is summarized in Tables~\ref{tab_k-multipole_D6e} and \ref{tab_k-multipole_D6o} in Appendix~\ref{sec:Electric multipoles and CEF parameters}.

\begin{table*}[htb!]
\caption{
Even-parity multipoles in momentum space up to $l=4$ in the cubic $O_{\rm h}$ group.
The higher-order expression with respect to $\bm{k}$ is also shown for $Q_0$ and $(M_{x}, M_{y}, M_{z})$ in the bracket. 
The multipole with the prime becomes active in multi-orbital systems as discussed in Appendix~\ref{sec:k-space multipoles in multi-orbital systems}. 
}
\label{tab_k-multipole-e}
\centering
\begingroup
\renewcommand{\arraystretch}{1.4}
 \begin{tabular}{ccccc}
 \hline \hline
rank & type & irrep. & symbol & definition \\ \hline
$0$ & E & ${\rm A}_{1g}^{+}$ & $Q_0$ & $\sigma_{0}$ [$(k_x^2+k_y^2+k_z^2)\sigma_0$] \\
    & MT & ${\rm A}^-_{1g}$ & $T_0'$ & $i \sigma_0$ \\ \hline
$1$ & M & ${\rm T}_{1g}^{-}$ & $M_x$, $M_y$, $M_z$ & $ \sigma_{x}$, $\sigma_{y}$, $\sigma_{z}$ $\displaystyle \left[(\bm{k}\cdot \bm{\sigma})k_x-\frac{1}{3}k^2\sigma_x, {\rm (cyclic)} \right]$ \\ 
    & ET  & ${\rm T}^+_{1g}$ & $G_x'$, $G_y'$, $G_z'$ & $i \sigma_x$, $i \sigma_y$, $i \sigma_z$ \\
\hline
$2$ & E & ${\rm E}_{g}^{+}$ & $Q_{u}$, $Q_{v}$ & $\displaystyle \frac{1}{2}(3k_{z}^2-k^2)\sigma_{0}$, $\displaystyle \frac{\sqrt{3}}{2}(k_{x}^2-k_{y}^2)\sigma_{0}$ \\
    &      & ${\rm T}_{2g}^{+}$ & $Q_{yz}$, $Q_{zx}$, $Q_{xy}$ & $\sqrt{3}k_{y}k_{z}\sigma_{0}$, (cyclic) \\
    & MT & ${\rm E}_{g}^{-}$ & $T_{u}$, $T_{v}$ & $3k_{z}Q_{z}-\bm{k}\cdot\bm{Q}$, $\sqrt{3}(k_{x}Q_{x}-k_{y}Q_{y})$ \\
    &     & ${\rm T}_{2g}^{-}$ & $T_{yz}$, $T_{zx}$, $T_{xy}$ & $\sqrt{3}(k_{y}Q_{z}+k_{z}Q_{y})$, (cyclic) \\ \hline
3 & M & ${\rm A}_{2g}^{-}$ & $M_{xyz}$ & $\sqrt{15}(k_{y}k_{z}\sigma_{x}+k_{z}k_{x}\sigma_{y}+k_{x}k_{y}\sigma_{z})$ \\
  &   & ${\rm T}_{1g}^{-}$ & $M_{x}^{\alpha}$, $M_{y}^{\alpha}$, $M_{z}^{\alpha}$ & $\displaystyle 3\left[\frac{1}{2}(3k_{x}^{2}-k^{2})\sigma_{x}-k_{x}(k_{y}\sigma_{y}+k_{z}\sigma_{z})\right]$, (cyclic) \\
  &   & ${\rm T}_{2g}^{-}$ & $M_{x}^{\beta}$, $M_{y}^{\beta}$, $M_{z}^{\beta}$ & $\displaystyle \sqrt{15}\left[\frac{1}{2}(k_{y}^{2}-k_{z}^{2})\sigma_{x}+k_{x}(k_{y}\sigma_{y}-k_{z}\sigma_{z})\right]$, (cyclic) \\
  & ET &${\rm A}^+_{2g}$ & $G_{xyz}'$ & $\sqrt{15}i(k_{y}k_{z}\sigma_{x}+k_{z}k_{x}\sigma_{y}+k_{x}k_{y}\sigma_{z})$ \\
  & & ${\rm T}^+_{1g}$ & $G_{x}^{\alpha'}$, $G_{y}^{\alpha'}$, $G_{z}^{\alpha'}$ & $\displaystyle 3i\left[\frac{1}{2}(3k_{x}^{2}-k^{2})\sigma_{x}-k_{x}(k_{y}\sigma_{y}+k_{z}\sigma_{z})\right]$, (cyclic) \\
  & & ${\rm T}^+_{2g}$ & $G_{x}^{\beta'}$, $G_{y}^{\beta'}$, $G_{z}^{\beta'}$ & $\displaystyle \sqrt{15}i\left[\frac{1}{2}(k_{y}^{2}-k_{z}^{2})\sigma_{x}+k_{x}(k_{y}\sigma_{y}-k_{z}\sigma_{z})\right]$, (cyclic) \\
\hline
$4$ & E & ${\rm A}_{1g}^{+}$ & $Q_{4}$ & $\displaystyle \frac{5\sqrt{21}}{12}\left(k_x^4+k_y^4+k_z^4-\frac{3}{5} k^4\right)\sigma_{0}$ \\
&   & ${\rm E}_{g}^{+}$ & $Q_{4u}$, $Q_{4v}$ & $\displaystyle \frac{7\sqrt{15}}{6}\left[k_z^4-\frac{k_x^4+k_y^4}{2}-\frac{3}{7} k^2(3k_z^2-k^2)\right]\sigma_{0}$, $\displaystyle \frac{7\sqrt{5}}{4}\left[k_x^4-k_y^4-\frac{6}{7}k^2 (k_x^2-k_y^2)\right]\sigma_{0}$ \\
    &   & ${\rm T}_{1g}^{+}$ & $Q^{\alpha}_{4x}$, $Q^{\alpha}_{4y}$, $Q^{\alpha}_{4z}$ & $\displaystyle \frac{\sqrt{35}}{2}k_y k_z(k_y^2-k_z^2)\sigma_{0}$, (cyclic) \\
        &     & ${\rm T}_{2g}^{+}$ & $Q^{\beta}_{4x}$, $Q^{\beta}_{4y}$, $Q^{\beta}_{4z}$ & $\displaystyle \frac{\sqrt{5}}{2}k_{y}k_{z}(7k_x^2-k^2)\sigma_{0}$, (cyclic) \\
 & MT  & ${\rm A}_{1g}^{-}$ & $T_{4}$ & $\displaystyle \frac{5\sqrt{21}}{3}\left[k_x \left( k_x^2-\frac{3}{5} k^2 \right)Q_x+k_y \left( k_y^2-\frac{3}{5} k^2 \right)Q_y+k_z \left( k_z^2-\frac{3}{5} k^2 \right)Q_z\right]$ \\
&   & ${\rm E}_{g}^{-}$ & $T_{4u}$ & $\displaystyle \frac{\sqrt{15}}{3}\left[ \sum_\alpha k_\alpha (5 k_\alpha^2-3k^2)Q_\alpha -3 k_x (k_x^2-3k_y^2)Q_x + 3 k_y (3k_x^2-k_y^2)Q_y\right]$ \\
&   & & $T_{4v}$ & $\displaystyle \sqrt{5}[k_x(k_x^2-3k_z^2)Q_x -k_y (k_y^2-3k_z^2)Q_y-3k_z (k_x^2-k_y^2)Q_z]$ \\
      &     & ${\rm T}_{1g}^{-}$ & $T^{\alpha}_{4x}$, $T^{\alpha}_{4y}$, $T^{\alpha}_{4z}$ & $\displaystyle \frac{\sqrt{35}}{2}\left[ k_y (k_y^2-3k_z^2)Q_z-k_z (k_z^2-3 k_y^2)Q_y \right]$, (cyclic) \\
      &     & ${\rm T}_{2g}^{-}$ & $T^{\beta}_{4x}$, $T^{\beta}_{4y}$, $T^{\beta}_{4z}$ & $\displaystyle \frac{\sqrt{5}}{2} [12 k_x k_y k_z Q_x +k_z \{(5k_z^2-3k^2)-3(k_z^2-3k_x^2)\}Q_y + k_y \{(5k_y^2-3k^2)-3(k_y^2-3k_x^2) \}Q_z]$, (cyclic)  \\
\hline\hline
\end{tabular}
\endgroup
\end{table*}

\begin{table*}[htb!]
\caption{
Odd-parity multipoles in momentum space up to $l=4$ in the cubic $O_{\rm h}$ group.
}
\label{tab_k-multipole-o}
\centering
\begingroup
\renewcommand{\arraystretch}{1.4}
 \begin{tabular}{ccccc}
 \hline \hline
rank & type & irrep. & symbol & definition \\ \hline
$0$ & M  & ${\rm A}^-_{1u}$ & $M_0'$ & $i \bm{k} \cdot \bm{\sigma}$ \\
    & ET & ${\rm A}_{1u}^{+}$ & $G_0$ & $\bm{k}\cdot\bm{\sigma}$ \\ \hline
$1$ & E  & ${\rm T}_{1u}^{+}$ & $Q_x$, $Q_y$, $Q_z$ & $k_{y}\sigma_{z}-k_{z}\sigma_{y}$, $k_{z}\sigma_{x}-k_{x}\sigma_{z}$, $k_{x}\sigma_{y}-k_{y}\sigma_{x}$ \\
    & MT & ${\rm T}_{1u}^{-}$ & $T_x$, $T_y$, $T_z$ & $k_{x}\sigma_{0}$, $k_{y}\sigma_{0}$, $k_{z}\sigma_{0}$ \\ \hline
$2$ & M  & ${\rm E}^-_u$ & $M_{u}'$, $M_{v}'$ & $i(3k_{z}\sigma_{z}-\bm{k}\cdot\bm{\sigma})$, $\sqrt{3}i (k_{x}\sigma_{x}-k_{y}\sigma_{y})$ \\
    &     & ${\rm T}^-_{2u}$ & $M_{yz}'$, $M_{zx}'$, $M_{xy}'$ & $\sqrt{3}i (k_{y}\sigma_{z}+k_{z}\sigma_{y})$, (cyclic) \\
    & ET & ${\rm E}_{u}^{+}$ & $G_{u}$, $G_{v}$ & $3k_{z}\sigma_{z}-\bm{k}\cdot\bm{\sigma}$, $\sqrt{3}(k_{x}\sigma_{x}-k_{y}\sigma_{y})$ \\
    &    & ${\rm T}_{2u}^{+}$ & $G_{yz}$, $G_{zx}$, $G_{xy}$ & $\sqrt{3}(k_{y}\sigma_{z}+k_{z}\sigma_{y})$, (cyclic) \\ \hline
3 & E & ${\rm A}_{2u}^{+}$ & $Q_{xyz}$ & $\sqrt{15}(k_{y}k_{z}Q_{x}+k_{z}k_{x}Q_{y}+k_{x}k_{y}Q_{z})$ \\
  &   & ${\rm T}_{1u}^{+}$ & $Q_{x}^{\alpha}$, $Q_{y}^{\alpha}$, $Q_{z}^{\alpha}$ & $\displaystyle 3\left[\frac{1}{2}(3k_{x}^{2}-k^{2})Q_{x}-k_{x}(k_{y}Q_{y}+k_{z}Q_{z})\right]$, (cyclic) \\
  &   & ${\rm T}_{2u}^{+}$ & $Q_{x}^{\beta}$, $Q_{y}^{\beta}$, $Q_{z}^{\beta}$ & $\displaystyle \sqrt{15}\left[\frac{1}{2}(k_{y}^{2}-k_{z}^{2})Q_{x}+k_{x}(k_{y}Q_{y}-k_{z}Q_{z})\right]$, (cyclic) \\
  & MT & ${\rm A}_{2u}^{-}$ & $T_{xyz}$ & $\sqrt{15}k_{x}k_{y}k_{z}\sigma_{0}$ \\
  &   & ${\rm T}_{1u}^{-}$ & $T_{x}^{\alpha}$, $T_{y}^{\alpha}$, $T_{z}^{\alpha}$ & $\displaystyle \frac{1}{2}k_{x}(5k_{x}^{2}-3k^{2})\sigma_{0}$, (cyclic) \\
  &   & ${\rm T}_{2u}^{-}$ & $T_{x}^{\beta}$, $T_{y}^{\beta}$, $T_{z}^{\beta}$ & $\displaystyle \frac{\sqrt{15}}{2}k_{x}(k_{y}^{2}-k_{z}^{2})\sigma_{0}$, (cyclic) \\
\hline
$4$ & M  & ${\rm A}^-_{1u}$ & $M_{4}'$ & $\displaystyle \frac{5\sqrt{21}}{3}i\left[k_x \left( k_x^2-\frac{3}{5} k^2 \right)\sigma_x+k_y \left( k_y^2-\frac{3}{5} k^2 \right)\sigma_y+k_z \left( k_z^2-\frac{3}{5} k^2 \right)\sigma_z\right]$ \\
&   &${\rm E}^-_u$ & $M_{4u}'$ & $\displaystyle \frac{\sqrt{15}}{3}i\left[ \sum_\alpha k_\alpha (5 k_\alpha^2-3k^2)\sigma_\alpha -3 k_x (k_x^2-3k_y^2)\sigma_x + 3 k_y (3k_x^2-k_y^2)\sigma_y\right]$\\
&   & & $M_{4v}'$ & $\displaystyle \sqrt{5}i[k_x(k_x^2-3k_z^2)\sigma_x -k_y (k_y^2-3k_z^2)\sigma_y-3k_z (k_x^2-k_y^2)\sigma_z]$\\
    &   & ${\rm T}^-_{1u}$ & $M^{\alpha'}_{4x}$, $M^{\alpha'}_{4y}$, $M^{\alpha'}_{4z}$ & $\displaystyle \frac{\sqrt{35}}{2}i \left[ k_y (k_y^2-3k_z^2)\sigma_z-k_z (k_z^2-3 k_y^2)\sigma_y \right]$, (cyclic) \\
        &     & ${\rm T}^-_{2u}$ & $M^{\beta'}_{4x}$, $M^{\beta'}_{4y}$, $M^{\beta'}_{4z}$ & $\displaystyle \frac{\sqrt{5}}{2}i [12 k_x k_y k_z \sigma_x +k_z \{(5k_z^2-3k^2)-3(k_z^2-3k_x^2)\}\sigma_y + k_y \{(5k_y^2-3k^2)-3(k_y^2-3k_x^2) \}\sigma_z]$, (cyclic) \\
    & ET  & ${\rm A}_{1u}^{+}$ & $G_{4}$ & $\displaystyle \frac{5\sqrt{21}}{3}\left[k_x \left( k_x^2-\frac{3}{5} k^2 \right)\sigma_x+k_y \left( k_y^2-\frac{3}{5} k^2 \right)\sigma_y+k_z \left( k_z^2-\frac{3}{5} k^2 \right)\sigma_z\right]$ \\
&   &  ${\rm E}_{u}^{+}$ & $G_{4u}$ & $\displaystyle \frac{\sqrt{15}}{3}\left[ \sum_\alpha k_\alpha (5 k_\alpha^2-3k^2)\sigma_\alpha -3 k_x (k_x^2-3k_y^2)\sigma_x + 3 k_y (3k_x^2-k_y^2)\sigma_y\right]$\\
&   &  & $G_{4v}$ & $\displaystyle \sqrt{5}[k_x(k_x^2-3k_z^2)\sigma_x -k_y (k_y^2-3k_z^2)\sigma_y-3k_z (k_x^2-k_y^2)\sigma_z]$ \\
    &  & ${\rm T}_{1u}^{+}$ & $G^{\alpha}_{4x}$, $G^{\alpha}_{4y}$, $G^{\alpha}_{4z}$ &  $\displaystyle \frac{\sqrt{35}}{2}\left[ k_y (k_y^2-3k_z^2)\sigma_z-k_z (k_z^2-3 k_y^2)\sigma_y \right]$, (cyclic) \\
       &     & ${\rm T}_{2u}^{+}$ & $G^{\beta}_{4x}$, $G^{\beta}_{4y}$, $G^{\beta}_{4z}$ & $\displaystyle \frac{\sqrt{5}}{2} [12 k_x k_y k_z \sigma_x +k_z \{(5k_z^2-3k^2)-3(k_z^2-3k_x^2)\}\sigma_y + k_y \{(5k_y^2-3k^2)-3(k_y^2-3k_x^2) \}\sigma_z]$, (cyclic) \\
\hline\hline
\end{tabular}
\endgroup
\end{table*}

\section{Physical properties in the presence of multipoles}
\label{sec:Physical properties under multipoles}

When a thermodynamic average of a multipole degree of freedom $X_{lm}\equiv\braket{\hat{X}_{lm}}$, which we call a multipole moment, becomes nonzero, it generates anisotropic electric and/or magnetic fields via electromagnetic potentials.
Moreover, it affects electronic states and lead to unconventional physical phenomena, such as the magneto-current effect through multipole-multipole interactions. 
In this section, we first give the formula for electric and magnetic fields in the presence of the multipole moment in Sec.~\ref{sec:Electromagnetic potential, and electric and magnetic fields}.
Then, we show how the multipole moment affects the electronic band structures in Sec.~\ref{sec:Electronic band structure under multipoles}. 
We also classify linear response tensors, such as magneto-electric and magneto-current tensors, from the multipole point of view in Sec.~\ref{sec:Linear response tensors}.

\subsection{Electromagnetic potential, and electric and magnetic fields}
\label{sec:Electromagnetic potential, and electric and magnetic fields}

By the definition of the multipole expansion of the scalar and vector potentials with the Coulomb gauge $\bm{\nabla}\cdot\bm{A}=0$\cite{hayami2018microscopic}, we obtain
\begin{align}
&
\phi(\bm{r})=\sum_{lm}\sqrt{\frac{4\pi}{2l+1}}Q_{lm}\frac{Y_{lm}(\hat{\bm{r}})}{r^{l+1}},
\\&
\bm{A}(\bm{r})=\sum_{lm}\biggl[
i\sqrt{\frac{4\pi(l+1)}{(2l+1)l}}M_{lm}\frac{\bm{Y}_{lm}^{l}(\hat{\bm{r}})}{r^{l+1}}
\cr&\quad\quad\quad\quad\quad
-\sqrt{4\pi(l+1)}T_{lm}\frac{\bm{Y}_{lm}^{l+1}(\hat{\bm{r}})}{r^{l+2}}
\biggr],
\end{align}
where $\bm{Y}_{lm}^{l'}(\hat{\bm{r}})$ ($l'=l-1,l,l+1$) is the vector spherical harmonics that transforms like $Y_{lm}(\hat{\bm{r}})$ under spatial rotation, and $l'$ is its orbital angular momentum~\cite{Kusunose_JPSJ.77.064710}.
By taking derivatives, we obtain the electric and magnetic fields in the presence of the multipole moment as
\begin{align}
&
\bm{E}(\bm{r})=-\bm{\nabla}\phi
=-\sum_{lm}\sqrt{4\pi(l+1)}Q_{lm}\frac{\bm{Y}_{lm}^{l+1}(\hat{\bm{r}})}{r^{l+2}},
\\&
\bm{B}(\bm{r})=\bm{\nabla}\times\bm{A}
=-\sum_{lm}\sqrt{4\pi(l+1)}M_{lm}\frac{\bm{Y}_{lm}^{l+1}(\hat{\bm{r}})}{r^{l+2}}.
\end{align}
Note that $\bm{E}(\bm{r})$ and $\bm{B}(\bm{r})$ have formally equivalent forms with $Q_{lm}$ and $M_{lm}$, and no toroidal moments $G_{lm}$ and $T_{lm}$ appear in these expressions.
Thus, the toroidal moments $G_{lm}$ and $T_{lm}$ affect physical quantities not through classical electromagnetic interactions but through the quantum-mechanical multipole-multipole interactions.
The latter also affects the phase of electrons through the vector potential, $e\bm{A}$. 

It is interesting to note that by using the dual nature of electric and magnetic quantities, we can introduce dual potentials with opposite space-time-parity counterparts as
\begin{align}
&
\phi^{*}(\bm{r})\equiv\sum_{lm}\sqrt{\frac{4\pi}{2l+1}}M_{lm}\frac{Y_{lm}(\hat{\bm{r}})}{r^{l+1}},
\\&
\bm{A}^{*}(\bm{r})\equiv\sum_{lm}\biggl[
i\sqrt{\frac{4\pi(l+1)}{(2l+1)l}}Q_{lm}\frac{\bm{Y}_{lm}^{l}(\hat{\bm{r}})}{r^{l+1}}
\cr&\quad\quad\quad\quad\quad
-\sqrt{4\pi(l+1)}G_{lm}\frac{\bm{Y}_{lm}^{l+1}(\hat{\bm{r}})}{r^{l+2}}
\biggr].
\end{align}
Then, the electric and magnetic fields are given by $\bm{E}=\bm{\nabla}\times\bm{A}^{*}$ and $\bm{B}=-\bm{\nabla}\phi^{*}$.
If a magnetic monopole charge $e^{*}$ were present, $G_{lm}$ could affect the phase of electrons via $e^{*}\bm{A}^{*}$.

\subsection{Electronic band structure}
\label{sec:Electronic band structure under multipoles}

By using the expressions of multipoles in momentum space in Sec.~\ref{sec:Multipoles in momentum space}, we find the effect of multipoles on the electronic band structure from the microscopic viewpoint. 
Conversely, we can deduce what type of multipoles are activated when the electronic band structure is determined from the first-principle band calculations or detected by the ARPES and de Haas-van Alphen measurements, and so on. 

A Hamiltonian in condensed matter physics must be totally symmetric and time-reversal even. 
Therefore, the one-electron Hamiltonian must be in the scalar-product form as 
\begin{align}
\label{eq:Hamkspace1}
\mathcal{H}=\sum_{X}^{Q,M,T,G}\sum_{\bm{k}\sigma\sigma'}\sum_{lm}X_{lm}^{\rm ext}X_{lm}^{\sigma\sigma'}(\bm{k})c^{\dagger}_{\bm{k}\sigma}c^{}_{\bm{k}\sigma'},
\end{align}
where $c^{\dagger}_{\bm{k}\sigma}$ ($c^{}_{\bm{k}\sigma}$) is the creation (annihilation) operator of a electron with the wave vector $\bm{k}$ and spin $\bm{\sigma}$.
Note that we assume $X_{lm}$ in real representation so that the Hermite conjugation is omitted in the scalar product.
For example, the lower-rank contributions read from Tables~\ref{tab_k-multipole-e} and \ref{tab_k-multipole-o} as
\begin{align}
\mathcal{H}=&\sum_{\bm{k}\sigma\sigma'}\biggl[
\frac{\hbar^{2}\bm{k}^{2}}{2m}\sigma_{0}
+\bm{Q}^{\rm ext}\cdot(\bm{k}\times\bm{\sigma})
+\bm{M}^{\rm ext}\cdot\bm{\sigma}
\cr&
+\bm{T}^{\rm ext}\cdot\bm{k}\,\sigma_{0}
+G_{0}^{\rm ext}(\bm{k}\cdot\bm{\sigma})
+\cdots
\biggr]_{\sigma\sigma'}c^{\dagger}_{\bm{k}\sigma}c^{}_{\bm{k}\sigma'},
\end{align}
where it has taken as $Q_{0}^{\rm ext}=\hbar^{2}/2m$.

$X^{\rm ext}_{lm}$ ($X=Q$, $M$, $T$, and $G$) represent ``symmetry-breaking'' fields, which bring about symmetry-breaking for the $c$-electron systems. 
The microscopic origins of $X^{\rm ext}_{lm}$ are the external fields applied to the systems, the crystalline electric field (CEF) from ligand ions, and molecular fields originating from multipole-multipole interactions by the spontaneous electronic orderings, and so on. 
For example, $\bm{M}^{\rm ext}$ arises from an external magnetic field $\bm{H}$ or molecular field of ferromagnetic ordering. 
The representative external fields are summarized in Table~\ref{tab_external fields}.

\begin{table}[t!]
\begin{center}
\caption{
Representative external fields.
$\bm{u}$ represents a displacement vector field. 
}
\label{tab_external fields}
\begingroup
\renewcommand{\arraystretch}{1.2}
 \begin{tabular}{ccccc}
 \hline \hline
multipole 
& $\mathcal{P}$ & $\mathcal{T}$ & external field & symbol
 \\
\hline
$\bm{G}^{\rm ext}$  & $+$ & $+$ & rotation & $\bm{\omega}=(\bm{\nabla}\times\bm{u})$/2 \\
$Q^{\rm ext}_{0}$, $Q^{\rm ext}_{2m}$  &     & $+$ & strain 
& $\varepsilon_{ij}=(\partial_{i}u_{j}+\partial_{j}u_{i})/2$ \\ 
$\bm{M}^{\rm ext}$       &     & $-$ & magnetic field & $\bm{H}$ \\ \hline
$\bm{Q}^{\rm ext}$       & $-$ & $+$ & electric field & $\bm{E}$ \\
$\bm{T}^{\rm ext}$       &     & $-$ & electric current & $\bm{J}$, $\bm{\nabla}\times\bm{H}$ \\
\hline\hline
\end{tabular}
\endgroup
\end{center}
\end{table}

In order to examine the effect on the band structure, the classification according to their spatial-time inversion properties is useful. 
In the single-band systems, the Hamiltonian in Eq.~(\ref{eq:Hamkspace1}) can be divided as 
\begin{align}
\label{eq:Hamkspace2}
\mathcal{H}=&\sum_{\bm{k}\sigma\sigma'} \biggl[ 
\varepsilon^{\rm S}(\bm{k})\delta_{\sigma\sigma'}+\varepsilon^{\rm A}(\bm{k})\delta_{\sigma\sigma'}
\cr&\quad\quad\quad\quad\quad
+f^{\rm S}_{\sigma\sigma'}(\bm{k})+f_{\sigma\sigma'}^{\rm A}(\bm{k}) 
\biggr] c^{\dagger}_{\bm{k}\sigma}c^{}_{\bm{k}\sigma'}, 
\end{align}
where $\varepsilon$($f_{\sigma\sigma'}$) is the charge(spin) sector and the superscript ${\rm S}$(${\rm A}$) represents symmetric(anti-symmetric) contribution with respect to $\bm{k}$.
As the even-rank $Q_{lm}(\bm{k})$, $T_{lm}(\bm{k})$, and the odd-rank $M_{lm}(\bm{k})$ are even function of $\bm{k}$, and other multipoles are odd function of $\bm{k}$, each coefficient is identified as 
\begin{align}
\label{eq:dispersion_sym_spinless}
\varepsilon^{\rm S}(\bm{k})
&= \sum_{lm}^{{\rm even}} Q^{\rm ext}_{lm} Q_{lm} (\bm{k}), \\
\label{eq:dispersion_asym_spinless}
\varepsilon^{\rm A}(\bm{k})
&= \sum_{lm}^{{\rm odd}} T^{\rm ext}_{lm} T_{lm} (\bm{k}), \\
\label{eq:dispersion_sym_spinful}
f^{\rm S}_{\sigma \sigma'}(\bm{k})
&= \sum_{lm}^{{\rm odd}} M^{\rm ext}_{lm} M_{lm}^{\sigma\sigma'} (\bm{k}) + \sum_{lm}^{{\rm even}}  T^{\rm ext}_{lm} T_{lm}^{\sigma\sigma'} (\bm{k}), \\
\label{eq:dispersion_asym_spinful}
f^{\rm A}_{\sigma \sigma'}(\bm{k})
&= \sum_{lm}^{{\rm even}} G^{\rm ext}_{lm} G_{lm}^{\sigma\sigma'} (\bm{k}) + \sum_{lm}^{{\rm odd}} Q^{\rm ext}_{lm} Q_{lm}^{\sigma\sigma'} (\bm{k}).  
\end{align}
When we consider the electronic band structure under symmetry-breaking fields, we apply symmetry operations only to the $c$-electron systems $X_{lm}(\bm{k})$ without acting on $X_{lm}^{\rm ext}$.
Thus, for single-band systems, the spatial inversion operation $\mathcal{P}$ corresponds to $(\bm{k},\sigma_{0},\bm{\sigma})\to(-\bm{k},\sigma_{0},\bm{\sigma})$, while the time-reversal operation $\mathcal{T}$ corresponds to $(\bm{k},\sigma_{0},\bm{\sigma})\to(-\bm{k},\sigma_{0},-\bm{\sigma})$.
The effects of the multipole moments on each part of the electronic structure are summarized in Table~\ref{tab_kmultipole_band}. 
We discuss coefficients in Eq.~(\ref{eq:dispersion_sym_spinless})-(\ref{eq:dispersion_asym_spinful}) one by one below.

\begin{table}[t!]
\begin{center}
\caption{
Interplay between the multipole moments and space-time classified part of electronic structure. 
$\mathcal{P}$ and $\mathcal{T}$ represent the spatial inversion and time-reversal operations corresponding to $(\bm{k},\sigma_{0},\bm{\sigma})\to(-\bm{k},\sigma_{0},\bm{\sigma})$ and $(\bm{k},\sigma_{0},\bm{\sigma})\to(-\bm{k},\sigma_{0},-\bm{\sigma})$, respectively.
S(A) represents that symmetric(anti-symmetric) contribution arises in the band structure, and 
SS is the abbreviation of the spin splitting. 
}
\label{tab_kmultipole_band}
\begingroup
\renewcommand{\arraystretch}{1.2}
 \begin{tabular}{lccccccc}
 \hline \hline
type  & $\mathcal{P}$ & $\mathcal{T}$ 
& multipole  & band structure
 \\
\hline
$\varepsilon^{\rm S}(\bm{k})$ & $+$ & $+$     & $Q_{lm}^{\rm ext}$ (even) & S w/o SS\\ 
$f^{\rm S}_{\sigma\sigma'}(\bm{k})$ &  & $-$     & $M_{lm}^{\rm ext}$ (odd), $T_{lm}^{\rm ext}$ (even)  &  S w/ SS    \\ \hline
$f^{\rm A}_{\sigma\sigma'}(\bm{k})$ & $-$  & $+$ & $G_{lm}^{\rm ext}$ (even), $Q_{lm}^{\rm ext}$ (odd)  &  A w/ SS  \\
$\varepsilon^{\rm A}(\bm{k})$ & & $-$ & $T_{lm}$ (odd) & A w/o SS \\ 
\hline\hline
\end{tabular}
\endgroup
\end{center}
\end{table}

\begin{figure*}[htb!]
\begin{center}
\includegraphics[width=0.7 \hsize]{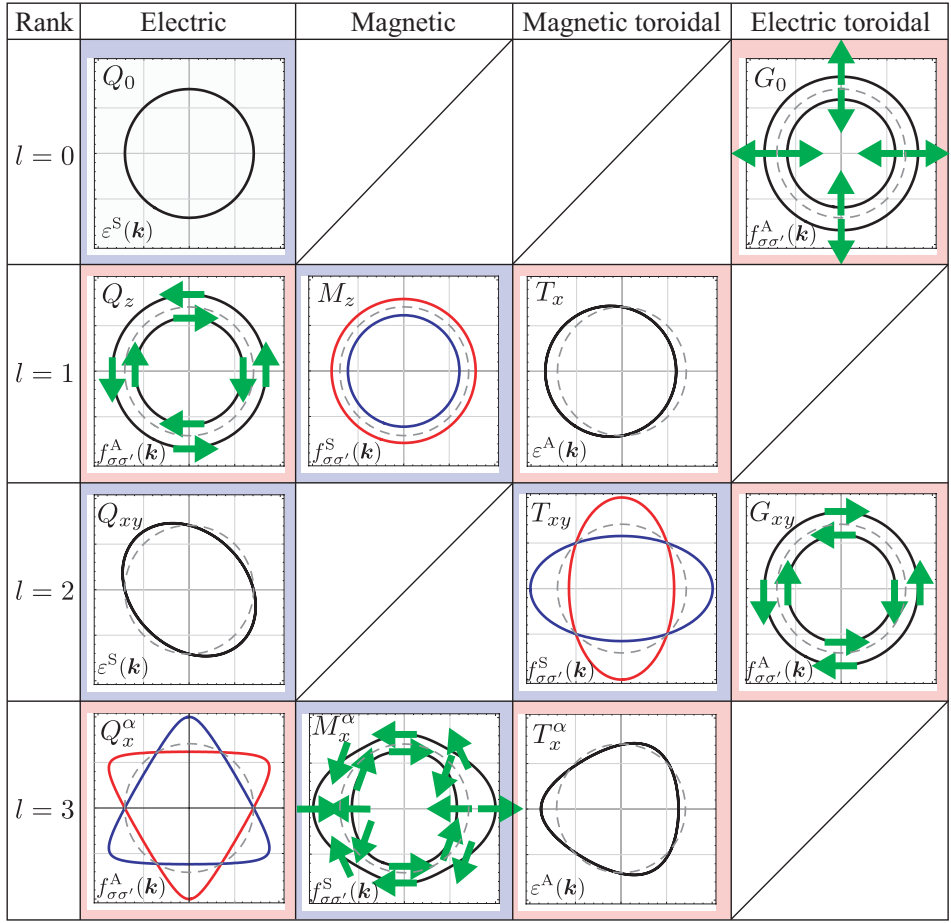}
\caption{
\label{Fig:band_table}
Examples of the typical electronic band modulations in the $k_x$-$k_y$ plane at $k_z=0$ in the presence of the E, M, MT, and ET multipoles up to the rank $l=3$ in the single-band picture. 
The blue (red) boxes stand for the band modulations by the even-parity (odd-parity) multipoles. 
The green arrows represent the in-plane spin moments at $\bm{k}$, while the red (blue) curves show the up(down)-spin moments along the out-of-plane direction. 
The dashed circles in each box represent the energy contours in the absence of the multipole, which corresponds to the energy contour for $Q_0^{\rm ext}$.
The superscript ``ext" of $X_{lm}^{\rm ext}$ is omitted for simplicity.
}
\end{center}
\end{figure*}

$\varepsilon^{\rm S}(\bm{k})$ in Eq.~(\ref{eq:dispersion_sym_spinless}) represents the symmetric-type band dispersions without the spin splitting.  
This band structure is present when there are both the spatial inversion and time-reversal symmetries. 
For example, the kinetic energy of free electron, which is given by $\varepsilon^{\rm S}(\bm{k})= \hbar^{2}\bm{k}^{2} /2m$, corresponds to the E monopole in terms of multipole terminology. 
The higher-rank quadrupole-type deformation in the band structure is attributed to the presence of $Q_{2m}^{\rm ext}=(Q_u^{\rm ext}, Q_v^{\rm ext}, Q_{yz}^{\rm ext}, Q_{zx}^{\rm ext}, Q_{xy}^{\rm ext})$.
The band deformation caused by $Q_{xy}^{\rm ext}$ is shown in Fig.~\ref{Fig:band_table} for instance. 
This quadrupole-type deformation corresponds to orbital (nematic) orderings, which have been discussed in iron-based superconductors~\cite{Fang_PhysRevB.77.224509,Fernandes_PhysRevLett.105.157003,Kruger_PhysRevB.79.054504,Lv_PhysRevB.80.224506,Lee_PhysRevLett.103.267001,Onari_PhysRevLett.109.137001} and Sr$_3$Ru$_2$O$_7$~\cite{Raghu_PhysRevB.79.214402,Lee_PhysRevB.80.104438,Tsuchiizu_PhysRevLett.111.057003}.

$f^{\rm S}_{\sigma \sigma'}(\bm{k})$ in Eq.~(\ref{eq:dispersion_sym_spinful}) represents the symmetric-type band dispersions with the spin splitting. 
This band structure appears once the time-reversal symmetry is broken while the spatial inversion symmetry is still present.  
The odd-rank M multipoles and even-rank MT multipoles are categorized in this symmetry group. 
The band structure shows the spin polarization owing to the time-reversal symmetry breaking. 
The typical example is the band structure in the ferromagnetic ordering; the M dipole $M_{z}$ leads to the symmetric spin splitting with the dispersions
\begin{align}
\varepsilon_{\sigma} (\bm{k})= 
\displaystyle \frac{\hbar^2 \bm{k}^2}{2m} + \sigma H_z,  
\end{align}
where $M_{z}^{\rm ext}=H_{z}$ and $\sigma=\pm 1$. 
The results indicate a uniform spin splitting in the band structure, as shown in Fig.~\ref{Fig:band_table}.
In the case of the higher-rank multipoles, e.g., the MT quadrupole $T_{xy}$, the band dispersions are given by 
\begin{align}
\varepsilon_\sigma (\bm{k})= 
\displaystyle \frac{\hbar^2 \bm{k}^2}{2m} + \sigma T^{\rm ext}_{xy}  \sqrt{(k_x^2-k_y^2)^2+ (k_x^2+k_y^2)k_z^2}, 
\end{align}
which indicates the quadrupole-type spin splitting, as shown in Fig.~\ref{Fig:band_table}.  

$f^{\rm A}_{\sigma \sigma'}(\bm{k})$ in Eq.~(\ref{eq:dispersion_asym_spinful}) represents the asymmetric-type band dispersions (the anti-symmetric contribution to the background symmetric band dispersions) with the spin splitting. 
The odd-rank E multipoles and even-rank ET multipoles contribute to $f^{\rm A}_{\sigma \sigma'}(\bm{k})$. 
This band structure is characterized by the breaking of the spatial inversion symmetry and is often realized in the presence of the spin-orbit coupling. 
For example, the Rashba-type spin-orbit coupling with the form of $k_y \sigma_x -k_x \sigma_y$ corresponds to the E dipole $Q_{z}$, while the Dresselhaus-type spin-orbit coupling $k_x (k_y^2-k_z^2)\sigma_x+k_y (k_z^2-k_x^2)\sigma_y+k_z (k_x^2-k_y^2)\sigma_z$ corresponds to E octupole $Q_{xyz}$, as shown in Table~\ref{tab_k-multipole-o}. 
In other words, the Rashba-type (Dresselhaus-type) spin-orbit coupling appears when the corresponding odd-parity $\bm{Q}^{\rm ext}$ ($Q_{xyz}^{\rm ext}$) is present due to the crystal structure. 
To be specific, the E dipole component in $f^{\rm A}_{\sigma \sigma'}(\bm{k})$ is induced by polar point groups ($C_{4{\rm v}}$, $C_4$, $C_{2{\rm v}}$, $C_2$, $C_{\rm s}$, $C_{6{\rm v}}$, $C_6$, $C_{3{\rm v}}$, $C_3$, $C_1$), while ET monopole, E dipole, or ET quadrupole component in $f^{\rm A}_{\sigma \sigma'}(\bm{k})$ is induced by gyrotropic point groups ($O$, $T$, $D_4$, $D_{2{\rm d}}$, $C_{4{\rm v}}$, $C_4$, $S_4$, $D_2$, $C_{2{\rm v}}$, $C_2$, $C_{\rm s}$, $D_6$, $C_{6{\rm v}}$, $C_6$, $D_3$, $C_{3{\rm v}}$, $C_3$, $C_1$), as discussed in Sec.~\ref{sec:Point-group irreducible representations}.

$\varepsilon^{\rm A}(\bm{k})$ in Eq.~(\ref{eq:dispersion_asym_spinless}) represents the asymmetric-type band dispersions with the spin degeneracy. 
This band structure is obtained when the systems lack both the spatial inversion and time-reversal symmetries. 
The multipoles in this category are the odd-rank MT ones, which exhibit the odd-order $\bm{k}$ dispersions. 
For example, the toroidal dipole $T_z$ order leads to the $\bm{k}$-linear dispersions, 
\begin{align}
\varepsilon (\bm{k})=\frac{\hbar^2 \bm{k}^2}{2m} + T^{\rm ext}_{z} k_z,  
\end{align}
whereas the MT octupole leads to the third-order modulations with respect to $\bm{k}$; $k_x (5 k_x^2-3 \bm{k}^2)$ for $T_x^{\alpha\,{\rm ext}}$, which is schematically shown in Fig.~\ref{Fig:band_table}.  

Note that asymmetric band modulations are also obtained by combining the multipoles belonging to $f^{\rm S}_{\sigma \sigma'}(\bm{k})$ and $f^{\rm A}_{\sigma \sigma'}(\bm{k})$.  
For example, the noncentrosymmetric polar ferromagnets where the M dipole moment lies in the $x$ direction ($M_x^{\rm ext}$) with the amplitude $H_x$ and the E dipole moment lies in the $z$ direction ($Q_z^{\rm ext}$) with the amplitude of $E_z$. 
In this case, the Hamiltonian is written as 
\begin{align}
\mathcal{H} 
= 
\left( 
\begin{array}{cc}
\displaystyle \frac{\hbar^2 \bm{k}^2}{2m} & -E_{z} (k_y+i k_x) + H_{x}   \\
-E_{z} (k_y-i k_x)+ H_{x}  & \displaystyle \frac{\hbar^2 \bm{k}^2}{2m} 
\end{array} \right),  
\label{eq:localeff}
\end{align}
where the eigenvalues are easily obtained as 
\begin{align}
\displaystyle \varepsilon_{\sigma} (\bm{k})= 
\frac{\hbar^2 \bm{k}^2}{2m} \pm \sqrt{E_{z}^2 (k_x^2+k_y^2) + H_{x}^2   -2 E_{z} H_{x} k_y}. 
\end{align}
The obtained band dispersions clearly show a shift of the band bottom along the $k_y$ direction when $E_{z}H_{x} \neq 0$, which indicates the emergence of the MT dipole along the $y$ direction $\bm{T}^{\rm ext}\equiv \bm{Q}^{\rm ext}\times\bm{M}^{\rm ext}=E_{z}H_{x}\bm{e}_y$ where $\bm{e}_y$ is the unit vector along the $y$ direction. 
Note that there is no spin degeneracy in this situation because the symmetry of the product $\mathcal{P}\mathcal{T}$ is broken.
This result is also confirmed by the multipole-operator aspect in momentum space, since the direct product of $M_x \propto \sigma_x$ and $Q_z \propto k_x \sigma_y - k_y \sigma_x$, which are given in Tables~\ref{tab_k-multipole-e} and \ref{tab_k-multipole-o} respectively, leads to the functional form of the MT dipole as $\{\sigma_{x},k_{x}\sigma_{y}-k_{y}\sigma_{x}\}=-2k_{y}\sigma_{0}\propto T_y$. 
The similar discussions can be straightforwardly extended to the antiferromagnets accompanied with the spatial-inversion symmetry breaking. 
For example, the staggered-type antiferromagnetic orderings on the one-dimensional zigzag and two-dimensional honeycomb structures are regarded as the MT dipole and octupole orders, respectively~\cite{Yanase_JPSJ.83.014703,Hayami_PhysRevB.90.024432,Hayami_doi:10.7566/JPSJ.84.064717,hayami2016emergent}.

\subsection{Linear-response tensors}
\label{sec:Linear response tensors}

Next, we discuss the cross-correlated phenomena in terms of multipole degrees of freedom. 
We examine what types of linear responses are expected in the presence of multipoles on the basis of the Kubo formula. 
In the linear response $\langle\hat{A}_{i}\rangle=\sum_{j}\chi_{ij}F_{j}$, where $F_{j}$ is an external field along the $j$ direction ($i,j=x,y,z$), the coupling with the external perturbation is given by $\mathcal{H}_{\rm ext}=-\hat{C}_{j}F_{j}$, and $\hat{B}_{j}=d\hat{C}_{j}/dt$, the linear response tensor is generally written by 
\begin{align}
\chi_{ij}^{\rm (2)}=-\frac{i}{V}\sum_{\bm{k},n,m}\frac{f_{n\bm{k}} -f_{m\bm{k}}}{\varepsilon_{n\bm{k}}-\varepsilon_{m\bm{k}}}
 \frac{ \langle \psi_{n\bm{k}}|\hat{A}_i|\psi_{m\bm{k}}\rangle \langle \psi_{m\bm{k}}|\hat{B}_j|\psi_{n\bm{k}}\rangle}{\varepsilon_{n\bm{k}}-\varepsilon_{m\bm{k}}+i\gamma },  
 \label{eq_chi}
\end{align}
where $V$ is the system volume, $\gamma$ is a broadening factor $f_{m\bm{k}}=f(\varepsilon_{m\bm{k}})$ is the Fermi distribution function, and $\varepsilon_{m\bm{k}}$ and $\psi_{m\bm{k}}$ are the eigenvalue and eigenstate of the Hamiltonian. $m$ and $n$ are the band indices.
This expression does not take account of the orbital motion due to the Lorentz force in the presence of the external magnetic field, which could be important for some response functions such as the normal Hall conductivity~\cite{Fukuyama_PTP.42.494,Fukuyama_PTP.42.1284}. 
$\hat{A}_i$ and $\hat{B}_j$ are rank-1 tensor and Hermite operators and $\chi_{ij}^{(2)}$ is rank-2 tensor [the superscript $(l)$ in $\chi_{ij}$ represents the rank-$l$ tensor], although the following discussion is straightforwardly extended to higher-rank operators, such as the E quadrupole representing the elastic property, as discussed in Appendix~\ref{sec:Other linear response tensors}.

We decompose $\chi_{ij}^{\rm (2)}$ in Eq.~(\ref{eq_chi}) into the energy degenerate (J: $\varepsilon_{n\bm{k}}=\varepsilon_{m\bm{k}}$) and non-degenerate (E: $\varepsilon_{n\bm{k}}\neq\varepsilon_{m\bm{k}}$) parts as 
\begin{align}
&\chi_{ij}^{\rm (2)}= \chi^{\rm (J)}_{ij} + \chi^{\rm (E)}_{ij},\cr 
&\quad
\chi^{\rm (J)}_{ij}=  - \frac{\gamma}{V}\sum_{\bm{k}nm}^{=}\frac{f_{n\bm{k}} -f_{m\bm{k}}}{\varepsilon_{n\bm{k}}-\varepsilon_{m\bm{k}}}
\frac{A_{i\bm{k}}^{nm}B_{j\bm{k}}^{mn}}{\left(\varepsilon_{n\bm{k}}-\varepsilon_{m\bm{k}}\right)^2+\gamma^2},
\label{eq:chi2} \\
&\quad
\chi^{\rm (E)}_{ij}=- \frac{i}{V}\sum_{\bm{k}nm}^{\neq}
\frac{\left(f_{n\bm{k}} -f_{m\bm{k}}\right)A_{i\bm{k}}^{nm}B_{j\bm{k}}^{mn}}{\left(\varepsilon_{n\bm{k}}-\varepsilon_{m\bm{k}}\right)^2+\gamma^2 },
\label{eq:chi1}
\end{align}
where $A_{i\bm{k}}^{nm}=\braket{\psi_{n\bm{k}}|\hat{A}_{i}|\psi_{m\bm{k}}}$ and $B_{j\bm{k}}^{mn}=\braket{\psi_{m\bm{k}}|\hat{B}_{j}|\psi_{n\bm{k}}}$, and $(f_{n\bm{k}}-f_{m\bm{k}})/(\varepsilon_{n\bm{k}}-\varepsilon_{m\bm{k}})\to\partial f/\partial\varepsilon_{n\bm{k}}$. 
$\chi_{ij}^{\rm (J)}$ is the dissipative (current driven) part, 
while $\chi_{ij}^{\rm (E)}$ is the non-dissipative (electric-field driven) part. 
As discussed in Appendix~\ref{sec:Space-time inversion operations of linear response tensor}~\cite{Watanabe_PhysRevB.96.064432}, $\chi^{\rm (J)}_{ij}$ and $\chi^{\rm (E)}_{ij}$ have different transformation property under the time-reversal operation, i.e., the latter (former) changes the sign when the time-reversal properties of $\hat{A}_{i}$ and $\hat{B}_{j}$ are the same (different).
On the other hand, the spatial inversion changes the sign of $\chi^{\rm (J,E)}_{ij}$ when the parities of $\hat{A}_{i}$ and $\hat{B}_{j}$ are different.
These facts indicate that different multipoles contribute to $\chi^{\rm (J)}_{ij}$ and $\chi^{\rm (E)}_{ij}$. 
The relation between the 2nd-rank linear-response tensors and relevant multipoles is summarized in Table~\ref{lrtensor}.

\begin{table*}[t!]
\caption{
Relation between the components of the linear-response rank-2 tensors 
$\chi^{(2)}_{ij}=\chi_{ij}^{\rm (J)}+\chi_{ij}^{\rm (E)}$ and relevant multipoles in the presence ($\bigcirc$) or absence ($\times$) of $(\mathcal{P}, \mathcal{T}, \mathcal{PT})$, where the superscript J (E) represents the current driven dissipative (electric-field driven non-dissipative) part. 
The superscripts S and A represent the symmetric and antisymmetric components of the linear-response tensor, respectively.
The symbol ``$=$'' represents that there are no additional contributions from that in the presence of both spatial inversion and time-reversal symmetries. 
Note that we restrict our discussion only to the electronic contributions in the thermal conductivity although phonons usually contribute to it as well.
The Seebeck coefficient corresponds to the symmetric components of $\beta_{ij}$.
}
\label{lrtensor}
\centering
\begin{tabular}{lcccccccc} \hline\hline
tensor & type & $(\mathcal{P},\mathcal{T},\mathcal{PT})=$ & 
$(\bigcirc,\bigcirc,\bigcirc)$ 
& $(\bigcirc,\times,\times)$ 
& $(\times,\bigcirc,\times)$ 
& $(\times,\times,\bigcirc)$ 
& $(\times,\times,\times)$ \\ \hline 
electric conductivity & polar & &  $\sigma_{ij}^{\rm (J,S)}$
& $+\sigma_{ij}^{\rm (E,A)}$
& = & = & $\sigma_{ij}^{\rm (J,S)}+\sigma_{ij}^{\rm (E,A)}$ 
  \\
thermoelectric conductivity~\cite{Luttinger_PhysRev.135.A1505,Deo_PhysRev.141.738,Mahan1990manyparticle} & polar & & $\beta_{ij}^{\rm (J)}$ 
& $+\beta_{ij}^{\rm (E)}$
& =  & = 
& $\beta_{ij}^{\rm (J)}+\beta_{ij}^{\rm (E)}$  \\ 
thermal conductivity~\cite{Luttinger_PhysRev.135.A1505,Deo_PhysRev.141.738,Mahan1990manyparticle,Shitade_ptu162} 
& polar & &  $\kappa_{ij}^{\rm (J)}$ 
& $+\kappa_{ij}^{\rm (E)}$
& = & = 
& $\kappa_{ij}^{\rm (J)}+\kappa_{ij}^{\rm (E)}$  \\ 
magneto-electric(current) tensor & axial & & --- 
& --- 
& $\alpha_{ij}^{\rm (J)}$
&  $\alpha_{ij}^{\rm (E)}$
& $\alpha_{ij}^{\rm (J)}$+$\alpha_{ij}^{\rm (E)}$ \\
\hline
symmetric (S) components & & & $Q_{0}$, $Q_{ij}$ 
& $T_{0}$, $T_{ij}$ 
& $G_{0}$, $G_{ij}$ 
& $M_{0}$, $M_{ij}$ 
&  
\\
anti-symmetric (A) components & & & $G_{i}$ 
& $M_{i}$ 
& $Q_{i}$ 
& $T_{i}$ 
& 
\\
\hline \hline
\end{tabular}
\end{table*}

The linear-response (rank-2) tensor $\chi_{ij}^{\rm (2)}$ has 9 independent components when $\hat{A}_i$ and $\hat{B}_j$ are rank-1 tensor. 
As will be discussed in Appendix~\ref{sec:Other linear response tensors}, each component is related with the multipole degree of freedom: $\chi_{ij}^{\rm (2)}$ is spanned by the multipoles in the following form, 
\begin{align} 
\label{eq:rank11tensor}
\hat{\chi}^{\rm (2)} = 
\begin{pmatrix}
X_{0}-X_{u}+X_{v}& X_{xy}+Y_{z} & X_{zx}-Y_{y} \\
X_{xy}-Y_{z} & X_{0}-X_{u}-X_{v} & X_{yz}+Y_{x} \\
X_{zx}+Y_{y} & X_{yz}-Y_{x} & X_{0}+2X_{u} 
\end{pmatrix}, 
\end{align}
where $X=Q$ or $T$ ($G$ or $M$) and $Y=G$ or $M$ ($Q$ or $T$) for the polar (axial) 2nd-rank tensor depending on their time-reversal property. 
Among 9 multipoles, rank-0 $(X_0)$, rank-1 $(Y_x, Y_y, Y_z)$, 
and rank-2 $(X_u, X_v, X_{yz}, X_{zx}, X_{xy})$ multipoles represent the isotropic component, and antisymmetric and symmetric traceless components, respectively.

Since $\chi_{ij}^{\rm (J)}$ and $\chi_{ij}^{\rm (E)}$ have definite parities under spatial inversion and time-reversal operations ($\chi_{ij}^{\rm (J)}$ and $\chi_{ij}^{\rm (E)}$ have opposite time-reversal property), the type of multipoles is determined so as to share the same parities. 
When the linear-response tensors $\chi^{\rm (J)}_{ij}$ or $\chi^{\rm (E)}_{ij}$ are both the time-reversal and spatial inversion even, i.e., time-reversal-even polar tensor, the relevant multipoles are the E monopole $Q_0$, ET dipole $G_i$, and E quadrupole $Q_{ij}$. 
When the time-reversal (spatial inversion) operation changes the sign of the linear-response tensors, i.e., time-reversal-odd polar tensor (time-reversal-even axial tensor), they are characterized by the MT monopole $T_0$, M dipole $M_i$, and MT quadrupole $T_{ij}$ (ET monopole $G_0$, E dipole $Q_i$, and ET quadrupole $G_{ij}$), which become finite in the absence of the time-reversal (spatial inversion) symmetry. 
Moreover, the M monopole $M_0$, MT dipole $T_i$, and M quadrupole $M_{ij}$ contribute to the time-reversal-odd axial tensors, and it can be finite in the case where there are no time-reversal and spatial inversion symmetries while their product is preserved. 
In the following, we discuss two fundamental examples of the linear-response tensors: one is the electrical conductivity tensor and the other is the magneto-electric (current) tensor. 
We also discuss the other linear-response tensors, such as the spin conductivity tensor and Piezo-electric(current) tensor in Appendix~\ref{sec:Other linear response tensors}.

\subsubsection{Electric conductivity tensor}
\label{sec:Electrical conductivity tensor}

First, we consider the polar 2nd-rank electric conductivity, 
\begin{align}
\bm{J}&=\hat{\sigma}\bm{E}
=\left(\hat{\sigma}^{\rm (J)}+\hat{\sigma}^{\rm (E)}\right) \bm{E},
\end{align}
namely, we adopt $\hat{A}_{i}=\hat{B}_{i}=\hat{J}_{i}$ in the general formula (\ref{eq_chi}).
As the electric current operator $\hat{J}_i$ is both time-reversal and spatial-inversion odd, the polar electric 
conductivity tensor $\hat{\sigma}^{\rm (E)}$ (electric-field driven non-dissipative part) becomes nonzero only when the time-reversal symmetry is broken, while there is no symmetry restriction for the current driven dissipative part $\hat{\sigma}^{\rm (J)}$. 
Meanwhile, from Eqs.~(\ref{eq:chi1}) and (\ref{eq:chi2}), $\hat{\sigma}^{\rm (E)}$ is the antisymmetric tensor $\sigma_{ij}^{\rm (E)}=-\sigma_{ji}^{\rm (E)}\equiv\sigma_{ij}^{\rm (E,A)}$ and $\hat{\sigma}^{\rm (J)}$ is the symmetric tensor $\sigma_{ij}^{\rm (J)}=\sigma_{ji}^{\rm (J)}\equiv\sigma_{ij}^{\rm (J,S)}$. 
According to these facts, the corresponding multipole degrees of freedom for $\hat{\sigma}^{\rm (J,S)}$ and $\hat{\sigma}^{\rm (E,A)}$ are
identified as 
\begin{align}
&
\hat{\sigma}^{\rm (J,S)}=
\begin{pmatrix}
Q_{0}-Q_{u}+Q_{v} & Q_{xy} & Q_{zx} \\
Q_{xy} & Q_{0}-Q_{u}-Q_{v} & Q_{yz} \\
Q_{zx} & Q_{yz} & Q_{0}+2Q_{u}
\end{pmatrix}, 
\cr&
\hat{\sigma}^{\rm (E,A)}=
\begin{pmatrix}
0 & M_{z} & -M_{y} \\
-M_{z} & 0 & M_{x} \\
M_{y} & -M_{x} & 0
\end{pmatrix}. 
\end{align}
The result shows that the anomalous Hall conductivity corresponding to the antisymmetric non-dissipative part becomes nonzero in the presence of the M dipole moment, and arises from the interband contribution.

\subsubsection{Magneto-electric(current) tensor}
\label{sec:Magneto-electric(current) tensor}

Let us consider the magneto-electric (current) effect where the magnetization is induced by the electric field (current), 
\begin{align}
\bm{M}&=\hat{\alpha}\bm{E}=\left(\hat{\alpha}^{\rm (J)}+\hat{\alpha}^{\rm (E)}\right)\bm{E}.
\end{align}
The axial 2nd-rank magneto-electric (current) tensor $\hat{\alpha}$ is obtained by using $\hat{A}_i = \hat{M}_i$ and $\hat{B}_j = \hat{J}_j$ in Eq.~(\ref{eq_chi}). 
As $\hat{M}_i$ is time-reversal odd and spatial inversion even, the dissipative part $\hat{\alpha}^{\rm (J)}$ becomes nonzero in the absence of only the spatial inversion symmetry.
On the other hand, the non-dissipative $\hat{\alpha}^{\rm (E)}$ becomes nonzero only in the absence of both the time-reversal and spatial inversion symmetries.
Thus, $\hat{\alpha}^{\rm (J)}$ and $\hat{\alpha}^{\rm (E)}$ are identified as  
\begin{align}
&
\hat{\alpha}^{\rm (J)}=
\begin{pmatrix}
G_{0}-G_{u}+G_{v} & G_{xy}+Q_{z} & G_{zx}-Q_{y} \\
G_{xy}-Q_{z} & G_{0}-G_{u}-G_{v} & G_{yz}+Q_{x} \\
G_{zx}+Q_{y} & G_{yz}-Q_{x} & G_{0}+2G_{u}
\end{pmatrix},
\cr&
\hat{\alpha}^{\rm (E)}=
\begin{pmatrix}
M_{0}-M_{u}+M_{v} & M_{xy}+T_{z} & M_{zx}-T_{y} \\
M_{xy}-T_{z} & M_{0}-M_{u}-M_{v} & M_{yz}+T_{x} \\
M_{zx}+T_{y} & M_{yz}-T_{x} & M_{0}+2M_{u}
\end{pmatrix}.
\end{align}
The isotropic longitudinal magneto-electric (current) response is realized in the presence of the M (ET) monopole, the antisymmetric transverse response in the presence of the MT (E) dipole, and the symmetric transverse and traceless longitudinal responses in the presence of the M (ET) quadrupoles.

\begin{figure}[htb!]
\begin{center}
\includegraphics[width=1.0 \hsize]{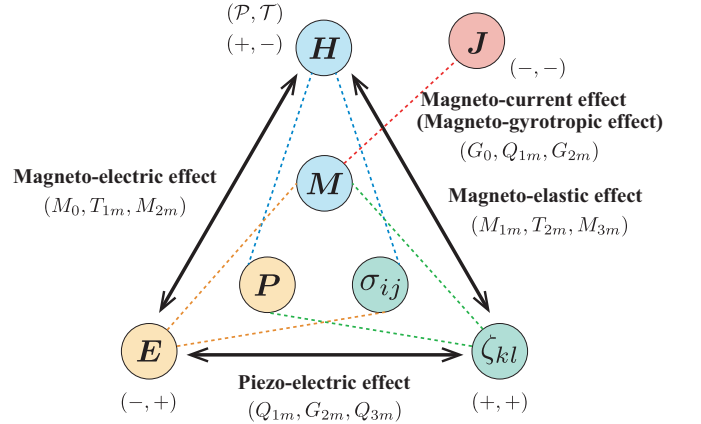}
\caption{
\label{Fig:Heckmann}
Schematic picture of a revised Heckmann diagram showing the relation among electric, magnetic, and mechanical properties in solids. 
$\bm{E}$, $\bm{H}$, $\bm{J}$, and $\zeta_{kl}=\varepsilon_{kl}+\sum_{m}\epsilon_{klm}\omega_{m}$ are external electric field, magnetic field, electric current, and strain-rotaion field, while $\bm{P}$, $\bm{M}$, and $\sigma_{ij}$ represent electric polarization, magnetization, and stress tensor,  
respectively. 
The relevant multipoles in the linear-response tensor for each cross-correlated coupling are also shown. 
}
\end{center}
\end{figure}

The non-dissipative $\hat{\alpha}^{\rm (E)}$ originates from the non-degenerate (interband) contributions, while the dissipative $\hat{\alpha}^{\rm (J)}$ mainly arises from the intraband contributions. 
In other words, $\hat{\alpha}^{\rm(E)}$ plays an important role in insulating systems, while $\hat{\alpha}^{\rm (J)}$ becomes important for metallic systems. 
Especially, in the presence of $\mathcal{PT}$ symmetry, the latter contribution is forbidden. 
From the above considerations, the former tensor is called magneto-electric tensor and the latter tensor is called magneto-current (magneto-gyrotropic) tensor, which has essentially the same origin as so-call 
the Edelstein effect~\cite{ivchenko1978new,edelstein1990spin}. 
The relations between the cross-correlated responses including the Piezo-electric effect and the relevant multipoles are summarized in Fig.~\ref{Fig:Heckmann}.

\section{Point-group irreducible representations}
\label{sec:Point-group irreducible representations}

In the crystal systems, the rotational symmetry and inversion symmetry in some cases, are lost.
In such cases, the components of the same rank (the irreducible representation of the rotational group) split into subgroups according to the point-group irreducible representation. 
In this section, we discuss such a reduction for multipoles.
First, we discuss the CEF potential under all the point groups in Sec.~\ref{sec:Generalization of crystalline-electric-field potential}, which is nothing but the sum of the E multipoles in the totally symmetric ${\rm A}_{1g}^{+}$ representation.
Then, we show how to classify multipoles under point-group irreducible representations 
in Sec.~\ref{sec:Active multipoles in solids}. 
Such analyses offer microscopic investigation of potential active multipole degrees of freedom in solids. 
We show several examples by considering specific basis functions in the tetragonal $D_{\rm 2d}$ group. 

Similar analysis can be applied to any point-group symmetry.
For such purpose, we provide various tables for the parent cubic $O_{\rm h}$ group in the main text and the hexagonal $D_{\rm 6h}$ group in Appendix~\ref{sec:Electric multipoles and CEF parameters}, and the compatible relations between these parent groups and subgroups in Tables~\ref{tab_multipoles_table1} and \ref{tab_multipoles_table2}.

\subsection{CEF potential} 
\label{sec:Generalization of crystalline-electric-field potential}

Under the point-group symmetry, some components of the E multipoles reduce to the totally symmetric ${\rm A}_{1g}^{+}$ representation.
These components constitute the CEF potentials.
Therefore, the CEF potential is represented by 
\begin{align}
V(\bm{r})=\sum_{lm}^{\in {\rm A}_{1g}^{+}}c_{lm}O_{lm}(\bm{r}),
\end{align}
where $c_{lm}$ is the CEF parameter. 
Since we consider the largest orbital-angular momentum $L=3$ corresponding to the $f$ orbital in the one-electron state, up to the rank $l=2L=6$ is sufficient in the summation. 
The even-parity (even-rank) CEF leads to the even-parity hybridizations, such as $s$-$d$ and $p$-$f$ orbitals, while the odd-parity (odd-rank) CEF leads to the odd-parity hybridizations for $s$-$p$, $p$-$d$, $d$-$f$, and $s$-$f$ orbitals 
at the same site. 
Note that the odd-parity CEF is present only in the lack of inversion center at lattice sites.

The nonzero CEF parameters under each point group except for triclinic crystals ($C_{1}$, $C_{i}$) are summarized in Table~\ref{tab_cefparam}, which is constructed by reading the E multipoles belonging to ${\rm A}_{1g}^{+}$ representation in the reduction rules in Tables~\ref{tab_multipoles_table1} and \ref{tab_multipoles_table2} in Appendix~\ref{sec:Electric multipoles and CEF parameters}.
The relevant harmonics are given in Tables~\ref{tab_Oh_CEFe}, \ref{tab_Oh_CEFo}, \ref{tab_D6h_CEFe}, and \ref{tab_D6h_CEFo}.

\begin{table*}
\caption{
Finite CEF parameters in terms of the tesseral harmonics under the point group except for triclinic crystals ($C_{1}$, $C_i$),
where $(lm)$, $(lm)'$, and $(l)$ means the presence of $O_{lm}^{\rm (c)}$, $O_{lm}^{\rm (s)}$, and $O_{l}$ (see Table~\ref{tab_Oh_CEFe} for the expressions of $O_l$).
The checkmark in the columns 
N, P, C, and G represent the noncentrosymmetric (no inversion operation), polar (presence of rank-1 CEFs, see also rank-1 multipoles in Table~\ref{tab_totalsymmetric_o}), chiral (no improper rotations, $\mathcal{P}\mathcal{C}_{n}$), and gyrotropic (presence of less than rank-2 odd-parity multipoles in Table~\ref{tab_totalsymmetric_o}, i.e., presence of $\chi_{ij}^{\rm A}$) point groups, respectively. 
We take the $x$ axis as the $C_{2}'$ rotation axis. 
For $C_{3 {\rm v}}$, we take the $zx$ plane as the $\sigma_v$ mirror plane.
For monoclinic crystals, the standard orientation is taken. 
}
\label{tab_cefparam}
\centering
\renewcommand{\arraystretch}{1.2}
\begin{tabular}{ccccccll} \hline\hline
crystal system & point group & N & P & C & G & even-parity CEF & odd-parity CEF  \\ \hline
cubic & $O_{\rm h}$ & & & & & $(4)$, $(6)$ & ---  \\
& $O$ & $\checkmark$ & & $\checkmark$ & $\checkmark$ & $(4)$, $(6)$ & ---  \\
& $T_{\rm d}$ & $\checkmark$ & & & & $(4)$, $(6)$ & $(32)'$ \\
& $T$ & $\checkmark$ & & $\checkmark$ & $\checkmark$  & $(4)$, $(6)$, $(6t)$ & $(32)'$ \\ 
& $T_{\rm h}$ & & & & & $(4)$, $(6)$, $(6t)$ & --- \\ \hline
tetragonal & $D_{\rm 4h}$ & & & & & $(20)$, $(40)$, $(44)$, $(60)$, $(64)$ & --- \\
& $D_{4}^{}$ & $\checkmark$ & & $\checkmark$ & $\checkmark$ & $(20)$, $(40)$, $(44)$, $(60)$, $(64)$ & $(54)'$ \\
& $D_{2{\rm d}}$ & $\checkmark$ & & & $\checkmark$ & $(20)$, $(40)$, $(44)$, $(60)$, $(64)$ & $(32)'$, $(52)'$ \\
& $C_{4{\rm v}}$ & $\checkmark$ & $\checkmark$ & & $\checkmark$ & $(20)$, $(40)$, $(44)$, $(60)$, $(64)$ & $(10)$, $(30)$, $(50)$, $(54)$ \\
& $C_{4{\rm h}}$ & & & & & $(20)$, $(40)$, $(44)$, $(60)$, $(64)$, $(44)'$, $(64)'$ & --- \\
& $C_{4}$ & $\checkmark$ & $\checkmark$ & $\checkmark$ & $\checkmark$ & $(20)$, $(40)$, $(44)$, $(60)$, $(64)$, $(44)'$, $(64)'$ & $(10)$, $(30)$, $(50)$, $(54)$, $(54)'$ \\
& $S_{4}$ & $\checkmark$ & & & $\checkmark$ & $(20)$, $(40)$, $(44)$, $(60)$, $(64)$, $(44)'$, $(64)'$ & $(32)$, $(52)$, $(32)'$, $(52)'$ \\ \hline
orthorhombic & $D_{\rm 2h}$ & & & & & $(20)$, $(22)$, $(40)$, $(42)$, $(44)$, $(60)$, $(62)$, $(64)$, $(66)$ & --- \\
& $D_{2}$ & $\checkmark$ & & $\checkmark$ & $\checkmark$ & $(20)$, $(22)$, $(40)$, $(42)$, $(44)$, $(60)$, $(62)$, $(64)$, $(66)$ & $(32)'$, $(52)'$, $(54)'$ \\
& $C_{2{\rm v}}$ & $\checkmark$ & $\checkmark$ & & $\checkmark$ & $(20)$, $(22)$, $(40)$, $(42)$, $(44)$, $(60)$, $(62)$, $(64)$, $(66)$ & $(10)$, $(30)$, $(32)$, $(50)$, $(52)$, $(54)$ \\\hline
monoclinic & $C_{\rm 2h}$ & & & & & $(20)$, $(22)$, $(40)$, $(42)$, $(44)$, $(60)$, $(62)$, $(64)$, $(66)$ & --- \\
 & & & & &  &$(22)'$, $(42)'$, $(44)'$, $(62)'$, $(64)'$, $(66)'$ &  \\
& $C_{2}$ & $\checkmark$ & $\checkmark$ & $\checkmark$ & $\checkmark$ & $(20)$, $(22)$, $(40)$, $(42)$, $(44)$, $(60)$, $(62)$, $(64)$, $(66)$ & $(10)$, $(30)$, $(32)$, $(50)$, $(52)$, $(54)$ \\
 & & & & &  &$(22)'$, $(42)'$, $(44)'$, $(62)'$, $(64)'$, $(66)'$ & $(32)'$, $(52)'$, $(54)'$ \\
& $C_{{\rm s}}$ & $\checkmark$ & $\checkmark$ &  & $\checkmark$ & $(20)$, $(22)$, $(40)$, $(42)$, $(44)$, $(60)$, $(62)$, $(64)$, $(66)$ & $(11)$, $(31)$, $(33)$, $(51)$, $(53)$, $(55)$ \\
 & & & & &  &$(22)'$, $(42)'$, $(44)'$, $(62)'$, $(64)'$, $(66)'$ & $(11)'$, $(31)'$, $(33)'$, $(51)'$, $(53)'$, $(55)'$ \\
\hline
hexagonal & $D_{6{\rm h}}$ & & & & & $(20)$, $(40)$, $(60)$, $(66)$ & --- \\
& $D_{6}$ & $\checkmark$ & & $\checkmark$ & $\checkmark$ & $(20)$, $(40)$, $(60)$, $(66)$ & --- \\
& $D_{3{\rm h}}$ & $\checkmark$ & & & & $(20)$, $(40)$, $(60)$, $(66)$ & $(33)$, $(53)$ \\
& $C_{6{\rm v}}$ & $\checkmark$ & $\checkmark$ & & $\checkmark$ & $(20)$, $(40)$, $(60)$, $(66)$ & $(10)$, $(30)$, $(50)$ \\
& $C_{6{\rm h}}$ & & & & & $(20)$, $(40)$, $(60)$, $(66)$, $(66)'$ & --- \\
& $C_{6}$ & $\checkmark$ & $\checkmark$ & $\checkmark$ & $\checkmark$ & $(20)$, $(40)$, $(60)$, $(66)$, $(66)'$ & $(10)$, $(30)$, $(50)$ \\
& $C_{3{\rm h}}$ & $\checkmark$ & & & & $(20)$, $(40)$, $(60)$, $(66)$, $(66)'$ & $(33)$, $(53)$, $(33)'$, $(53)'$ \\
 \hline
trigonal & $D_{3{\rm d}}$ & & & & & $(20)$, $(40)$, $(60)$, $(66)$, $(43)'$, $(63)'$ & --- \\
& $D_{3}$ & $\checkmark$ & & $\checkmark$ & $\checkmark$ & $(20)$, $(40)$, $(60)$, $(66)$, $(43)'$, $(63)'$ & $(33)$, $(53)$ \\
& $C_{3{\rm v}}$ & $\checkmark$ & $\checkmark$ & & $\checkmark$ & $(20)$, $(40)$, $(43)$, $(60)$, $(63)$, $(66)$ & $(10)$, $(30)$, $(33)$, $(50)$, $(53)$ \\
& $C_{3{\rm i}}$ & & & & & $(20)$, $(40)$, $(43)$, $(60)$, $(63)$, $(66)$, & --- \\
 & & & & & & $(43)'$, $(63)'$, $(66)'$ &  \\ 
& $C_{3}$ & $\checkmark$ & $\checkmark$ & $\checkmark$ & $\checkmark$ & $(20)$, $(40)$, $(43)$, $(60)$, $(63)$, $(66)$, & $(10)$, $(30)$, $(33)$, $(50)$, $(53)$, \\
 & & & & & & $(43)'$, $(63)'$, $(66)'$ & $(33)'$, $(53)'$ \\ 
\hline\hline
\end{tabular}
\end{table*}

\begin{table*}
\caption{
Multipoles under cubic, tetragonal, orthorhombic, and monoclinic crystals. 
The upper and lower columns represent even-parity and odd-parity multipoles, respectively.  
We take the $x$ axis as the $C_{2}'$ rotation axis.
}
\label{tab_multipoles_table1}
\centering
\begin{tabular}{llll|ccccc|ccccccc|ccc|ccc} \hline\hline
E & ET & M & MT &
$O_{\rm h}$ & $O$ & $T_{\rm d}$ & $T_{\rm h}$ & $T$ &
$D_{\rm 4h}$ & $D_{4}$ & $C_{\rm 4h}$ & $D_{\rm 2d}$ & $C_{\rm 4v}$ & $C_{4}$ & $S_{4}$ &
$D_{\rm 2h}$ & $D_{2}$ & $C_{\rm 2v}$ & $C_{\rm 2h}$ & $C_{\rm 2}$ & $C_{\rm s}$\\ \hline
$Q_{0}$, $Q_{4}$ & --- & --- & $T_{0}$, $T_{4}$ &
A$_{1g}$ & A$_{1}$ & A$_{1}$ & A$_{g}$ & A &
A$_{1g}$ & A$_{1}$ & A$_{g}$ & A$_{1}$ & A$_{1}$ & A & A &
A$_{g}$ & A & A$_{1}$ & A$_{g}$ & A & A$'$ \\
--- & $G_{xyz}$ & $M_{xyz}$ & --- &
A$_{2g}$ & A$_{2}$ & A$_{2}$ & A$_{g}$ & A &
B$_{1g}$ & B$_{1}$ & B$_{g}$ & B$_{1}$ & B$_{1}$ & B & B &
A$_{g}$ & A & A$_{1}$ & A$_{g}$ & A & A$'$ \\
$Q_{u}$, $Q_{4u}$ & --- & --- & $T_{u}$, $T_{4u}$ &
E$_{g}$ & E & E & E$_{g}$ & E &
A$_{1g}$ & A$_{1}$ & A$_{g}$ & A$_{1}$ & A$_{1}$ & A & A &
A$_{g}$ & A & A$_{1}$ & A$_{g}$ & A & A$'$ \\
$Q_{v}$, $Q_{4v}$ & --- & --- & $T_{v}$, $T_{4v}$ &
& & & & &
B$_{1g}$ & B$_{1}$ & B$_{g}$ & B$_{1}$ & B$_{1}$ & B & B &
A$_{g}$ & A & A$_{1}$ & A$_{g}$ & A & A$'$ \\
$Q_{4x}^{\alpha}$ & $G_{x}$, $G_{x}^{\alpha}$  & $M_{x}$, $M_{x}^{\alpha}$ & $T_{4x}^{\alpha}$ &
T$_{1g}$ & T$_{1}$ & T$_{1}$ & T$_{g}$ & T &
E$_{g}$ & E & E$_{g}$ & E & E & E & E &
B$_{3g}$ & B$_{3}$ & B$_{2}$ & B$_{g}$ & B & A$''$ \\
$Q_{4y}^{\alpha}$ & $G_{y}$, $G_{y}^{\alpha}$ & $M_{y}$, $M_{y}^{\alpha}$ & $T_{4y}^{\alpha}$ &
& & & & &
& & & & & & &
B$_{2g}$ & B$_{2}$ & B$_{1}$ & B$_{g}$ & B & A$''$ \\
$Q_{4z}^{\alpha}$ & $G_{z}$, $G_{z}^{\alpha}$ & $M_{z}$, $M_{z}^{\alpha}$ & $T_{4z}^{\alpha}$ &
& & & & &
A$_{2g}$ & A$_{2}$ & A$_{g}$ & A$_{2}$ & A$_{2}$ & A & A &
B$_{1g}$ & B$_{1}$ & A$_{2}$ & A$_{g}$ & A & A$'$ \\
$Q_{yz}$, $Q_{4x}^{\beta}$ & $G_{x}^{\beta}$ &$M_{x}^{\beta}$ &  $T_{yz}$, $T_{4x}^{\beta}$ &
T$_{2g}$ & T$_{2}$ & T$_{2}$ & T$_{g}$ & T &
E$_{g}$ & E & E$_{g}$ & E & E & E & E &
B$_{3g}$ & B$_{3}$ & B$_{2}$ & B$_{g}$ & B & A$''$ \\
$Q_{zx}$, $Q_{4y}^{\beta}$ & $G_{y}^{\beta}$ & $M_{y}^{\beta}$ & $T_{zx}$, $T_{4y}^{\beta}$ &
& & & & &
& & & & & & &
B$_{2g}$ & B$_{2}$ & B$_{1}$ & B$_{g}$ & B & A$''$ \\
$Q_{xy}$, $Q_{4z}^{\beta}$ & $G_{z}^{\beta}$ & $M_{z}^{\beta}$ & $T_{xy}$, $T_{4z}^{\beta}$ &
& & & & &
B$_{2g}$ & B$_{2}$ & B$_{g}$ & B$_{2}$ & B$_{2}$ & B & B &
B$_{1g}$ & B$_{1}$ & A$_{2}$ & A$_{g}$ & A & A$'$ \\ \hline
--- & $G_{0}$, $G_{4}$ & $M_{0}$, $M_{4}$ & --- &
A$_{1u}$ & A$_{1}$ & A$_{2}$ & A$_{u}$ & A &
A$_{1u}$ & A$_{1}$ & A$_{u}$ & B$_{1}$ & A$_{2}$ & A & B &
A$_{u}$ & A & A$_{2}$ & A$_{u}$ & A & A$''$ \\
$Q_{xyz}$ & --- & --- & $T_{xyz}$ &
A$_{2u}$ & A$_{2}$ & A$_{1}$ & A$_{u}$ & A &
B$_{1u}$ & B$_{1}$ & B$_{u}$ & A$_{1}$ & B$_{2}$ & B & A &
A$_{u}$ & A & A$_{2}$ & A$_{u}$ & A & A$''$ \\
--- & $G_{u}$, $G_{4u}$ & $M_{u}$, $M_{4u}$ & --- &
E$_{u}$ & E & E & E$_{u}$ & E &
A$_{1u}$ & A$_{1}$ & A$_{u}$ & B$_{1}$ & A$_{2}$ & A & B &
A$_{u}$ & A & A$_{2}$ & A$_{u}$ & A & A$''$ \\
--- & $G_{v}$, $G_{4v}$ & $M_{v}$, $M_{4v}$ & --- &
& & & & &
B$_{1u}$ & B$_{1}$ & B$_{u}$ & A$_{1}$ & B$_{2}$ & B & A &
A$_{u}$ & A & A$_{2}$ & A$_{u}$ & A & A$''$ \\
$Q_{x}$, $Q_{x}^{\alpha}$ & $G_{4x}^{\alpha}$ & $M_{4x}^{\alpha}$ & $T_{x}$, $T_{x}^{\alpha}$ &
T$_{1u}$ & T$_{1}$ & T$_{2}$ & T$_{u}$ & T &
E$_{u}$ & E & E$_{u}$ & E & E & E & E &
B$_{3u}$ & B$_{3}$ & B$_{1}$ & B$_{u}$ & B & A$'$ \\
$Q_{y}$, $Q_{y}^{\alpha}$ & $G_{4y}^{\alpha}$ & $M_{4y}^{\alpha}$ & $T_{y}$, $T_{y}^{\alpha}$ &
& & & & &
& & & & & & &
B$_{2u}$ & B$_{2}$ & B$_{2}$ & B$_{u}$ & B & A$'$ \\
$Q_{z}$, $Q_{z}^{\alpha}$ & $G_{4z}^{\alpha}$ & $M_{4z}^{\alpha}$ & $T_{z}$, $T_{z}^{\alpha}$ &
& & & & &
A$_{2u}$ & A$_{2}$ & A$_{u}$ & B$_{2}$ & A$_{1}$ & A & B &
B$_{1u}$ & B$_{1}$ & A$_{1}$ & A$_{u}$ & A & A$''$ \\
$Q_{x}^{\beta}$ & $G_{yz}$, $G_{4x}^{\beta}$ & $M_{yz}$, $M_{4x}^{\beta}$ & $T_{x}^{\beta}$ &
T$_{2u}$ & T$_{2}$ & T$_{1}$ & T$_{u}$ & T &
E$_{u}$ & E & E$_{u}$ & E & E & E & E &
B$_{3u}$ & B$_{3}$ & B$_{1}$ & B$_{u}$ & B & A$'$ \\
$Q_{y}^{\beta}$ & $G_{zx}$, $G_{4y}^{\beta}$ & $M_{zx}$, $M_{4y}^{\beta}$ & $T_{y}^{\beta}$ &
& & & & &
& & & & & & &
B$_{2u}$ & B$_{2}$ & B$_{2}$ & B$_{u}$ & B & A$'$ \\
$Q_{z}^{\beta}$ & $G_{xy}$, $G_{4z}^{\beta}$ & $M_{xy}$, $M_{4z}^{\beta}$ & $T_{z}^{\beta}$ &
& & & & &
B$_{2u}$ & B$_{2}$ & B$_{u}$ & A$_{2}$ & B$_{1}$ & B & A &
B$_{1u}$ & B$_{1}$ & A$_{1}$ & A$_{u}$ & A & A$''$ \\
\hline\hline
\end{tabular}
\end{table*}

Now, let us consider a typical example by considering the system with the $p$-$d$ hybridized orbitals under the $D_{2{\rm d}}$ group.  
From Table~\ref{tab_cefparam}, the atomic CEF Hamiltonian is represented by 
\begin{align}
\label{eq:HamCEF_ex1}
\mathcal{H}_{\rm CEF}&= \mathcal{H}^{\rm even}_{\rm CEF} + \mathcal{H}^{\rm odd}_{\rm CEF} + \mathcal{H}_{\Delta}, \\
\mathcal{H}^{\rm even}_{\rm CEF} &=c_{20} O^{\rm (c)}_{20}
+c_{40} O^{\rm (c)}_{40}
+c_{44} O^{\rm (c)}_{44}, \\
\mathcal{H}^{\rm odd}_{\rm CEF}&=c'_{32} O^{\rm (s)}_{32}. 
\end{align}
Note that $c_{60}$, $c_{64}$, and $c'_{52}$ become zero for the $p$-$d$ hybridized systems, as the maximum rank is given by $2\,{\rm max}(L_{1}=1,L_{2}=2)=4$. 
$\mathcal{H}^{\rm even}_{\rm CEF}$ represents the even-parity CEF Hamiltonian, which split three $p$ orbitals $(\phi_x, \phi_y, \phi_z)$ into two orbitals $(\phi_x, \phi_y)$ with the irreducible representation E and a orbital $\phi_z$ with the representation B$_{2}$. 
The five $d$ orbitals $(\phi_{u}, \phi_{v}, \phi_{yz}, \phi_{zx}, \phi_{xy})$ into three single orbitals $\phi_{u}$ with A$_{1}$, $\phi_{v}$ with B$_{1}$, and $\phi_{xy}$ with B$_{2}$ and two orbitals $(\phi_{yz}, \phi_{zx})$ with E. 
Meanwhile, $\mathcal{H}^{\rm odd}_{\rm CEF}$ leads to the odd-parity hybridization between $p$ and $d$ orbitals belonging to the same irreducible representation. 
In the present case, $\phi_z$ hybridizes with $\phi_{xy}$, and $(\phi_x, \phi_y)$ hybridize with $(\phi_{yz}, \phi_{zx})$.   
$\mathcal{H}_{\Delta}$ represents the atomic energy level for $p$ and $d$ orbitals where the energy difference is taken as $\Delta=\varepsilon_{p}-\varepsilon_{d}$. 

\begin{figure}[h!]
\begin{center}
\includegraphics[width=1.0 \hsize]{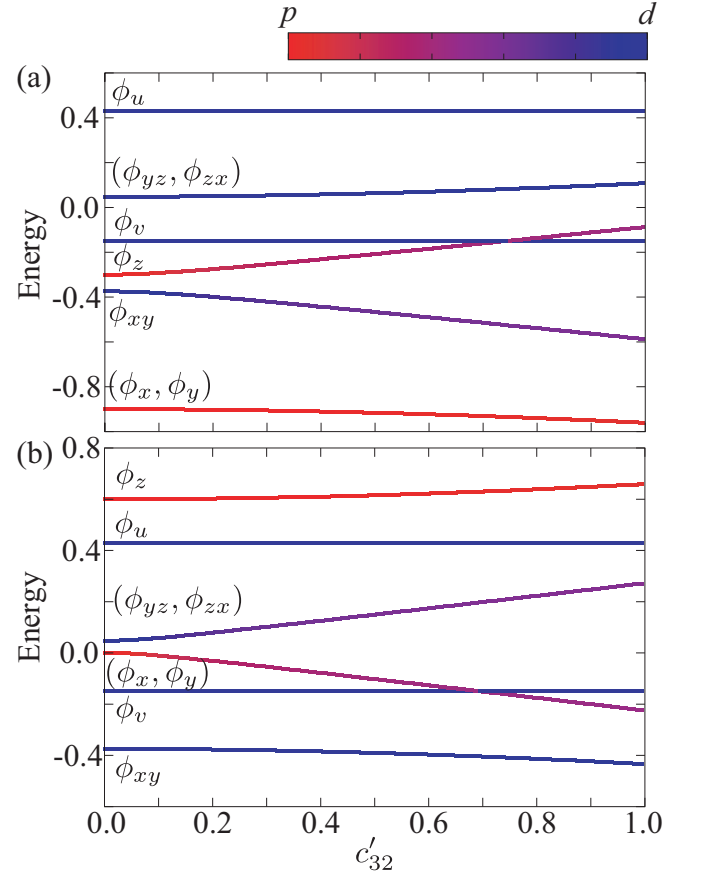} 
\caption{
\label{Fig:CEF}
Odd-parity CEF dependences of the atomic energy level for $p$-$d$ hybrid orbitals under the tetragonal $D_{\rm 2d}$ group. 
The colors show the weights for $p$ and $d$ orbitals. 
The difference of the atomic levels for $p$ and $d$ orbitals is taken as (a) $\Delta=-0.7$ and (b) $\Delta=0.2$, and the CEF parameters are $c_{20}=1$, $c_{40}=0.5$, and $c_{44}=0.4$. 
}
\end{center}
\end{figure}

The CEF levels in Eq.~(\ref{eq:HamCEF_ex1}) are shown in Fig.~\ref{Fig:CEF}, which largely depends on the model parameters.  
The strong odd-parity hybridization between $\phi_{z}$ and $\phi_{xy}$ is expected for $\Delta=-0.7$ in Fig.~\ref{Fig:CEF}(a), while that between $(\phi_x, \phi_y)$ and $(\phi_{yz}, \phi_{zx})$ occurs for $\Delta=0.2$ in Fig.~\ref{Fig:CEF}(b). 
These odd-parity hybridizations make odd-parity multipole degrees of freedom active, as discussed in the subsequent subsection.

\subsection{Active multipoles under point groups}
\label{sec:Active multipoles in solids}

When a CEF splitting is large and one of CEF multiplets dominates low-energy physics, the point-group irreducible representation is suitable to classify multipole moments. 
The classification is done by using the reduction rules, which are summarized in Tables~\ref{tab_multipoles_table1} and \ref{tab_multipoles_table2} in Appendix~\ref{sec:Electric multipoles and CEF parameters} for 32 point groups in 7 crystal systems. 

Tables~\ref{tab_multipoles_table1} and \ref{tab_multipoles_table2} are useful to investigate what type of multipoles are activated in the specific crystal structures, which stimulate microscopic understanding of physical phenomena induced by multipoles, as discussed in Sec.~\ref{sec:Physical properties under multipoles}. 
For example, the magneto-current effect, where the uniform magnetization is induced by electric current as was discussed in Sec.~\ref{sec:Linear response tensors}, can occur in the point groups ($O$, $T$, $D_{4}$, $C_{4}$, $D_{6}$, $C_{6}$, $D_{3}$, $C_{3}$, $D_{2}$, $D_{2{\rm d}}$, $C_{4{\rm v}}$, $S_{4}$, $C_{2}$, $C_{\rm s}$, $C_{6{\rm v}}$, $C_{3{\rm v}}$, $C_{2{\rm v}}$, $C_{1}$), which is so-called the gyrotropic point groups.
This is because any of the relevant multipoles for the magneto-current effect [$G_0$, $(Q_x, Q_y, Q_z)$, and $(G_u, G_v, G_{yz}, G_{zx}, G_{xy})$] belong to the totally symmetric representation in the gyrotropic point groups.  

Moreover, tables~\ref{tab_multipoles_table1} and \ref{tab_multipoles_table2} show what multipoles are potential order parameters in the systems. 
Note that as the E (M) and ET (MT) multipoles belong to different irreducible representations in some point groups, the distinction of them should be essential in such point groups.

Let us demonstrate how to identify active multipole degrees of freedom in the low-energy multiplets by using the $p$-$d$ hybridized CEF states in the tetragonal $D_{\rm 2d}$ group as was discussed in the previous subsection. 
First, we consider the basis functions $(\phi_{yz}, \phi_{zx})$ in the representation E under the $D_{\rm 4h}$ group, which is the parent group of $D_{\rm 2d}$ and does not show orbital hybridization with different parities ($c_{32}'=0$). 
By taking the direct product of the basis functions, the active multipole degrees of freedom is obtained as 
\begin{align}
{\rm E}_{g} \otimes {\rm E}_{g} = {\rm A}^+_{1g} \oplus {\rm A}^-_{2g} \oplus {\rm B}^+_{1g} \oplus {\rm B}^+_{2g},  
\end{align}
where the superscripts $+$ ($-$) represent time-reversal even (odd), and all the multipoles are even-parity (the subscript $g$). 
Note that the symmetrized (anti-symmetrized) product of bases corresponds to time-reversal even(odd) operators when the basis functions are spinless (the angular momentum is an integer).
On the other hand, the symmetrized (anti-symmetrized) product of the basis functions correspond to time-reversal odd(even) operators when the basis functions are spinful (the angular momentum is a half integer). 
The above irreducible decomposition for the spinless basis in the tetragonal crystal system is summarized in Table~\ref{tab_tetra_less}.

The corresponding multipoles are given by 
\begin{align}
{\rm A}^+_{1g}: \quad &  \hat{\rho}_0 \leftrightarrow Q_0, \\
{\rm A}^-_{2g}: \quad &  \hat{\rho}_y \leftrightarrow M_z, \\
{\rm B}^+_{1g}: \quad &  \hat{\rho}_z \leftrightarrow Q_v, \\
{\rm B}^+_{2g}: \quad &  \hat{\rho}_x \leftrightarrow Q_{xy},
\end{align}
where $\hat{\rho}_{0}$ and $\hat{\bm{\rho}}$ are $2\times 2$ unit and Pauli matrices for doubly degenerate orbitals $(\phi_{yz}, \phi_{zx})$. 
Obviously, only even-parity E and M multipoles are active in this CEF multiplets.

\begin{table*}[t!]
\caption{
Multipoles under the tetragonal crystal system ($D_{4\rm h}$ and related groups) in the spinless basis. 
We introduce the abbreviation, $(g,u)^{\pm}={\rm A}^{\pm}_{1g,u}\oplus{\rm A}^{\mp}_{2g,u}\oplus{\rm B}^{\pm}_{1g,u}\oplus{\rm B}^{\pm}_{2g,u}$. 
The subscripts $g,u$ are omitted for the $D_{4}$ group. 
The subscripts $1,2$ are omitted for the $C_{4{\rm h}}$ group. 
A$_{1u}$ $\leftrightarrow$ B$_{1u}$ and A$_{2u}$ $\leftrightarrow$ B$_{2u}$, and then, the subscripts $g,u$ are omitted for the $D_{2{\rm d}}$ group. 
A$_{1u}$ $\leftrightarrow$ B$_{1u}$ and A$_{2u}$ $\leftrightarrow$ B$_{2u}$, and then, the subscripts $g,u$ and $1,2$ are omitted for the $S_{4}$ group. 
A$_{1u}$ $\leftrightarrow$ A$_{2u}$ and B$_{1u}$ $\leftrightarrow$ B$_{2u}$, and then, the subscripts $g,u$ are omitted for the $C_{4{\rm v}}$ group. 
A$_{1u}$ $\leftrightarrow$ A$_{2u}$ and B$_{1u}$ $\leftrightarrow$ B$_{2u}$, and then, the subscripts $g,u$ and $1,2$ are omitted for the $C_{4}$ group. 
}
\label{tab_tetra_less}
\centering
\begin{tabular}{c|c|cc|cccc|ccccc}
& (s) A$_{1g}$ & (p) A$_{2u}$ & E$_{u}$ & (d) A$_{1g}$ & B$_{1g}$ & B$_{2g}$ & E$_{g}$ & (f) B$_{1u}$ & A$_{2u}$ & E$_{u}$ & B$_{2u}$ & E$_{u}$ \\ \hline
(s) A$_{1g}$ & A$^+_{1g}$ & A$^{\pm}_{2u}$ & E$^{\pm}_{u}$ & A$^{\pm}_{1g}$ & B$^{\pm}_{1g}$ & B$^{\pm}_{2g}$ & E$^{\pm}_{g}$ & B$^{\pm}_{1u}$ & A$^{\pm}_{2u}$ &E$^{\pm}_{u}$ & B$^{\pm}_{2u}$ & E$^{\pm}_{u}$ \\ \hline
(p) A$_{2u}$ & & A$^+_{1g}$ & E$^{\pm}_{g}$ & A$^{\pm}_{2u}$ & B$^{\pm}_{2u}$ & B$^{\pm}_{1u}$ & E$^{\pm}_{u}$ & B$^{\pm}_{2g}$ & A$^{\pm}_{1g}$ & $E^{\pm}_{g}$ & B$^{\pm}_{1g}$ & E$^{\pm}_{g}$ \\
E$_{u}$ & & & $(g)^+$ & E$^{\pm}_{u}$ & E$^{\pm}_{u}$ & E$^{\pm}_{u}$ & $(u)^{\pm}$ & E$^{\pm}_{g}$ & E$^{\pm}_{g}$ & $(g)^{\pm}$ & E$^{\pm}_{g}$ & $(g)^{\pm}$ \\ \hline
(d) A$_{1g}$ & & & & A$^+_{1g}$ & B$^{\pm}_{1g}$ & B$^{\pm}_{2g}$ & E$^{\pm}_{g}$ & B$^{\pm}_{1u}$ & A$^{\pm}_{2u}$ & E$^{\pm}_{u}$ & B$^{\pm}_{2u}$ & E$^{\pm}_{u}$ \\
B$_{1g}$ & & & & & A$^+_{1g}$ & A$^{\pm}_{2g}$ & E$^{\pm}_{g}$ & A$^{\pm}_{1u}$ & B$^{\pm}_{2u}$ & E$^{\pm}_{u}$ & A$^{\pm}_{2u}$ & E$^{\pm}_{u}$ \\
B$_{2g}$ & & & & & & A$^+_{1g}$ & E$^{\pm}_{g}$ & A$^{\pm}_{2u}$ & B$^{\pm}_{1u}$ &E$^{\pm}_{u}$ & A$^{\pm}_{1u}$ & E$^{\pm}_{u}$ \\
E$_{g}$ & & & & & & & $(g)^+$ & E$^{\pm}_{u}$ & E$^{\pm}_{u}$ & $(u)^{\pm}$ & E$^{\pm}_{u}$ & $(u)^{\pm}$ \\ \hline
(f) B$_{1u}$ & & & & & & & & A$^+_{1g}$ & B$^{\pm}_{2g}$ & E$^{\pm}_{g}$ & A$^{\pm}_{2g}$ & E$^{\pm}_{g}$ \\
A$_{2u}$ & & & & & & & & & A$^+_{1g}$ & E$^{\pm}_{g}$ & B$^{\pm}_{1g}$ & E$^{\pm}_{g}$ \\
E$_{u}$ & & & & & & & & & & $(g)^+$ & E$^{\pm}_{g}$ & $(g)^{\pm}$ \\
B$_{2u}$ & & & & & & & & & & & A$^+_{1g}$ & E$^{\pm}_{g}$ \\
E$_{u}$ & & & & & & & & & & & & $(g)^+$
\end{tabular}
\end{table*}

Next, let us consider the basis functions $(\phi_z, \phi_{xy})$ belonging to the representation B$_2$ under the $D_{2{\rm d}}$ group. 
The multipole degrees of freedom are obtained by decomposing the direct product of the basis functions (using Table~\ref{tab_tetra_less}) as 
\begin{align}
({\rm B}_{2}\oplus {\rm B}_{2}) \otimes ({\rm B}_{2}\oplus {\rm B}_{2})= (2 {\rm A}^+_{1})_{\rm intra}\oplus ({\rm A}^+_{1} \oplus {\rm A}^-_{1})_{\rm inter}, 
\end{align}
where $(\cdots)_{\rm intra}$ and $(\cdots)_{\rm inter}$ represent the multipoles activated in non-hybrid and hybrid orbitals, respectively. 
A set of independent operators in the basis $(\phi_z, \phi_{xy})$ is given by 
\begin{align}
{\rm A}^+_{1}: \quad & \{\hat{\tau}_0, \hat{\tau}_z \} \leftrightarrow \{Q_{0}, Q_{u} \},  \quad \hat{\tau}_x \leftrightarrow G_{v},   \\
{\rm A}^-_{1}: \quad &\hat{\tau}_y \leftrightarrow M_{v}, 
\end{align} 
where $\hat{\tau}_{0}$ and $\hat{\bm{\tau}}$ are $2\times 2$ unit and Pauli matrices in the basis $(\phi_z, \phi_{xy})$ and $\{ \cdots \}$ represent the appropriate linear combination. 
$\tau_0$ and $\tau_z$ represent the intra-orbital degree of freedom, while $\tau_x$ and $\tau_y$ represent the inter-orbital degree of freedom.

The expected linear responses are uniquely found once the multipole degrees of freedom are identified. 
In this case, the magneto-current effect is induced by the emergence of the ET quadrupole $G_{v}\propto k_x \sigma_x -k_y \sigma_y$; the induced magnetization $M_x$ by the electric current in the $x$ direction and $M_y$ by the electric current in the $y$ direction should have opposite sign and the same magnitude. 
The result implies that larger magneto-current response is expected for larger ET quadrupole $G_{v}$. 
In a similar way, the magneto-electric effect is anticipated once the time-reversal symmetry breaking occurs, as the anti-symmetrized representation is the M quadrupole $M_{v}$.

In the spinful basis, possible multipoles are obtained by changing the irreducible representation of the basis function as ${\rm B}_2 \otimes {\rm E}_{1/2} \to {\rm E}_{3/2}$, and the decomposition according to Table~\ref{tab_tetra_ful} gives 
\begin{align}
&({\rm E}_{3/2}\oplus {\rm E}_{3/2}) \otimes ({\rm E}_{3/2}\oplus {\rm E}_{3/2}) \cr
&\quad
= (2 {\rm A}^+_{1} \oplus 2{\rm A}^-_{2}\oplus 2 {\rm E}^-)_{\rm intra} \cr
&\quad\quad
\oplus ({\rm A}^+_{1}\oplus {\rm A}^+_{2} \oplus {\rm E}^+ \oplus {\rm A}^-_{1} \oplus {\rm A}^-_{2} \oplus {\rm E}^-)_{\rm inter}.   
\end{align}
A set of independent operators in the basis $(\phi_z, \phi_{xy})$ is given by
\begin{align}
{\rm A}^+_{1}: \quad &\{ \hat{\tau}_0 \hat{\sigma}_0, \hat{\tau}_z  \hat{\sigma}_0 \} \leftrightarrow \{ Q_{0}, Q_{u} \}, \quad \hat{\tau}_x  \hat{\sigma}_0 \leftrightarrow G_{v},   \\
{\rm A}^+_{2}: \quad & \hat{\tau}_y \hat{\sigma}_z \leftrightarrow G_{xy}, \\
{\rm E}^+: \quad & (\hat{\tau}_y \hat{\sigma}_x, \hat{\tau}_y \hat{\sigma}_y) \leftrightarrow (Q_x, Q_y),  \\
{\rm A}^-_{1}: \quad &\hat{\tau}_y \hat{\sigma}_0 \leftrightarrow M_{v}, \\
{\rm A}^-_{2}: \quad & \{ \hat{\tau}_0 \hat{\sigma}_z, \hat{\tau}_z \hat{\sigma}_z \} \leftrightarrow \{ M_z, M^{\alpha}_z \}, \quad \hat{\tau}_x \hat{\sigma}_z \leftrightarrow M_{xy}, \\
{\rm E}^-: \quad &\{ (\hat{\tau}_0 \hat{\sigma}_x, \hat{\tau}_0 \hat{\sigma}_y), (\hat{\tau}_z \hat{\sigma}_x, \hat{\tau}_z \hat{\sigma}_y) \} \cr&\qquad\qquad\qquad
 \leftrightarrow \{ (M_x, M_y), (M^{\alpha}_x, M^{\alpha}_y) \},  \cr
& (\hat{\tau}_x \hat{\sigma}_x, \hat{\tau}_x \hat{\sigma}_y) \leftrightarrow (T_x, T_y), 
\end{align} 

\begin{table*}[t!]
\caption{
Multipoles under the tetragonal crystal system ($D_{4\rm h}$ and related groups) in the spinful basis. 
We introduce the abbreviation, 
$(g,u)^{\pm}={\rm A}^{\pm}_{1g,u}\oplus{\rm A}^{\mp}_{2g,u}\oplus{\rm E}^{\mp}_{g,u}$, and $\braket{g,u}^{\pm}={\rm B}^{\pm}_{1g,u}\oplus{\rm B}^{\pm}_{2g,u}\oplus{\rm E}^{\pm}_{g,u}$.
The subscripts $g,u$ are omitted for the $D_{4}$ group. 
The subscripts $1,2$ are omitted for the $C_{4{\rm h}}$ group. 
A$_{1u}$ $\leftrightarrow$ B$_{1u}$, A$_{2u}$ $\leftrightarrow$ B$_{2u}$, and E$_{1/2u}$ $\leftrightarrow$ E$_{3/2u}$, and then, the subscripts $g,u$ are omitted for the $D_{2{\rm d}}$ group. 
A$_{1u}$ $\leftrightarrow$ B$_{1u}$, A$_{2u}$ $\leftrightarrow$ B$_{2u}$, and E$_{1/2u}$ $\leftrightarrow$ E$_{3/2u}$, and then, the subscripts $g,u$ and $1,2$ are omitted for the $S_{4}$ group. 
A$_{1u}$ $\leftrightarrow$ A$_{2u}$ and B$_{1u}$ $\leftrightarrow$ B$_{2u}$, and then, the subscripts $g,u$ are omitted for the $C_{4{\rm v}}$ group. 
A$_{1u}$ $\leftrightarrow$ A$_{2u}$ and B$_{1u}$ $\leftrightarrow$ B$_{2u}$, and then, the subscripts $g,u$ and $1,2$ are omitted for the $C_{4}$ group. 
}
\label{tab_tetra_ful}
\centering
\scalebox{0.85}{
\begin{tabular}{cc|c|ccc|ccccc|ccccccc}
& & (s) E$_{1/2g}$ & (p) E$_{1/2u}$ & E$_{1/2u}$ & E$_{3/2u}$ & (d) E$_{1/2g}$ & E$_{3/2g}$ & E$_{3/2g}$ & E$_{1/2g}$ & E$_{3/2g}$ & (f) E$_{3/2u}$ & E$_{1/2u}$ & E$_{1/2u}$ & E$_{3/2u}$ & E$_{3/2u}$ & E$_{1/2u}$ & E$_{3/2u}$ \\ \hline
(s) A$_{1g}\otimes$ E$_{1/2g} \to$ & E$_{1/2g}$ & $(g)^+$ & $(u)^{\pm}$ & $(u)^{\pm}$ & $\braket{u}^{\pm}$ & $(g)^{\pm}$ & $\braket{g}^{\pm}$ & $\braket{g}^{\pm}$ & $(g)^{\pm}$ & $\braket{g}^{\pm}$ & $\braket{u}^{\pm}$ & $(u)^{\pm}$ & $(u)^{\pm}$ & $\braket{u}^{\pm}$ & $\braket{u}^{\pm}$ & $(u)^{\pm}$ & $\braket{u}^{\pm}$  \\ \hline
(p) A$_{2u}\otimes$ E$_{1/2g} \to$ & E$_{1/2u}$ & & $(g)^+$ & $(g)^{\pm}$ & $\braket{g}^{\pm}$ & $(u)^{\pm}$ & $\braket{u}^{\pm}$ &  $\braket{u}^{\pm}$ & $(u)^{\pm}$ & $\braket{u}^{\pm}$ & $\braket{g}^{\pm}$ & $(g)^{\pm}$ & $(g)^{\pm}$ & $\braket{g}^{\pm}$ & $\braket{g}^{\pm}$ & $(g)^{\pm}$ & $\braket{g}^{\pm}$ \\
E$_{u}\otimes$ E$_{1/2g} \to$ & E$_{1/2u}$ & & & $(g)^+$ & $\braket{g}^{\pm}$ & $(u)^{\pm}$ & $\braket{u}^{\pm}$ & $\braket{u}^{\pm}$ & $(u)^{\pm}$ & $\braket{u}^{\pm}$ & $\braket{g}^{\pm}$ & $(g)^{\pm}$ & $(g)^{\pm}$ & $\braket{g}^{\pm}$ & $\braket{g}^{\pm}$ & $(g)^{\pm}$ & $\braket{g}^{\pm}$\\
 & E$_{3/2u}$ & & & & $(g)^+$ & $\braket{u}^{\pm}$ & $(u)^{\pm}$ & $(u)^{\pm}$ & $\braket{u}^{\pm}$ & $(u)^{\pm}$ & $(g)^{\pm}$ & $\braket{g}^{\pm}$ & $\braket{g}^{\pm}$ & $(g)^{\pm}$ & $(g)^{\pm}$ & $\braket{g}^{\pm}$ & $(g)^{\pm}$ \\ \hline
(d) A$_{1g}\otimes$ E$_{1/2g} \to$ & $E_{1/2g}$ & & & & & $(g)^+$ & $\braket{g}^{\pm}$ & $\braket{g}^{\pm}$ & $(g)^{\pm}$ & $\braket{g}^{\pm}$ & $\braket{u}^{\pm}$ & $(u)^{\pm}$ & $(u)^{\pm}$ & $\braket{u}^{\pm}$ & $\braket{u}^{\pm}$ & $(u)^{\pm}$ & $\braket{u}^{\pm}$ \\
B$_{1g}\otimes$ E$_{1/2g} \to$ & E$_{3/2g}$ & & & & & & $(g)^+$ & $(g)^{\pm}$ & $\braket{g}^{\pm}$ & $(g)^{\pm}$ & $(u)^{\pm}$ & $\braket{u}^{\pm}$ & $\braket{u}^{\pm}$ & $(u)^{\pm}$ & $(u)^{\pm}$ & $\braket{u}^{\pm}$ & $(u)^{\pm}$ \\
B$_{2g}\otimes$ E$_{1/2g} \to$ & E$_{3/2g}$ & & & & & & & $(g)^+$ & $\braket{g}^{\pm}$ & $(g)^{\pm}$ & $(u)^{\pm}$ & $\braket{u}^{\pm}$ & $\braket{u}^{\pm}$ & $(u)^{\pm}$ & $(u)^{\pm}$ & $\braket{u}^{\pm}$ & $(u)^{\pm}$ \\
E$_{g}\otimes$ E$_{1/2g} \to$ & E$_{1/2g}$ & & & & & & & & $(g)^+$ & $\braket{g}^{\pm}$ & $\braket{u}^{\pm}$ & $(u)^{\pm}$ & $(u)^{\pm}$ & $\braket{u}^{\pm}$ & $\braket{u}^{\pm}$ & $(u)^{\pm}$ & $\braket{u}^{\pm}$ \\
& E$_{3/2g}$ & & & & & & & & & $(g)^+$ & $(u)^{\pm}$ & $\braket{u}^{\pm}$ & $\braket{u}^{\pm}$ & $(u)^{\pm}$ & $(u)^{\pm}$ & $\braket{u}^{\pm}$ & $(u)^{\pm}$ \\ \hline
(f) B$_{1u}\otimes$ E$_{1/2g} \to$ & E$_{3/2u}$ & & & & & & & & & & $(g)^+$ & $\braket{g}^{\pm}$ & $\braket{g}^{\pm}$ & $(g)^{\pm}$ & $(g)^{\pm}$ & $\braket{g}^{\pm}$ & $(g)^{\pm}$ \\
A$_{2u}\otimes$ E$_{1/2g} \to$ & E$_{1/2u}$ & & & & & & & & & & & $(g)^+$ & $(g)^{\pm}$ & $\braket{g}^{\pm}$ & $\braket{g}^{\pm}$ & $(g)^{\pm}$ & $\braket{g}^{\pm}$ \\
E$_{u}\otimes$ E$_{1/2g} \to$ & E$_{1/2u}$ & & & & & & & & & & & & $(g)^+$ & $\braket{g}^{\pm}$ & $\braket{g}^{\pm}$ & $(g)^{\pm}$ & $\braket{g}^{\pm}$ \\
& E$_{3/2u}$ & & & & & & & & & & & & & $(g)^+$ & $(g)^{\pm}$ & $\braket{g}^{\pm}$ & $(g)^{\pm}$ \\
B$_{2u}\otimes$ E$_{1/2g} \to$ & E$_{3/2u}$ & & & & & & & & & & & & & & $(g)^+$ & $\braket{g}^{\pm}$ & $(g)^{\pm}$ \\
E$_{u}\otimes$ E$_{1/2g} \to$& E$_{1/2u}$ & & & & & & & & & & & & & & & $(g)^+$ & $\braket{g}^{\pm}$\\
& E$_{3/2u}$ & & & & & & & & & & & & & & & & $(g)^+$
\end{tabular}
}
\end{table*}

\begin{table*}
\caption{
Even-parity multipoles belonging to totally symmetric representation in the point groups except for triclinic ones $(C_1, C_{\rm i})$. 
$\chi^{\rm P}_{ij}$ and $\chi^{\rm A}_{ijk}$ represent rank-2 polar and rank-3 axial tensors, respectively. 
$\chi^{\rm P}_{ij}$ ($\chi_{ij}^{\rm (2)}$) has 9 components and $\chi^{\rm A}_{ijk}$ has 27 components at most. 
The components of $\chi^{\rm A}_{ijk}$ are decomposed into symmetric monopole response denoted by 1 with 3 components ($\chi_{i}^{\rm (1)}$), antisymmetric dipole response denoted by 2 with 9 components ($\chi_{ij}^{\rm (2)}$), and symmetric quadrupole response denoted by 3 with 15 components ($\chi_{ijk}^{\rm (3)}$), respectively. 
The independent number of components in $\chi^{\rm P}_{ij}$ and $\chi^{\rm A}_{ijk}$ is shown for each point group. 
Responses in the absence of time-reversal symmetry, such as the magnetoelectric effect and Nernst effect, are obtained by replacing $(Q_{lm}, G_{lm})$ with $(T_{lm}, M_{lm})$. 
}
\label{tab_totalsymmetric_e}
\centering
\renewcommand{\arraystretch}{1.2}
\begin{tabular}{cccccccccccccccc} \hline\hline
crystal system & point group& N & P & C & G & \multicolumn{5}{c}{totally symmetric rep.} & $\chi^{\rm P}_{ij}$ & \multicolumn{3}{c}{$\chi^{\rm A}_{ijk}$} \\ 
& & & & & & rank 0& rank 1 & rank 2 & rank 3 & rank 4 & 2  & 1 & 2 & 3   \\ \hline
cubic & $O_{\rm h}$&  &  &  &  & $Q_0$& & & & $Q_4$ & 1 & & 1 &   \\
& $O$& \checkmark &  & \checkmark & \checkmark & $Q_0$ & & & & $Q_4$ & 1  & & 1 &  \\
& $T_{\rm d}$& \checkmark & &  &  & $Q_0$ & & & & $Q_4$ & 1 & & 1 &   \\
& $T$ & \checkmark &  & \checkmark & \checkmark& $Q_0$ & & & $G_{xyz}$ & $Q_4$ & 1 & & 1 & 1  \\ 
& $T_{\rm h}$&  &  &  &  & $Q_0$ & & & $G_{xyz}$ & $Q_4$ & 1 & & 1 & 1  \\ \hline
tetragonal & $D_{\rm 4h}$&  &  &  &  & $Q_0$ & & $Q_u$ & & $Q_4$, $Q_{4u}$ & 2 & & 2 & 1  \\
& $D_{4}^{}$& \checkmark &  & \checkmark & \checkmark & $Q_0$ & & $Q_u$ & & $Q_4$, $Q_{4u}$ & 2 & & 2  & 1  \\
& $D_{2{\rm d}}$& \checkmark &  &  & \checkmark &  $Q_0$ & & $Q_u$ & &$Q_4$, $Q_{4u}$ & 2 & & 2 & 1  \\
& $C_{4{\rm v}}$& \checkmark & \checkmark &  & \checkmark &  $Q_0$ & & $Q_u$& & $Q_4$, $Q_{4u}$ & 2 & & 2 & 1  \\
& $C_{4{\rm h}}$&  &  &  &  &  $Q_0$ & $G_z$ & $Q_u$ & $G_z^\alpha$ & $Q_4$, $Q_{4u}$, $Q_{4z}^{\alpha}$ & 3 & 1 & 3 & 3 \\
& $C_{4}$& \checkmark & \checkmark & \checkmark & \checkmark &  $Q_0$ & $G_z$ & $Q_u$ & $G_z^\alpha$ &$Q_4$, $Q_{4u}$, $Q_{4z}^{\alpha}$ & 3 & 1 & 3  & 3 \\
& $S_{4}$& \checkmark &  &  & \checkmark & $Q_0$ & $G_z$ & $Q_u$ & $G_z^\alpha$ & $Q_4$, $Q_{4u}$, $Q_{4z}^{\alpha}$ & 3 & 1 & 3 & 3  \\ \hline
orthorhombic & $D_{\rm 2h}$&  &  &  &  & $Q_0$ & & $Q_u$, $Q_v$ & $G_{xyz}$ & $Q_4$, $Q_{4u}$, $Q_{4v}$ & 3 & & 3 & 3  \\
& $D_{2}$& \checkmark &  & \checkmark & \checkmark & $Q_0$ & & $Q_u$, $Q_v$ & $G_{xyz}$ & $Q_4$, $Q_{4u}$, $Q_{4v}$ & 3 & & 3 & 3  \\
& $C_{2{\rm v}}$& \checkmark & \checkmark &  & \checkmark & $Q_0$ & & $Q_u$, $Q_v$ & $G_{xyz}$ & $Q_4$, $Q_{4u}$, $Q_{4v}$ & 3 & & 3 & 3  \\\hline
monoclinic & $C_{\rm 2h}$&  &  &  &  & $Q_0$ & $G_z$ & $Q_u$, $Q_v$, $Q_{xy}$ & $G_{xyz}$, $G_z^\alpha$, $G_z^\beta$ & $Q_4$, $Q_{4u}$, $Q_{4v}$, $Q_{4z}^{\alpha}$, $Q_{4z}^\beta$ & 5 &1 & 5 & 7   \\
& $C_{2}$& \checkmark & \checkmark & \checkmark & \checkmark & $Q_0$ & $G_z$ & $Q_u$, $Q_v$, $Q_{xy}$ & $G_{xyz}$, $G_z^\alpha$, $G_z^\beta$ & $Q_4$, $Q_{4u}$, $Q_{4v}$, $Q_{4z}^{\alpha}$, $Q_{4z}^\beta$ & 5 & 1 & 5 & 7 \\
& $C_{{\rm s}}$& \checkmark & \checkmark &  & \checkmark & $Q_0$ & $G_z$ & $Q_u$, $Q_v$, $Q_{xy}$ & $G_{xyz}$, $G_z^\alpha$, $G_z^\beta$ & $Q_4$, $Q_{4u}$, $Q_{4v}$, $Q_{4z}^{\alpha}$, $Q_{4z}^\beta$ & 5 & 1 & 5 & 7  \\
\hline
hexagonal & $D_{6{\rm h}}$&  &  &  &  & $Q_0$ & & $Q_u$ & & $Q_{40}$ & 2 & & 2 & 1  \\
& $D_{6}$& \checkmark &  & \checkmark & \checkmark &  $Q_0$ & & $Q_u$ & & $Q_{40}$ & 2 & & 2 & 1  \\
& $D_{3{\rm h}}$& \checkmark &  &  &  &  $Q_0$ & & $Q_u$ & & $Q_{40}$ & 2 & & 2 & 1   \\
& $C_{6{\rm v}}$& \checkmark & \checkmark &  & \checkmark &  $Q_0$ & & $Q_u$ & & $Q_{40}$ & 2 & & 2  & 1  \\
& $C_{6{\rm h}}$&  &  &  &  & $Q_0$ & $G_z$ & $Q_u$ & $G_z^\alpha$ & $Q_{40}$ & 3 & 1 & 3 & 3  \\
& $C_{6}$& \checkmark & \checkmark & \checkmark & \checkmark & $Q_0$ & $G_z$ & $Q_u$ & $G_z^\alpha$ & $Q_{40}$ & 3 & 1 & 3 & 3  \\
& $C_{3{\rm h}}$& \checkmark &  &  &  &  $Q_0$ & $G_z$ & $Q_u$ & $G_z^\alpha$ & $Q_{40}$ & 3 & 1 & 3 & 3  \\
 \hline
trigonal & $D_{3{\rm d}}$&  &  &  &  & $Q_0$ & & $Q_u$ & $G_{3a}$ & $Q_{40}$, $Q_{4a}$ & 2 & & 2 &2   \\
& $D_{3}$& \checkmark &  & \checkmark & \checkmark &  $Q_0$ & & $Q_u$ & $G_{3a}$ & $Q_{40}$, $Q_{4a}$ & 2 & & 2 & 2  \\
& $C_{3{\rm v}}$& \checkmark & \checkmark &  & \checkmark &  $Q_0$ & & $Q_u$ & $G_{3b}$ & $Q_{40}$, $Q_{4b}$ & 2 & & 2 & 2  \\
& $C_{3{\rm i}}$&  &  &  &  &  $Q_0$ & $G_z$ & $Q_u$ & $G_z^\alpha$, $G_{3a}$, $G_{3b}$ & $Q_{40}$, $Q_{4a}$, $Q_{4b}$ & 3 & 1 & 3 & 5  \\
& $C_{3}$& \checkmark & \checkmark & \checkmark & \checkmark &  $Q_0$ & $G_z$ & $Q_u$ & $G_z^\alpha$, $G_{3a}$, $G_{3b}$ & $Q_{40}$, $Q_{4a}$, $Q_{4b}$ & 3 & 1 & 3 & 5  \\
\hline\hline
\end{tabular}
\end{table*}

\begin{table*}
\caption{
Odd-parity multipoles belonging to totally symmetric representation in the point groups except for triclinic ones $(C_1, C_{\rm i})$ and other centrosymmetric point groups. 
$\chi^{\rm A}_{ij}$ and $\chi^{\rm P}_{ijk}$ represent rank-2 axial and rank-3 polar tensors, respectively. 
$\chi^{\rm A}_{ij}$ has 9 components and $\chi^{\rm P}_{ijk}$ has 27 components at most. 
Note that the polar point groups (P) have rank-1 E multipoles ($\bm{Q}$), the chiral point groups (C) have rank-0 ET multipoles ($G_{0}$), and the gyrotropic point groups (G) have less than or equal to rank-2 E/ET multipoles ($G_{0}, \bm{Q}, G_{2m}$) yielding nonzero $\chi_{ij}^{\rm A}$. 
The gyrotropic point groups with only rank-1 E dipoles ($\bm{Q}$) in $(C_{4{\rm v}}, C_{6{\rm v}}, C_{3{\rm v}})$ are called the weak gyrotropic point groups, which do not show a natural optical rotation.
}
\label{tab_totalsymmetric_o}
\centering
\renewcommand{\arraystretch}{1.2}
\begin{tabular}{cccccccccccccccc} \hline\hline
crystal system & point group& N & P & C & G & \multicolumn{5}{c}{totally symmetric rep.} & $\chi^{\rm A}_{ij}$ & \multicolumn{3}{c}{$\chi^{\rm P}_{ijk}$} \\ 
& &  &  &  &  & rank 0& rank 1 & rank 2 & rank 3 & rank 4 & 2 & 1& 2 & 3  \\ \hline
cubic &  $O$& \checkmark &  & \checkmark & \checkmark  & $G_0$ & & & & $G_4$ & 1 & & 1 &   \\
& $T_{\rm d}$& \checkmark &  &  &   &  & & & $Q_{xyz}$ & & &  & & 1   \\
& $T$& \checkmark &  & \checkmark & \checkmark  & $G_0$ & & & $Q_{xyz}$ & $G_4$ & 1 & & 1 & 1  \\ 
\hline
tetragonal  & $D_{4}^{}$& \checkmark &  & \checkmark & \checkmark  & $G_0$ & & $G_u$ & & $G_4$, $G_{4u}$ & 2 & & 2 & 1  \\
& $D_{2{\rm d}}$& \checkmark &  &  & \checkmark  &   & & $G_v$ & $Q_{xyz}$ & $G_{4v}$ & 1 & & 1 & 2  \\
& $C_{4{\rm v}}$& \checkmark & \checkmark &  & \checkmark  &   & $Q_z$ & & $Q_z^\alpha$ & $G_{4z}^{\alpha}$ & 1 & 1 & 1 & 2  \\
& $C_{4}$& \checkmark & \checkmark & \checkmark & \checkmark  &  $G_0$ & $Q_z$ & $G_u$ & $Q_z^\alpha$ & $G_4$, $G_{4u}$, $G_{4z}^{\alpha}$ & 3 & 1 & 3 & 3   \\
& $S_{4}$& \checkmark &  &  & \checkmark  &  & & $G_v$, $G_{xy}$ & $Q_{xyz}$, $Q_z^\beta$ &  $G_{4v}$, $G_{4z}^\beta$ & 2 & & 2 & 4  \\ \hline
orthorhombic & $D_{2}$& \checkmark &  & \checkmark & \checkmark  & $G_0$ & & $G_u$, $G_v$ & $Q_{xyz}$ &  $G_4$, $G_{4u}$, $G_{4v}$ & 3 & & 3 & 3   \\
& $C_{2{\rm v}}$& \checkmark & \checkmark &  & \checkmark  &  & $Q_z$ & $G_{xy}$ & $Q_z^\alpha$, $Q_z^\beta$ &  $G_{4z}^\alpha$, $G_{4z}^\beta$ & 2 & 1 & 2 & 4  \\\hline
monoclinic &  $C_{2}$& \checkmark & \checkmark & \checkmark & \checkmark  & $G_0$ & $Q_z$ & $G_u$, $G_v$, $G_{xy}$ & $Q_{xyz}$, $Q_z^\alpha$, $Q_z^\beta$  & $G_4$, $G_{4u}$, $G_{4v}$, $G_{4z}^{\alpha}$, $G_{4z}^\beta$ & 5 & 1 & 5 & 7  \\
& $C_{{\rm s}}$& \checkmark & \checkmark &  & \checkmark  &  & $Q_x$, $Q_y$ & $G_{yz}$, $G_{zx}$  & $Q_x^\alpha$, $Q_y^\alpha$, $Q_x^\beta$, $Q_y^\beta$ & $G_{4x}^\alpha$, $G_{4y}^\alpha$, $G_{4x}^\beta$, $G_{4y}^\beta$ & 4 & 2 & 4 & 8  \\
\hline
hexagonal &  $D_{6}$& \checkmark &  & \checkmark & \checkmark  &  $G_0$ & & $G_u$ & & $G_{40}$ & 2 & & 2 & 1  \\
&  $D_{3{\rm h}}$& \checkmark &  &  &   &  & &  & $Q_{3a}$ & $G_{4a}$ & & & & 1  \\
& $C_{6{\rm v}}$& \checkmark & \checkmark &  & \checkmark  &   & $Q_z$ &  & $Q_z^\alpha$ &  & 1 & 1 & 1 & 2  \\
& $C_{6}$& \checkmark & \checkmark & \checkmark & \checkmark  & $G_0$ & $Q_z$ & $G_u$ & $Q_z^\alpha$ & $G_{40}$ & 3 & 1 & 3 & 3 \\
& $C_{3{\rm h}}$& \checkmark &  &  &   &   &  & & $Q_{3a}$, $Q_{3b}$ & $G_{4a}$, $G_{4b}$ & & & & 2  \\
 \hline
trigonal & $D_{3}$& \checkmark &  & \checkmark & \checkmark  &  $G_0$ & & $G_u$ & $Q_{3a}$ & $G_{40}$, $G_{4a}$ & 2 & & 2 & 2  \\
& $C_{3{\rm v}}$& \checkmark & \checkmark &  & \checkmark  &   & $Q_z$ & & $Q_z^\alpha$, $Q_{3a}$ &  $G_{4a}$ & 1 & 1 & 1 & 3  \\
& $C_{3}$& \checkmark & \checkmark & \checkmark & \checkmark  & $G_0$  & $Q_z$ & $G_u$ & $Q_z^\alpha$, $Q_{3a}$, $Q_{3b}$ & $G_{40}$, $G_{4a}$, $G_{4b}$ & 3 & 1 & 3 & 5  \\
\hline\hline
\end{tabular}
\end{table*}

\begin{table}[t!]
\begin{center}
\caption{
Representative polar (P) and axial (A) tensors in condensed matter physics.
The odd-parity tensors in the lower rows can be finite only when the spatial inversion symmetry is broken.
}
\label{tab_tensors}
\begingroup
\renewcommand{\arraystretch}{1.2}
 \begin{tabular}{lccl}
 \hline \hline
tensor  & rank  &$\mathcal{P}$ & 
 \\
\hline
$\chi_{i}^{\rm A}$ & 1 & $+$ & magneto-caloric coefficient 
\\
$\chi_{ij}^{\rm P}$ & 2  &     & (thermo)electric/thermal conductivity
\\ 
$\chi_{ijk}^{\rm A}$ & 3 &           & spin conductivity, Nernst coefficient 
\\ 
$\chi_{ijkl}^{\rm P}$ & 4 &           & elastic stiffness tensor
\\ \hline
$\chi_{i}^{\rm P}$ & 1 & $-$       & electro-caloric coefficient 
\\
$\chi_{ij}^{\rm A}$ & 2 &       & linear magneto-electric tensor 
\\
$\chi_{ijk}^{\rm P}$ & 3  &           & 
Piezo-electric tensor 
\\
$\chi_{ijkl}^{\rm A}$ & 4  &           & third-order magneto-electric tensor 
\\ 
\hline\hline
\end{tabular}
\endgroup
\end{center}
\end{table}

In the case of the spinful basis functions, several odd-parity multipole degrees of freedom are potentially activated in addition to $M_{v}$ and $G_{v}$ in the spinless basis functions. 
For example, the MT dipole $(T_x, T_y)$, which is the origin of the magneto-electric effect, can be a primary order parameter when the thermodynamic averages of $(\hat{\tau}_x \hat{\sigma}_x, \hat{\tau}_x \hat{\sigma}_y)$ become nonzero.
Interestingly, it is also possible to activate the MT dipole $(T_x, T_y)$ by an external magnetic field or spontaneous magnetic ordering in the $xy$ plane, since the M dipoles $(M_x, M_y)$ belong to the same irreducible representation as $(T_x, T_y)$. 
This fact indicates that E dipole $\bm{Q}$ is not necessary for the emergence of the MT dipole $\bm{T}$, which has never been clarified in previous interpretations of the MT dipole. 
Similarly, the M quadrupole $M_{xy}$ becomes active under an external magnetic field along the $z$ direction.

Final example is the basis 
function as $(\phi_x, \phi_{y}, \phi_{yz}, \phi_{zx})$ belonging to the E representation under the $D_{\rm 2d}$ group. 
The direct product of the basis functions is given by 
\begin{align}
&({\rm E}\oplus {\rm E}) \otimes ({\rm E}\oplus {\rm E})= \cr
&\quad\quad
 (2 {\rm A}^+_{1} \oplus 2 {\rm A}^-_{2} \oplus 2{\rm B}^+_{1} \oplus 2{\rm B}^+_{2})_{\rm intra} \cr
&\quad\quad
\oplus ({\rm A}^+_{1} \oplus {\rm A}^+_{2} \oplus {\rm B}^+_{1} \oplus {\rm B}^+_{2}\oplus {\rm A}^-_{1} \oplus {\rm A}^-_{2}
\cr&\quad\quad\quad\quad\quad
 \oplus {\rm B}^-_{1} \oplus {\rm B}^-_{2})_{\rm inter}.
\end{align}
This gives us a set of independent operators as
\begin{align}
{\rm A}^+_{1}: \quad & \{\hat{\rho}_{0} \hat{\tau}_0 , \hat{\rho}_{0}\hat{\tau}_z  \} \leftrightarrow \{Q_{0}, Q_{u} \},  \quad \hat{\rho}_0\hat{\tau}_x \leftrightarrow G_{v},   \\
{\rm A}^+_{2}: \quad & \hat{\rho}_y\hat{\tau}_y \leftrightarrow G_{xy},   \\
{\rm B}^+_{1}: \quad & \{\hat{\rho}_{z} \hat{\tau}_0 , \hat{\rho}_{z}\hat{\tau}_z  \} \leftrightarrow \{Q_{v}, Q_{4v} \},  \quad \hat{\rho}_z\hat{\tau}_x \leftrightarrow G_{u},   \\
{\rm B}^+_{2}: \quad & \{\hat{\rho}_{x} \hat{\tau}_0 , \hat{\rho}_{x}\hat{\tau}_z  \} \leftrightarrow \{Q_{xy}, Q^{\beta}_{4z} \},  \quad \hat{\rho}_x\hat{\tau}_x \leftrightarrow Q_{z},   \\
{\rm A}^-_{1}: \quad & \hat{\rho}_0\hat{\tau}_y \leftrightarrow M_{v},   \\
{\rm A}^-_{2}: \quad & \{\hat{\rho}_{y} \hat{\tau}_0 , \hat{\rho}_{y}\hat{\tau}_z  \} \leftrightarrow \{M_{z}, M^{\alpha}_{z} \},  \quad \hat{\rho}_y\hat{\tau}_x \leftrightarrow M_{xy},   \\
{\rm B}^-_{1}: \quad & \hat{\rho}_z\hat{\tau}_y \leftrightarrow M_{u},   \\
{\rm B}^-_{2}: \quad & \hat{\rho}_x\hat{\tau}_y \leftrightarrow T_{z},   
\end{align} 
where $\hat{\tau}_0$ and $\hat{\bm{\tau}}$ are the unit and Pauli matrices acting on the different parity orbitals. 
In the case of the basis functions $(\phi_x, \phi_{y}, \phi_{yz}, \phi_{zx})$, there are several odd-parity multipole degrees of freedom even in the spinless basis functions, such as the E dipole $Q_z$ and MT dipole $T_z$. 

\subsection{Linear response tensors}

Finally, we show nonzero rank-2 and rank-3 tensors in each point group in terms of totally symmetric even-parity multipoles in Table~\ref{tab_totalsymmetric_e} and odd-parity multipoles in Table~\ref{tab_totalsymmetric_o}. 
As described in Sec.~\ref{sec:Linear response tensors}, the rank-2 tensor $\chi_{ij}^{\rm (2)}$ has 9 independent components, which are characterized by the rank-0, rank-1, and rank-2 multipoles. 
Similarly, as discussed in Appendix~\ref{sec:Other linear response tensors}, the rank-2 tensor response by the rank-1 external field is described by the rank-3 tensor $\chi_{ijk}$ with 27 components, which are decomposed into a symmetric monopole response tensor with 3 components $\chi_{i}^{\rm (1)}$, antisymmetric dipole response tensor with 9 components $\chi_{ij}^{\rm (2)}$, and symmetric quadrupole response tensor with 15 components $\chi_{ijk}^{\rm (3)}$. 
Moreover, the components of $\hat{\chi}^{\rm (1,2,3)}$ are characterized by multipoles. 
For example, in the case of $D_{\rm 2d}$ group, one component in the rank-2 axial tensor and three components of the rank-3 polar tensor become nonzero, as odd-parity multipoles $G_v$ and $Q_{xyz}$ belong to totally symmetric representation in $D_{\rm 2d}$ group as shown in Table~\ref{tab_totalsymmetric_o}. 
Nonzero $\chi^{\rm A}_{ij}$ and $\chi^{\rm P}_{ijk}$ imply the emergence of the magnetoelectric effect, piezoelectric tensor, and so on. 
We summarize representative tensors in condensed matter physics in Table~\ref{tab_tensors}.

\section{Summary}
\label{sec:Summary}
In summary, we have investigated a general description of multipoles from the microscopic viewpoint. 
We have presented a definition of four multipoles in both real and momentum spaces, and how to apply to 32 point groups in 7 crystal systems. 
We have demonstrated which multipole degrees of freedom become active in the tetragonal $D_{\rm 2d}$ group as an example. 
We showed that physical properties in electron systems, such as the electromagnetic fields, band structures, and linear responses, are closely related with the multipoles, and hence, the multipole formulation gives a comprehensive and systematic understanding of physical phenomena in condensed matter physics at microscopic level. 
Such a comprehensive investigation of multipoles could stimulate an identification of unknown order parameters, such as MT and ET multipoles, and a further exploration of cross-correlated couplings through multipole-multipole interactions, since we present more than 30 tables as a useful reference in order to cover most cases in all point groups. 

\begin{acknowledgments}
We thank H. Harima and Y. Motome for fruitful discussions. 
This research was supported by JSPS KAKENHI Grants Numbers 15K05176, 15H05885, 18H04296 (J-Physics), 16H06590, and 18K13488. 
\end{acknowledgments}

\appendix

\section{Atomic basis wave functions}
\label{sec:Atomic basis wave functions}

We show the atomic spinless basis functions for $s$, $p$, $d$, and $f$ orbitals as a function of angles $\hat{\bm{r}}/r$, which are used in the main text.  
The basis functions are 
given by 
\begin{align}
\phi_{0}=\frac{1}{\sqrt{4\pi}}, 
\end{align}
for an $s$ orbital, 
\begin{align}
\phi_{x}=\sqrt{\frac{3}{4\pi}}\frac{x}{r},
\quad
\phi_{y}=\sqrt{\frac{3}{4\pi}}\frac{y}{r},
\quad
\phi_{z}=\sqrt{\frac{3}{4\pi}}\frac{z}{r},
\end{align}
for three $p$ orbitals, 
\begin{align}
&
\phi_{u}=\sqrt{\frac{5}{4\pi}}\frac{1}{2}\frac{3z^{2}-r^{2}}{r^2} ,
\quad
\phi_{v}=\sqrt{\frac{5}{4\pi}}\frac{\sqrt{3}}{2}\frac{x^{2}-y^{2}}{r^2},
\cr&
\phi_{yz}=\sqrt{\frac{5}{4\pi}}\sqrt{3}\frac{yz}{r^2},
\quad
\phi_{zx}=\sqrt{\frac{5}{4\pi}}\sqrt{3}\frac{zx}{r^2},
\cr&
\phi_{xy}=\sqrt{\frac{5}{4\pi}}\sqrt{3}\frac{xy}{r^2},
\end{align}
for five $d$ orbitals, and
\begin{align}
&
\phi_{xyz}=\sqrt{\frac{7}{4\pi}}\sqrt{15}\frac{xyz}{r^3},
\quad
\phi_{x}^{\alpha}=\sqrt{\frac{7}{4\pi}}\frac{1}{2}\frac{x(5x^{2}-3r^{2})}{r^3},
\cr&
\phi_{y}^{\alpha}=\sqrt{\frac{7}{4\pi}}\frac{1}{2}\frac{y(5y^{2}-3r^{2})}{r^3},
\quad
\phi_{z}^{\alpha}=\sqrt{\frac{7}{4\pi}}\frac{1}{2}\frac{z(5z^{2}-3r^{2})}{r^3},
\cr&
\phi_{x}^{\beta}=\sqrt{\frac{7}{4\pi}}\frac{\sqrt{15}}{2}\frac{x(y^{2}-z^{2})}{r^3},
\quad
\phi_{y}^{\beta}=\sqrt{\frac{7}{4\pi}}\frac{\sqrt{15}}{2}\frac{y(z^{2}-x^{2})}{r^3},
\cr&
\phi_{z}^{\beta}=\sqrt{\frac{7}{4\pi}}\frac{\sqrt{15}}{2}\frac{z(x^{2}-y^{2})}{r^3}.
\end{align}
for seven $f$ orbitals. 
They are proportional to E multipoles in the cubic $O_{\mathrm{h}}$ representation in Tables~\ref{tab_Oh_CEFe} and \ref{tab_Oh_CEFo}.

\section{Momentum-space multipoles in multi-orbital systems}
\label{sec:k-space multipoles in multi-orbital systems}

As mentioned in Sec.~\ref{sec:Multipoles in momentum space}, the extension of the momentum-space multipoles to multi-orbital systems is not unique.
Here, we discuss one possible extension.

As was discussed, in the single-band systems the odd-rank M and even-rank ET multipoles are given by $\bm{\sigma}\cdot\bm{\nabla}_{\bm{k}}O_{lm}(\bm{k})$ in Eqs.~(\ref{eq:mmom}) and (\ref{eq:etmom}).
By applying the time-reversal-conversion operator $(\bm{\sigma}\times\bm{k})\cdot\bm{\nabla}_{\bm{k}}$, we obtain the expressions for the odd-rank ET and even-rank M multipoles in multi-orbital systems as 
\begin{align*}
G_{lm} &= (\bm{\sigma}\times \bm{k}) \cdot \bm{\nabla}_{\bm{k}} [\bm{\sigma}\cdot \bm{\nabla}_{\bm{k}}O_{lm} (\bm{k})] \quad {\rm (odd)}, \\
M_{lm} &= (\bm{\sigma}\times \bm{k}) \cdot \bm{\nabla}_{\bm{k}} [\bm{\sigma}\cdot \bm{\nabla}_{\bm{k}}O_{lm}(\bm{k})] \quad {\rm (even)}.
\end{align*}
These expressions are simplified by simple algebra as
\begin{align}
M_{lm} &= \begin{cases}
i\bm{k}\cdot\bm{\sigma} & (l=0) \\
i\bm{k}\cdot \bm{\nabla}_{\bm{k}}[\bm{\sigma}\cdot \bm{\nabla}_{\bm{k}}O_{lm} (\bm{k})] & (l=2,4,\cdots)
\end{cases} \\
G_{lm} &= i\bm{k}\cdot\bm{\nabla}_{\bm{k}}[\bm{\sigma}\cdot \bm{\nabla}_{\bm{k}}O_{lm} (\bm{k})] \quad (l=1,3,\cdots),
\end{align}
where we have used $\bm{\nabla}^{2}O_{lm}=0$.
The specific form of the even-rank M and odd-rank ET multipoles up to $l=4$ and the MT monopole 
in momentum space is shown in Tables~\ref{tab_k-multipole-e} and \ref{tab_k-multipole-o}.

\section{Space-time inversion operations of linear response tensor}
\label{sec:Space-time inversion operations of linear response tensor}

We discuss the space-time inversion properties of linear response tensors, which is also discussed in Ref.~\onlinecite{Watanabe_PhysRevB.96.064432}. 
As discussed in Sec.~\ref{sec:Linear response tensors}, the linear response tensors in Eq.~(\ref{eq_chi}) are generally decomposed into 
\begin{align}
&\chi= \chi^{\rm (J)} + \chi^{\rm (E)},\cr
&
\chi^{\rm (J)}=  - \frac{\gamma}{V}\sum_{\bm{k}nm}^{=}\frac{f_{n\bm{k}} -f_{m\bm{k}}}{\varepsilon_{n\bm{k}}-\varepsilon_{m\bm{k}}}
 \frac{A_{\bm{k}}^{nm}B_{\bm{k}}^{mn}}{\left(\varepsilon_{n\bm{k}}-\varepsilon_{m\bm{k}}\right)^2+\gamma^2 },
\label{eq:chi2app}
\\
&
\chi^{\rm (E)}=- \frac{i}{V}\sum_{\bm{k}nm}^{\neq}\frac{\left(f_{n\bm{k}} -f_{m\bm{k}}\right)
A_{\bm{k}}^{nm}B_{\bm{k}}^{mn}}{\left(\varepsilon_{n\bm{k}}-\varepsilon_{m\bm{k}}\right)^2+\gamma^2 },
\label{eq:chi1app}
\end{align}
where $\hat{A}$ and $\hat{B}$ are Hermite operators and these equations correspond to Eqs.~(\ref{eq:chi2}) and (\ref{eq:chi1}), respectively. 
We assume that $\hat{A}$ and $\hat{B}$ have definite parties with respect to the spatial inversion and time-reversal operations as 
\begin{align}
&
\mathcal{P}(\hat{A})=s\hat{A},
\quad
\mathcal{T}(\hat{A})=t\hat{A},
\quad
(s,t=\pm1),
\\&
\mathcal{P}(\hat{B})=s'\hat{B},
\quad
\mathcal{T}(\hat{B})=t'\hat{B},
\quad
(s',t'=\pm1).
\end{align}
Moreover, for the Bloch states, $\mathcal{P}\ket{n\bm{k}}=\ket{n-\bm{k}}$ and $\mathcal{T}\ket{n\bm{k}}=\ket{\bar{n}-\bm{k}}$, where $\bar{n}$ is the time-reversal partner of $n$ state.
As $\mathcal{T}$ is anti-unitary, these relations lead to $A_{\bm{k}}^{nm}=sA_{-\bm{k}}^{nm}$, $A_{\bm{k}}^{nm}=tA_{-\bm{k}}^{\bar{m}\bar{n}}$, $B_{\bm{k}}^{nm}=s'B_{-\bm{k}}^{nm}$, and $B_{\bm{k}}^{mn}=t'B_{-\bm{k}}^{\bar{n}\bar{m}}$.

Thus, in the presence of the spatial inversion symmetry, we show that
\begin{align*}
\chi^{\rm (J)}&
=
- \frac{\gamma}{V}\sum_{\bm{k}nm}^{=}\frac{f_{n\bm{k}} -f_{m\bm{k}}}{\varepsilon_{n\bm{k}}-\varepsilon_{m\bm{k}}}
 \frac{sA_{-\bm{k}}^{nm}s'B_{-\bm{k}}^{mn}}{\left(\varepsilon_{n\bm{k}}-\varepsilon_{m\bm{k}}\right)^2+\gamma^2 }
\cr&
=
- \frac{\gamma}{V}\sum_{\bm{k}nm}^{=}\frac{f_{n-\bm{k}} -f_{m-\bm{k}}}{\varepsilon_{n-\bm{k}}-\varepsilon_{m-\bm{k}}}
 \frac{sA_{\bm{k}}^{nm}s'B_{\bm{k}}^{mn}}{\left(\varepsilon_{n-\bm{k}}-\varepsilon_{m-\bm{k}}\right)^2+\gamma^2 }
\cr&
=
- \frac{\gamma}{V}\sum_{\bm{k}nm}^{=}\frac{f_{n\bm{k}} -f_{m\bm{k}}}{\varepsilon_{n\bm{k}}-\varepsilon_{m\bm{k}}}
 \frac{A_{\bm{k}}^{nm}B_{\bm{k}}^{mn}}{\left(\varepsilon_{n\bm{k}}-\varepsilon_{m\bm{k}}\right)^2+\gamma^2 }ss'
\cr&=
ss'\chi^{\rm (J)},
\end{align*}
where we have used $\varepsilon_{n-\bm{k}}=\varepsilon_{n\bm{k}}$.
A similar discussion holds for $\chi^{\rm (E)}$, and we obtain
\begin{align}
\chi^{\rm (J)}=ss'\chi^{\rm (J)},
\quad
\chi^{\rm (E)}=ss'\chi^{\rm (E)}.
\end{align}
On the other hand, in the presence of the time-reversal symmetry, we have
\begin{align*}
\chi^{\rm (J)}
&=
- \frac{\gamma}{V}\sum_{\bm{k}nm}^{=}\frac{f_{n\bm{k}} -f_{m\bm{k}}}{\varepsilon_{n\bm{k}}-\varepsilon_{m\bm{k}}}
 \frac{tA_{-\bm{k}}^{\bar{m}\bar{n}}t'B_{-\bm{k}}^{\bar{n}\bar{m}}}{\left(\varepsilon_{n\bm{k}}-\varepsilon_{m\bm{k}}\right)^2+\gamma^2 }
\\&=
- \frac{\gamma}{V}\sum_{\bm{k}nm}^{=}\frac{f_{\bar{m}-\bm{k}} -f_{\bar{n}-\bm{k}}}{\varepsilon_{\bar{m}-\bm{k}}-\varepsilon_{\bar{n}-\bm{k}}}
 \frac{tA_{\bm{k}}^{nm}t'B_{\bm{k}}^{mn}}{\left(\varepsilon_{\bar{m}-\bm{k}}-\varepsilon_{\bar{n}-\bm{k}}\right)^2+\gamma^2 }
\\&=
- \frac{\gamma}{V}\sum_{\bm{k}nm}^{=}\frac{f_{m\bm{k}} -f_{n\bm{k}}}{\varepsilon_{m\bm{k}}-\varepsilon_{n\bm{k}}}
 \frac{A_{\bm{k}}^{nm}B_{\bm{k}}^{mn}}{\left(\varepsilon_{m\bm{k}}-\varepsilon_{n\bm{k}}\right)^2+\gamma^2 }tt'
\\&
=tt'\chi^{\rm (J)},
\end{align*}
where we have used $\varepsilon_{\bar{n}-\bm{k}}=\varepsilon_{n\bm{k}}$.
Similarly,
\begin{align*}
\chi^{\rm (E)}
&=
- \frac{i}{V}\sum_{\bm{k}nm}^{\neq}
 \frac{(f_{n\bm{k}} -f_{m\bm{k}})tA_{-\bm{k}}^{\bar{m}\bar{n}}t'B_{-\bm{k}}^{\bar{n}\bar{m}}}{\left(\varepsilon_{n\bm{k}}-\varepsilon_{m\bm{k}}\right)^2+\gamma^2 }
\\&=
- \frac{i}{V}\sum_{\bm{k}nm}^{\neq}
 \frac{(f_{\bar{m}-\bm{k}} -f_{\bar{n}-\bm{k}})tA_{\bm{k}}^{nm}t'B_{\bm{k}}^{mn}}{\left(\varepsilon_{\bar{m}-\bm{k}}-\varepsilon_{\bar{n}-\bm{k}}\right)^2+\gamma^2 }
\\&=
- \frac{i}{V}\sum_{\bm{k}nm}^{\neq}
 \frac{(f_{m\bm{k}} -f_{n\bm{k}})A_{\bm{k}}^{nm}B_{\bm{k}}^{mn}}{\left(\varepsilon_{m\bm{k}}-\varepsilon_{n\bm{k}}\right)^2+\gamma^2 }tt'
\\&
=-tt'\chi^{\rm (E)}.
\end{align*}
Thus, we obtain
\begin{align}
\chi^{\rm (J)}=tt'\chi^{\rm (J)},
\quad
\chi^{\rm (E)}=-tt'\chi^{\rm (E)}.
\end{align}

Due to the above symmetry properties, in the case of $s\ne s'$ a breaking of the spatial inversion is necessary to obtain a finite $\chi^{\rm (J,E)}$.
Similarly, in the case of $t\ne t'$ ($t=t'$) a breaking of time-reversal symmetry is necessary to obtain a finite $\chi^{\rm (J)}$ ($\chi^{\rm (E)})$.

\section{Other linear response tensors}
\label{sec:Other linear response tensors}

In this section, we show a natural extension of linear-response tensors as discussed in Sec.~\ref{sec:Linear response tensors}. 
In the main text, we discuss the rank-2 linear-response tensors $\chi_{ij}^{\rm (2)}$ with 9 independent components when $\hat{A}_i$ and $\hat{B}_j$ are both rank-1 tensors. 
The extension of the linear response tensors when $\hat{A}_i$ and $\hat{B}_j$ are higher-rank tensors is straightforward.

Let us consider the rank-2 response to the rank-1 external field, namely,
\begin{align}
A_{ij}=\sum_{k}\chi_{ijk}B_{k}.
\end{align}
We can decompose $A_{ij}$ into the monopole, dipole, and quadrupole contributions as
\begin{align}
A_{ij}=\delta_{ij}A^{\rm M}+\sum_{l}\epsilon_{ijl}A_{l}^{\rm D}+A_{ij}^{\rm Q},
\end{align}
where $\epsilon_{ijl}$ is the totally antisymmetric tensor (Levi-Civita symbol), and $A_{ij}^{\rm Q}$ is symmetric and traceless.
Note that $A_{i}^{\rm D}$ is polar (axial) when others are axial (polar) since $\epsilon_{ijk}$ and $\delta_{ij}$ are axial and polar tensors, respectively, and $A^{\rm M}$, $A_{i}^{\rm D}$, and $A_{ij}^{\rm Q}$ eventually have the same parity.
For each contribution, the linear-response tensor is introduced as
\begin{align}
&
A^{\rm M}=\frac{1}{3}\sum_{i}A_{ii}=\sum_{k}\chi_{k}^{\rm (1)}B_{k},
\\&
A_{i}^{\rm D}=\frac{1}{2}\sum_{jk}\epsilon_{ijk}A_{jk}=\sum_{k}\chi_{ik}^{\rm (2)}B_{k}
\\&
A_{ij}^{\rm Q}=\sum_{k}\chi_{ijk}^{\rm (3)}B_{k},
\end{align}
where $\chi_{ijk}^{\rm (3)}$ is symmetric and traceless with respect to $(i,j)$, i.e., $\chi_{ijk}^{\rm (3)}=\chi_{jik}^{\rm (3)}$ and $\sum_{i}\chi_{iik}^{\rm (3)}=0$.
Note that $\chi_{ij}^{\rm (2)}$ corresponds to Eq.~(\ref{eq:rank11tensor}) in the main text.
By inserting these expressions into the decomposition of $A_{ij}$, we obtain the relation to $\chi_{ijk}$ as
\begin{align}
A_{ij}=\sum_{k}\left(
\delta_{ij}\chi_{k}^{\rm (1)}+\sum_{l}\epsilon_{ijl}\chi_{lk}^{\rm (2)}+\chi_{ijk}^{\rm (3)}
\right)B_{k}.
\end{align}

Each linear-response function can be decomposed as
\begin{align}
&
\chi_{k}^{\rm (1)}=\chi_{k}^{\rm D}
\\&
\chi_{ij}^{\rm (2)}=\delta_{ij}\chi^{\rm M}+\sum_{k}\epsilon_{ijk}\chi_{k}^{\rm D'}+\chi_{ij}^{\rm Q},
\\&
\chi_{ijk}^{\rm (3)}=\delta_{ij}\chi_{k}^{\rm D''}+\sum_{lm}(\epsilon_{ikl}\delta_{jm}+\epsilon_{jkl}\delta_{im})\chi_{lm}^{\rm Q'}+\chi_{ijk}^{\rm O},
\end{align}
where $\chi_{ij}^{\rm Q}$ is symmetric and traceless, and $\chi_{ijk}^{\rm O}$ is totally symmetric with respect to any permutations of indices and $\sum_{i}\chi_{iik}^{\rm O}=-3\chi_{k}^{\rm D''}$.

Since $\chi^{\rm M}$, $\chi_{k}^{\rm D}$, $\chi_{ij}^{\rm Q}$, $\chi_{ijk}^{\rm O}$ are characterized by $Y_{0}$ (monopole), $X_{i}$ (dipole), $Y_{ij}$ (quadrupole), and $X_{ijk}$ (octupole), respectively, where $X$ is polar (axial) and $Y$ is axial (polar) when $\chi_{ijk}^{\rm (3)}$ is polar (axial), we explicitly express the linear-response functions in terms of multipoles as
\begin{align}
&
\chi^{\rm (1)}=\begin{pmatrix} X_{x} & X_{y}& X_{z} \end{pmatrix},
\\&
\chi^{\rm (2)}=
\begin{pmatrix}
Y_{0}-Y_{u}+Y_{v} & Y_{xy}+X_{z} & Y_{zx}-X_{y} \\
Y_{xy}-X_{z} & Y_{0}-Y_{u}-Y_{v} & Y_{yz}+X_{x} \\
Y_{zx}+X_{y} & Y_{yz}-X_{x} & Y_{0}+2Y_{u}
\end{pmatrix},
\end{align}
\begin{widetext}
\begin{align}
\chi^{\rm (3)}=
\begin{pmatrix}
X_{x}+Y_{yz}-X_{x}^{\alpha}-X_{x}^{\beta} & X_{y}-Y_{zx}-X_{y}^{\alpha}+X_{y}^{\beta} & -2X_{z}+2X_{z}^{\alpha} \\
-3X_{x}+Y_{yz}+3X_{x}^{\alpha}-X_{x}^{\beta} & 3X_{y}+Y_{zx}-3X_{y}^{\alpha}-X_{y}^{\beta} & -2Y_{xy}+2X_{z}^{\beta} \\
-3Y_{u}-Y_{v}+X_{xyz} & -3X_{z}-Y_{xy}-2X_{z}^{\alpha}-2X_{z}^{\beta} & -3X_{y}+Y_{zx}-2X_{y}^{\alpha}+2X_{y}^{\beta} \\
-3X_{z}+Y_{xy}-2X_{z}^{\alpha}+2X_{z}^{\beta} & 3Y_{u}-Y_{v}+X_{xyz} & -3X_{x}-Y_{yz}-2X_{x}^{\alpha}-2X_{x}^{\beta} \\
-3X_{y}-Y_{zx}-2X_{y}^{\alpha}-2X_{y}^{\beta} & -3X_{x}+Y_{yz}-2X_{x}^{\alpha}+2X_{x}^{\beta} & 2Y_{v}+X_{xyz}
\end{pmatrix},
\end{align}
\end{widetext}
where we have put $\chi^{\rm M}=Y_{0}$, 
$\chi_{i}^{\rm D}=\chi_{i}^{\rm D'}=X_{i}$, $\chi_{ij}^{\rm Q}=\chi_{ij}^{\rm Q'}=Y_{ij}$, 
$\chi_{ijk}^{\rm O}=X_{ijk}$, and $\chi_{i}^{\rm D''}=5X_{i}$ for notational simplicity.
We have also used the relations, 
\begin{align}
&
\sum_{i}Y_{ii}=0,
\quad
Y_{u}=\frac{1}{6}(2Y_{zz}-Y_{xx}-Y_{yy}),
\label{eq:quadeuclid}
\cr&
Y_{v}=\frac{1}{2}(Y_{xx}-Y_{yy}),
\end{align}
for quadrupoles, and
\begin{align}
&
\sum_{i}X_{iik}=-15X_{k}\quad(k=x,y,z),
\cr&
X_{x}^{\alpha}=\frac{1}{20}(2X_{xxx}-3X_{yyx}-3X_{zzx}),
\cr&
X_{y}^{\alpha}=\frac{1}{20}(2X_{yyy}-3X_{zzy}-3X_{xxy}),
\cr&
X_{z}^{\alpha}=\frac{1}{20}(2X_{zzz}-3X_{xxz}-3X_{yyz}),
\cr&
X_{x}^{\beta}=\frac{1}{4}(X_{yyx}-X_{zzx}),
\cr&
X_{y}^{\beta}=\frac{1}{4}(X_{zzy}-X_{xxy}),
\cr&
X_{z}^{\beta}=\frac{1}{4}(X_{xxz}-X_{yyz}),
\end{align}
for octupoles.
The prefactors in the right hand sides are chosen in order to simplify the resultant expressions in the linear-response tensors. 
In the expression of $\chi^{\rm (3)}$, the rows correspond to $(i,j)=(u,v,yz,zx,xy)$, e.g., $\chi_{uk}^{\rm (3)}=(2\chi_{zzk}^{\rm (3)}-\chi_{xxk}^{\rm (3)}-\chi_{yyk}^{\rm (3)})/6$ according to Eq.~(\ref{eq:quadeuclid}).

In order to complete the comprehensive lists, we also give the expression of $\chi^{\rm (3)}$ in the hexagonal $D_{\rm 6h}$ group as 
\begin{widetext}
\begin{align}
\chi^{\rm (3)}=
\begin{pmatrix}
X_{x}+Y_{yz}+2X_{3u} & X_{y}-Y_{zx}+2X_{3v} & -2X_{z}+2X_{z}^{\alpha} \\
-3X_{x}+Y_{yz}+X_{3a}-X_{3u} & 3X_{y}+Y_{zx}+X_{3b}+X_{3v} & -2Y_{xy}+2X_{z}^{\beta} \\
-3Y_{u}-Y_{v}+X_{xyz} & -3X_{z}-Y_{xy}-2X_{z}^{\alpha}-2X_{z}^{\beta} & -3X_{y}+Y_{zx}-2X_{y}^{\alpha}+4X_{3v} \\
-3X_{z}+Y_{xy}-2X_{z}^{\alpha}+2X_{z}^{\beta} & 3Y_{u}-Y_{v}+X_{xyz} & -3X_{x}-Y_{yz}-2X_{x}^{\alpha}+4X_{3u} \\
-3X_{y}-Y_{zx}+X_{3b}-X_{3v} & -3X_{x}+Y_{yz}-X_{3a}-X_{3u} & 2Y_{v}+X_{xyz}
\end{pmatrix},
\end{align}
\end{widetext}
where we have used the relations, 
\begin{align}
&
X_{3u}=-\frac{1}{2}(X^{\alpha}_{x}+X^\beta_x),
\cr&
X_{3v}=-\frac{1}{2}(X^{\alpha}_{y}-X^\beta_y),
\cr&
X_{3a}=\frac{1}{2}(5X^{\alpha}_{x}-3X^\beta_x),
\cr&
X_{3b}=-\frac{1}{2}(5X^{\alpha}_{y}+3X^\beta_y). 
\end{align}

We present two examples in the following: spin conductivity and Piezo-electric tensors.

\subsection{Spin conductivity tensor}
\label{sec:Spin conductivity tensor}

We here consider the axial 3rd-rank spin conductivity tensor, which is defined as 
\begin{align}
J^{\rm s}_{ij}=\sum_{k}\sigma^{\rm s}_{ijk} E_k, 
\end{align}
where $\hat{A}_{ij}=\hat{J}^{\rm s}_{ij}$ and $\hat{B}_j=\hat{J}_j$ in the general formula in Eq.~(\ref{eq_chi}). 
Note that there are 9 independent components for the spin current tensor, as it is characterized by the product of rank-1 tensors, $\hat{J}_i$ and $\hat{\sigma}_j$ ($\hat{\sigma}$ is the spin degree of freedom). 
From the symmetry point of view, the spin conductivity tensor is decomposed as 
\begin{align}
J^{{\rm s}}_{ij} \equiv J_{i} \sigma_{j} = \delta_{ij}J^{\rm s,M}+\sum_{k}\epsilon_{ijk}J_{k}^{\rm s,D}+J_{ij}^{\rm s,Q}, 
\end{align}
where
\begin{align}
&
J^{{\rm s,M}} = \frac{1}{3}\bm{J} \cdot \bm{\sigma},
\cr&
\bm{J}^{\rm s,D} = 
\frac{1}{2}(\bm{J}\times\bm{\sigma}), 
\cr&
\hat{J}^{\rm s,Q} =\frac{1}{2}
\begin{pmatrix}
[2 J_z \sigma_z - J_x \sigma_x- J_y \sigma_y]/3 \\
J_x \sigma_x - J_y \sigma_y \\
J_{y} \sigma_z + J_z \sigma_y \\
J_z \sigma_x + J_x \sigma_z \\
J_x \sigma_y + J_y \sigma_x 
\end{pmatrix}.
\end{align}
$J^{\rm s, M}$, $\bm{J}^{\rm s,D}$, and $\hat{J}^{\rm s,Q}$ possess the same symmetry as ET monopole, E dipole, and ET quadrupole, respectively. 
The corresponding tensors $\hat{\sigma}^{\rm s(1)}$, $\hat{\sigma}^{\rm s(2)}$, and $\hat{\sigma}^{\rm s(3)}$ are expressed as 
\begin{align}
&
\hat{\sigma}^{\rm s(1,J)}=
\begin{pmatrix}
M_{x} & M_{y} & M_{z} \\
\end{pmatrix}, 
\cr&
\hat{\sigma}^{\rm s(1,E)}=
\begin{pmatrix}
G_{x} & G_{y} & G_{z} \\
\end{pmatrix},
\end{align}
\begin{align}
&
\hat{\sigma}^{\rm s(2,J)}=
\begin{pmatrix}
T_{0}-T_{u}+T_{v} & T_{xy}+M_{z} & T_{zx}-M_{y} \\
T_{xy}-M_{z} & T_{0}-T_{u}-T_{v} & T_{yz}+M_{x} \\
T_{zx}+M_{y} & T_{yz}-M_{x} & T_{0}+2T_{u}
\end{pmatrix}, 
\cr&
\hat{\sigma}^{\rm s(2,E)}=
\begin{pmatrix}
Q_{0}-Q_{u}+Q_{v} & Q_{xy}+G_{z} & Q_{zx}-G_{y} \\
Q_{xy}-G_{z} & Q_{0}-Q_{u}-Q_{v} & Q_{yz}+G_{x} \\
Q_{zx}+G_{y} & Q_{yz}-G_{x} & Q_{0}+2Q_{u}
\end{pmatrix},
\end{align}
and 
\begin{widetext}
\begin{align}
&
\hat{\sigma}^{\rm s(3,J)}=
\begin{pmatrix}
M_{x}+T_{yz}-M_{x}^{\alpha}-M_{x}^{\beta} & M_{y}-T_{zx}-M_{y}^{\alpha}+M_{y}^{\beta} & -2M_{z}+2M_{z}^{\alpha} \\
-3M_{x}+T_{yz}+3M_{x}^{\alpha}-M_{x}^{\beta} & 3M_{y}+T_{zx}-3M_{y}^{\alpha}-M_{y}^{\beta} & -2T_{xy}+2M_{z}^{\beta} \\
-3T_{u}-T_{v}+M_{xyz} & -3M_{z}-T_{xy}-2M_{z}^{\alpha}-2M_{z}^{\beta} & -3M_{y}+T_{zx}-2M_{y}^{\alpha}+2M_{y}^{\beta} \\
-3M_{z}+T_{xy}-2M_{z}^{\alpha}+2M_{z}^{\beta} & 3T_{u}-T_{v}+M_{xyz} & -3M_{x}-T_{yz}-2M_{x}^{\alpha}-2M_{x}^{\beta} \\
-3M_{y}-T_{zx}-2M_{y}^{\alpha}-2M_{y}^{\beta} & -3M_{x}+T_{yz}-2M_{x}^{\alpha}+2M_{x}^{\beta} & 2T_{v}+M_{xyz}
\end{pmatrix},
\cr&
\hat{\sigma}^{\rm s(3,E)}=
\begin{pmatrix}
G_{x}+Q_{yz}-G_{x}^{\alpha}-G_{x}^{\beta} & G_{y}-Q_{zx}-G_{y}^{\alpha}+G_{y}^{\beta} & -2G_{z}+2G_{z}^{\alpha} \\
-3G_{x}+Q_{yz}+3G_{x}^{\alpha}-G_{x}^{\beta} & 3G_{y}+Q_{zx}-3G_{y}^{\alpha}-G_{y}^{\beta} & -2Q_{xy}+2G_{z}^{\beta} \\
-3Q_{u}-Q_{v}+G_{xyz} & -3G_{z}-Q_{xy}-2G_{z}^{\alpha}-2G_{z}^{\beta} & -3G_{y}+Q_{zx}-2G_{y}^{\alpha}+2G_{y}^{\beta} \\
-3G_{z}+Q_{xy}-2G_{z}^{\alpha}+2G_{z}^{\beta} & 3Q_{u}-Q_{v}+G_{xyz} & -3G_{x}-Q_{yz}-2G_{x}^{\alpha}-2G_{x}^{\beta} \\
-3G_{y}-Q_{zx}-2G_{y}^{\alpha}-2G_{y}^{\beta} & -3G_{x}+Q_{yz}-2G_{x}^{\alpha}+2G_{x}^{\beta} & 2Q_{v}+G_{xyz}
\end{pmatrix}.
\end{align}
\end{widetext}

\subsection{Piezo-electric tensor}
\label{sec:Electric(current)-elastic tensor}

Next, we consider a Piezo-electric effect where the strain and rotation is induced by the electric field (or current). 
The polar 3rd-rank Piezo-electric tensor $d_{ijk}$ is given by 
\begin{align}
\zeta_{ij}
=\sum_{k} d_{ijk} E_k. 
\end{align}
$\zeta_{ij}\equiv \partial u_{i}/\partial x_{j}$, where $\bm{u}$ is a displacement vector field.
It can be decomposed as
\begin{align}
\zeta_{ij}=\delta_{ij}\varepsilon_{0}+\sum_{k}\epsilon_{ijk}\,\omega_{k}+\tilde{\varepsilon}_{ij},
\end{align}
where $\varepsilon_{0}=(\bm{\nabla}\cdot\bm{u})/3$, $\bm{\omega}=(\bm{\nabla}\times\bm{u})/2$, and $\varepsilon_{ij}=(\zeta_{ij}+\zeta_{ji})/2=\delta_{ij}\varepsilon_{0}+\tilde{\varepsilon}_{ij}$ represent the bulk modulus, rotation, and symmetric strain, respectively.
Since $\varepsilon_{0}$, $(\omega_x, \omega_y, \omega_z)$, and $(\varepsilon_{u},\varepsilon_{v}, \varepsilon_{yz}, \varepsilon_{zx},\varepsilon_{xy})$ are the same symmetry as E monopole, ET dipole, and E quadrupole, respectively, the corresponding tensors $\hat{d}^{\rm (1)}$, $\hat{d}^{\rm (2)}$, and $\hat{d}^{(3)}$ are given by 
\begin{align}
&
\hat{d}^{\rm (1,J)}=
\begin{pmatrix}
T_{x} & T_{y} & T_{z} \\
\end{pmatrix}, 
\cr&
\hat{d}^{\rm (1,E)}=
\begin{pmatrix}
Q_{x} & Q_{y} & Q_{z} \\
\end{pmatrix},
\end{align}
\begin{align}
&
\hat{d}^{\rm (2,J)}=
\begin{pmatrix}
M_{0}-M_{u}+M_{v} & M_{xy}+T_{z} & M_{zx}-T_{y} \\
M_{xy}-T_{z} & M_{0}-M_{u}-M_{v} & M_{yz}+T_{x} \\
M_{zx}+T_{y} & M_{yz}-T_{x} & M_{0}+2M_{u}
\end{pmatrix}, 
\cr&
\hat{d}^{\rm (2,E)}=
\begin{pmatrix}
G_{0}-G_{u}+G_{v} & G_{xy}+Q_{z} & G_{zx}-Q_{y} \\
G_{xy}-Q_{z} & G_{0}-G_{u}-G_{v} & G_{yz}+Q_{x} \\
G_{zx}+Q_{y} & G_{yz}-Q_{x} & G_{0}+2G_{u}
\end{pmatrix},
\end{align}
and 
\begin{widetext}
\begin{align}
&
\hat{d}^{\rm (3,J)}=
\begin{pmatrix}
T_{x}+M_{yz}-T_{x}^{\alpha}-T_{x}^{\beta} & T_{y}-M_{zx}-T_{y}^{\alpha}+T_{y}^{\beta} & -2T_{z}+2T_{z}^{\alpha} \\
-3T_{x}+M_{yz}+3T_{x}^{\alpha}-T_{x}^{\beta} & 3T_{y}+M_{zx}-3T_{y}^{\alpha}-T_{y}^{\beta} & -2M_{xy}+2T_{z}^{\beta} \\
-3M_{u}-M_{v}+T_{xyz} & -3T_{z}-M_{xy}-2T_{z}^{\alpha}-2T_{z}^{\beta} & -3T_{y}+M_{zx}-2T_{y}^{\alpha}+2T_{y}^{\beta} \\
-3T_{z}+M_{xy}-2T_{z}^{\alpha}+2T_{z}^{\beta} & 3M_{u}-M_{v}+T_{xyz} & -3T_{x}-M_{yz}-2T_{x}^{\alpha}-2T_{x}^{\beta} \\
-3T_{y}-M_{zx}-2T_{y}^{\alpha}-2T_{y}^{\beta} & -3T_{x}+M_{yz}-2T_{x}^{\alpha}+2T_{x}^{\beta} & 2M_{v}+T_{xyz}
\end{pmatrix},
\cr&
\hat{d}^{\rm (3,E)}=
\begin{pmatrix}
Q_{x}+G_{yz}-Q_{x}^{\alpha}-Q_{x}^{\beta} & Q_{y}-G_{zx}-Q_{y}^{\alpha}+Q_{y}^{\beta} & -2Q_{z}+2Q_{z}^{\alpha} \\
-3Q_{x}+G_{yz}+3Q_{x}^{\alpha}-Q_{x}^{\beta} & 3Q_{y}+G_{zx}-3Q_{y}^{\alpha}-Q_{y}^{\beta} & -2G_{xy}+2Q_{z}^{\beta} \\
-3G_{u}-G_{v}+Q_{xyz} & -3Q_{z}-G_{xy}-2Q_{z}^{\alpha}-2Q_{z}^{\beta} & -3Q_{y}+G_{zx}-2Q_{y}^{\alpha}+2Q_{y}^{\beta} \\
-3Q_{z}+G_{xy}-2Q_{z}^{\alpha}+2Q_{z}^{\beta} & 3G_{u}-G_{v}+Q_{xyz} & -3Q_{x}-G_{yz}-2Q_{x}^{\alpha}-2Q_{x}^{\beta} \\
-3Q_{y}-G_{zx}-2Q_{y}^{\alpha}-2Q_{y}^{\beta} & -3Q_{x}+G_{yz}-2Q_{x}^{\alpha}+2Q_{x}^{\beta} & 2G_{v}+Q_{xyz}
\end{pmatrix}.
\end{align}
\end{widetext}
For example, the rotation by the electric field, i.e., nonzero $\hat{d}^{\rm (2,E)}$, occurs for the gyrotropic point groups, 
and the symmetric strain by the electric field, i.e., nonzero $\hat{d}^{\rm (3,E)}$, occurs for the noncentrosymetric point groups, as shown in Table~\ref{tab_totalsymmetric_o}.

\section{$O_{\rm h}$ and its subgroups}
\label{sec:Oh and its subgroups}

We summarize the active multipoles for the basis for $s$, $p$, $d$, and $f$ orbitals with/without the spin degree of freedom for the $O_{\rm h}$ and its subgroups. 
The active multipoles for cubic and orthorhombic crystals in the spinless (spinful) basis are shown in Table~\ref{tab_cubic_less} (\ref{tab_cubic_ful}) and \ref{tab_ortho_less} (\ref{tab_ortho_ful}), respectively.
The active multipoles for tetragonal crystals in the spinless (spinful) basis are shown in Table~\ref{tab_tetra_less} (\ref{tab_tetra_ful}) in the main text.

\begin{table*}[h]
\caption{
Multipoles under the cubic crystal system ($O_{\rm h}$) in the spinless basis. 
$(g,u)^{\pm}={\rm A}^{\pm}_{1g,u} \oplus {\rm E}^{\pm}_{g,u} \oplus {\rm T}^{\mp}_{1g,u} \oplus {\rm T}^{\pm}_{2g,u}$; $\braket{g,u}^{\pm}= {\rm A}^{\pm}_{2g,u} \oplus {\rm E}^{\pm}_{g,u} \oplus {\rm T}^{\pm}_{1g,u} \oplus {\rm T}^{\pm}_{2g,u}$. 
The superscript $^{\pm}$ includes both time-reversal-even ($^{+}$) and time-reversal-odd ($^{-}$) operators. 
The subscripts $g,u$ are omitted for the $O$ group. 
A$_{1u}$ $\leftrightarrow$ A$_{2u}$ and T$_{1u}$ $\leftrightarrow$ T$_{2u}$ for the $T_{\rm d}$ group. 
The subscripts $1,2$ are omitted for the $T_{\rm h}$ group. 
The subscripts $g,u$, $1,2$ are omitted for the $T$ group. 
}
\label{tab_cubic_less}
\centering
\begin{tabular}{c|c|c|cc|ccc}
& (s) A$_{1g}$ & (p) T$_{1u}$ & (d) E$_{g}$ & T$_{2g}$ & (f) A$_{2u}$ & T$_{1u}$ & T$_{2u}$ \\ \hline
(s) A$_{1g}$ & A$^+_{1g}$ & T$^{\pm}_{1u}$ & E$^{\pm}_{g}$ & T$^{\pm}_{2g}$ & A$^{\pm}_{2u}$ & T$^{\pm}_{1u}$ & T$^{\pm}_{2u}$ \\ \hline
(p) T$_{1u}$ & & $(g)^+$ & T$^{\pm}_{1u}$, T$^{\pm}_{2u}$ & $\braket{u}^{\pm}$ & T$^{\pm}_{2g}$ & $(g)^{\pm}$ & $\braket{g}^{\pm}$ \\ \hline
(d) E$_{g}$ & & & A$^+_{1g}$, A$^-_{2g}$, E$^+_{g}$ & T$^{\pm}_{1g}$, T$^{\pm}_{2g}$ & E$^{\pm}_{u}$ & T$^{\pm}_{1u}$, T$^{\pm}_{2u}$ & T$^{\pm}_{1u}$, T$^{\pm}_{2u}$ \\
T$_{2g}$ & & & & $(g)^+$ & T$^{\pm}_{1u}$ & $\braket{u}^{\pm}$ & $(u)^{\pm}$ \\ \hline
(f) A$_{2u}$ & & & & & A$^+_{1g}$ & T$^{\pm}_{2g}$ & T$^{\pm}_{1g}$ \\
T$_{1u}$ & & & & & & $(g)^+$ & $\braket{g}^{\pm}$ \\
T$_{2u}$ & & & & & & & $(g)^+$
\end{tabular}
\end{table*}

\begin{table*}[h]
\caption{
Multipoles under the cubic crystal system ($O_{\rm h}$) in the spinful basis. 
$(g,u)^{\pm}=A^{\pm}_{1g,u} \oplus {\rm E}^{\pm}_{g,u} \oplus {\rm T}^{\pm}_{2g,u} \oplus {\rm A}^{\mp}_{2g,u} \oplus 2{\rm T}^{\mp}_{1g,u} \oplus {\rm T}^{\mp}_{2g,u}$; $\braket{g,u}^{\pm}={\rm E}^{\pm}_{g,u} \oplus {\rm T}^{\pm}_{1g,u} \oplus {\rm T}^{\pm}_{2g,u}$. 
The subscripts $g,u$ are omitted for the $O$ group. 
A$_{1u}$ $\leftrightarrow$ A$_{2u}$, T$_{1u}$ $\leftrightarrow$ T$_{2u}$, and E$_{1/2u}$ $\leftrightarrow$ E$_{5/2u}$ for the $T_{\rm d}$ group. 
The subscripts $1,2$ are omitted for the $T_{\rm h}$ group. 
The subscripts $g,u$ and $1,2$ are omitted for the $T$ group. 
}
\label{tab_cubic_ful}
\centering
\scalebox{0.95}{
\begin{tabular}{cc|c|cc|ccc|ccccc}
& & (s) E$_{1/2g}$ & (p) E$_{1/2u}$ &G$_{3/2u}$ & (d) G$_{3/2g}$ &E$_{5/2g}$&G$_{3/2g}$ & (f) E$_{5/2u}$ & E$_{1/2u}$ & G$_{3/2u}$& E$_{5/2u}$& G$_{3/2u}$ \\ \hline
(s) A$_{1g}\otimes$ E$_{1/2g} \to$& $E_{1/2g}$ & A$^+_{1g}$, T$^-_{1g}$&A$^{\pm}_{1u}$, T$^{\pm}_{1u}$ & $\braket{u}^{\pm}$ & $\braket{g}^{\pm}$ & A$^{\pm}_{2g}$, T$^{\pm}_{2g}$ & $\braket{g}^{\pm}$ & A$^{\pm}_{2u}$, T$^{\pm}_{2u}$ & A$^{\pm}_{1u}$, T$^{\pm}_{1u}$&$\braket{u}^{\pm}$&A$^{\pm}_{2u}$, T$^{\pm}_{2u}$ &$\braket{u}^{\pm}$\\ \hline
(p) T$_{1u}\otimes$ E$_{1/2g}\to$&$E_{1/2u}$ & & A$^+_{1g}$, T$^-_{1g}$ & $\braket{g}^{\pm}$ &$\braket{u}^{\pm}$ & A$^{\pm}_{2u}$, T$^{\pm}_{2u}$&$\braket{u}^{\pm}$ &A$^{\pm}_{2g}$, T$^{\pm}_{2g}$&A$^{\pm}_{1g}$, T$^{\pm}_{1g}$&$\braket{g}^{\pm}$& A$^{\pm}_{2g}$, T$^{\pm}_{2g}$&$\braket{g}^{\pm}$\\
&$G_{3/2u}$ & & &$(g)^+$& $(u)^{\pm}$ & $\braket{u}^{\pm}$ & $(u)^{\pm}$ & $\braket{g}^{\pm}$ &$\braket{g}^{\pm}$& $(g)^{\pm}$ &$\braket{g}^{\pm}$&$(g)^{\pm}$ \\ \hline
(d) E$_{g}\otimes$ E$_{1/2g}\to$ & $G_{3/2g}$ & & & & $(g)^+$ & $\braket{g}^{\pm}$ &$(g)^{\pm}$ & $\braket{u}^{\pm}$ & $\braket{u}^{\pm}$&$(u)^{\pm}$&$\braket{u}^{\pm}$&$(u)^{\pm}$\\
T$_{2g}\otimes$ E$_{1/2g}\to$ & $E_{5/2g}$ & & & & & A$^+_{1g}$, T$^-_{1g}$ & $\braket{g}^{\pm}$ &A$^{\pm}_{1u}$, T$^{\pm}_{1u}$  & A$^{\pm}_{2u}$, T$^{\pm}_{2u}$ & $\braket{u}^{\pm}$ &A$^{\pm}_{1u}$, T$^{\pm}_{1u}$&$\braket{u}^{\pm}$ \\
& $G_{3/2g}$ & & & & & & $(g)^+$ &$\braket{u}^{\pm}$ & $\braket{u}^{\pm}$&$(u)^{\pm}$&$\braket{u}^{\pm}$&$(u)^{\pm}$  \\ \hline
(f) A$_{2u}\otimes$ E$_{1/2g}\to$ &$E_{5/2u}$ & & & & & & &A$^+_{1g}$, T$^-_{1g}$&A$^{\pm}_{2g}$, T$^{\pm}_{2g}$& $\braket{g}^{\pm}$ &A$^{\pm}_{1g}$, T$^{\pm}_{1g}$&$\braket{g}^{\pm}$ \\
T$_{1u}\otimes$ E$_{1/2g}\to$ &$E_{1/2u}$ & & & & & & & & A$^+_{1g}$, T$^-_{1g}$ &$\braket{g}^{\pm}$&A$^{\pm}_{2g}$, T$^{\pm}_{2g}$ &$\braket{g}^{\pm}$\\
&$G_{3/2u}$ & & & & & & & & & $(g)^+$ &$\braket{g}^{\pm}$&$(g)^{\pm}$ \\
T$_{2u}\otimes$ E$_{1/2g}\to$ &$E_{5/2u}$ & & & & & & & & & & A$^+_{1g}$, T$^-_{1g}$&$\braket{g}^{\pm}$ \\
 &$G_{3/2u}$ & & & & & & & & & & & $(g)^+$
\end{tabular}
}
\end{table*}

\begin{table*}[h]
\caption{
Multipoles under the orthorhombic crystal system ($D_{2\rm h}$) in the spinless basis. 
The subscripts $g,u$ are omitted for the $D_{2}$ group. 
A$_{g}$, B$_{1u}$ $\to$ A$_{1}$, A$_{u}$, B$_{1g}$ $\to$ A$_{2}$, B$_{3g,u}$ $\to$ B$_{1}$, and B$_{2g,u}$ $\to$ B$_{2}$ for the $C_{2{\rm v}}$ group. 
}
\label{tab_ortho_less}
\centering
\scalebox{1.0}{
\begin{tabular}{c|c|ccc|ccccc|ccccccc}
& (s) A$_{g}$ & (p) B$_{3u}$ & B$_{2u}$ & B$_{1u}$ & (d) A$_{g}$ & A$_{g}$ & B$_{3g}$ & B$_{2g}$ & B$_{1g}$ & (f) A$_{u}$ & B$_{3u}$ & B$_{2u}$ & B$_{1u}$ & B$_{3u}$ & B$_{2u}$ & B$_{1u}$ \\ \hline
(s) A$_{g}$ & A$^+_{g}$ & B$^{\pm}_{3u}$ & B$^{\pm}_{2u}$ & B$^{\pm}_{1u}$ & A$^{\pm}_{g}$ & A$^{\pm}_{g}$ & B$^{\pm}_{3g}$ & B$^{\pm}_{2g}$ & B$^{\pm}_{1g}$ & A$^{\pm}_{u}$ & B$^{\pm}_{3u}$ & B$^{\pm}_{2u}$ & B$^{\pm}_{1u}$ & B$^{\pm}_{3u}$ & B$^{\pm}_{2u}$ & B$^{\pm}_{1u}$ \\ \hline
(p) B$_{3u}$ & & A$^+_{g}$ & B$^{\pm}_{1g}$ & B$^{\pm}_{2g}$ & B$^{\pm}_{3u}$ & B$^{\pm}_{3u}$ & A$^{\pm}_{u}$ & B$^{\pm}_{1u}$ & B$^{\pm}_{2u}$ & B$^{\pm}_{3g}$ & A$^{\pm}_{g}$ &  B$^{\pm}_{1g}$ &  B$^{\pm}_{2g}$ &  A$^{\pm}_{g}$ &  B$^{\pm}_{1g}$ &  B$^{\pm}_{2g}$ \\
B$_{2u}$ & & & A$^+_{g}$ & B$^{\pm}_{3g}$ & B$^{\pm}_{2u}$ & B$^{\pm}_{2u}$ & B$^{\pm}_{1u}$ & A$^{\pm}_{u}$ & B$^{\pm}_{3u}$ & B$^{\pm}_{2g}$ & B$^{\pm}_{1g}$ & A$^{\pm}_{g}$ & B$^{\pm}_{3g}$ & B$^{\pm}_{1g}$ & A$^{\pm}_{g}$ & B$^{\pm}_{3g}$ \\
B$_{1u}$ & & & & A$^+_{g}$ & B$^{\pm}_{1u}$ & B$^{\pm}_{1u}$ & B$^{\pm}_{2u}$ & B$^{\pm}_{3u}$ & A$^{\pm}_{u}$ & B$^{\pm}_{1g}$ & B$^{\pm}_{2g}$ & B$^{\pm}_{3g}$ & A$^{\pm}_{g}$ & B$^{\pm}_{2g}$ & B$^{\pm}_{3g}$ & A$^{\pm}_{g}$ \\ \hline
(d) A$_{g}$ & & & & & A$^+_{g}$ & A$^{\pm}_{g}$ & B$^{\pm}_{3g}$ & B$^{\pm}_{2g}$ & B$^{\pm}_{1g}$ & A$^{\pm}_{u}$ & B$^{\pm}_{3u}$ & B$^{\pm}_{2u}$ & B$^{\pm}_{1u}$ & B$^{\pm}_{3u}$ & B$^{\pm}_{2u}$ & B$^{\pm}_{1u}$ \\
A$_{g}$ & & & & & & A$^+_{g}$ & B$^{\pm}_{3g}$ & B$^{\pm}_{2g}$ & B$^{\pm}_{1g}$ & A$^{\pm}_{u}$ & B$^{\pm}_{3u}$ & B$^{\pm}_{2u}$ & B$^{\pm}_{1u}$ & B$^{\pm}_{3u}$ & B$^{\pm}_{2u}$ & B$^{\pm}_{1u}$ \\
B$_{3g}$ & & & & & & & A$^+_{g}$ & B$^{\pm}_{1g}$ & B$^{\pm}_{2g}$ & B$^{\pm}_{3u}$ & A$^{\pm}_{u}$ &  B$^{\pm}_{1u}$ &  B$^{\pm}_{2u}$ &  A$^{\pm}_{u}$ &  B$^{\pm}_{1u}$ &  B$^{\pm}_{2u}$ \\
B$_{2g}$ & & & & & & & & A$^+_{g}$ & B$^{\pm}_{3g}$ & B$^{\pm}_{2u}$ & B$^{\pm}_{1u}$ & A$^{\pm}_{u}$ & B$^{\pm}_{3u}$ & B$^{\pm}_{1u}$ & A$^{\pm}_{u}$ & B$^{\pm}_{3u}$ \\
B$_{1g}$ & & & & & & & & & A$^+_{g}$ & B$^{\pm}_{1u}$ & B$^{\pm}_{2u}$ & B$^{\pm}_{3u}$ & A$^{\pm}_{u}$ & B$^{\pm}_{2u}$ & B$^{\pm}_{3u}$ & A$^{\pm}_{u}$ \\ \hline
(f) A$_{u}$ & & & & & & & & & & A$^+_{g}$ & B$^{\pm}_{3g}$ & B$^{\pm}_{2g}$ & B$^{\pm}_{1g}$ & B$^{\pm}_{3g}$ & B$^{\pm}_{2g}$ & B$^{\pm}_{1g}$ \\
B$_{3u}$ & & & & & & & & & & & A$^+_{g}$ &  B$^{\pm}_{1g}$ &  B$^{\pm}_{2g}$ &  A$^{\pm}_{g}$ &  B$^{\pm}_{1g}$ &  B$^{\pm}_{2g}$ \\
B$_{2u}$ & & & & & & & & & & & & A$^+_{g}$ & B$^{\pm}_{3g}$ & B$^{\pm}_{1g}$ & A$^{\pm}_{g}$ & B$^{\pm}_{3g}$ \\
B$_{1u}$ & & & & & & & & & & & & & A$^+_{g}$ & B$^{\pm}_{2g}$ & B$^{\pm}_{3g}$ & A$^{\pm}_{g}$ \\
B$_{3u}$ & & & & & & & & & & & & & &  A$^+_{g}$ &  B$^{\pm}_{1g}$ &  B$^{\pm}_{2g}$ \\
B$_{2u}$ & & & & & & & & & & & & & & & A$^+_{g}$ & B$^{\pm}_{3g}$ \\
B$_{1u}$ & & & & & & & & & & & & & & & & A$^+_{g}$ 
\end{tabular}
}
\end{table*}

\begin{table*}[h]
\caption{
Multipoles under the orthorhombic crystal system ($D_{2\rm h}$) in the spinful basis. 
$(g,u)^{\pm}={\rm A}^{\pm}_{g,u} \oplus {\rm B}^{\mp}_{1g,u} \oplus {\rm B}^{\mp}_{2g,u} \oplus {\rm B}^{\mp}_{3g,u}$. 
The subscripts $g,u$ are omitted for the $D_{2}$ group. 
A$_{g}$, B$_{1u}$ $\to$ A$_{1}$, A$_{u}$, B$_{1g}$ $\to$ A$_{2}$, B$_{3g,u}$ $\to$ B$_{1}$, and B$_{2g,u}$ $\to$ B$_{2}$ for the $C_{2{\rm v}}$ group. 
}
\label{tab_ortho_ful}
\centering
\scalebox{0.85}{
\begin{tabular}{cc|c|ccc|ccccc|ccccccc}
& & (s) E$_{1/2g}$ & (p) E$_{1/2u}$ & E$_{1/2u}$ & E$_{1/2u}$ & (d) E$_{1/2g}$ & E$_{1/2g}$ & E$_{1/2g}$ & E$_{1/2g}$ & E$_{1/2g}$ & (f) E$_{1/2u}$ & E$_{1/2u}$ & E$_{1/2u}$ & E$_{1/2u}$ & E$_{1/2u}$ & E$_{1/2u}$ & E$_{1/2u}$ \\ \hline
(s) A$_{g}\otimes$ E$_{1/2g} \to$ & E$_{1/2g}$ & $(g)^+$ & $(u)^{\pm}$ & $(u)^{\pm}$ & $(u)^{\pm}$ & $(g)^{\pm}$ & $(g)^{\pm}$ & $(g)^{\pm}$ & $(g)^{\pm}$ & $(g)^{\pm}$ & $(u)^{\pm}$ & $(u)^{\pm}$ & $(u)^{\pm}$ & $(u)^{\pm}$ & $(u)^{\pm}$ & $(u)^{\pm}$ & $(u)^{\pm}$ \\ \hline
(p) B$_{3u}\otimes$ E$_{1/2g} \to$& E$_{1/2u}$ & & $(g)^+$ & $(g)^{\pm}$ & $(g)^{\pm}$ & $(u)^{\pm}$ & $(u)^{\pm}$ & $(u)^{\pm}$ & $(u)^{\pm}$ & $(u)^{\pm}$ & $(g)^{\pm}$ & $(g)^{\pm}$ &  $(g)^{\pm}$ &  $(g)^{\pm}$ &  $(g)^{\pm}$ &  $(g)^{\pm}$ &  $(g)^{\pm}$ \\
B$_{2u}\otimes$ E$_{1/2g} \to$& E$_{1/2u}$ & & & $(g)^+$ & $(g)^{\pm}$ & $(u)^{\pm}$ & $(u)^{\pm}$ & $(u)^{\pm}$ & $(u)^{\pm}$ & $(u)^{\pm}$ & $(g)^{\pm}$ & $(g)^{\pm}$ & $(g)^{\pm}$ & $(g)^{\pm}$ & $(g)^{\pm}$ & $(g)^{\pm}$ & $(g)^{\pm}$ \\
B$_{1u}\otimes$ E$_{1/2g} \to$& E$_{1/2u}$ & & & & $(g)^+$ & $(u)^{\pm}$ & $(u)^{\pm}$ & $(u)^{\pm}$ & $(u)^{\pm}$ & $(u)^{\pm}$ & $(g)^{\pm}$ & $(g)^{\pm}$ & $(g)^{\pm}$ & $(g)^{\pm}$ & $(g)^{\pm}$ & $(g)^{\pm}$ & $(g)^{\pm}$ \\ \hline
(d) A$_{g}\otimes$ E$_{1/2g} \to$& E$_{1/2g}$ & & & & & $(g)^+$ & $(g)^{\pm}$ & $(g)^{\pm}$ & $(g)^{\pm}$ & $(g)^{\pm}$ & $(u)^{\pm}$ & $(u)^{\pm}$ & $(u)^{\pm}$ & $(u)^{\pm}$ & $(u)^{\pm}$ & $(u)^{\pm}$ & $(u)^{\pm}$ \\
A$_{g}\otimes$ E$_{1/2g} \to$ & E$_{1/2g}$ & & & & & & $(g)^+$ & $(g)^{\pm}$ & $(g)^{\pm}$ & $(g)^{\pm}$ & $(u)^{\pm}$ & $(u)^{\pm}$ & $(u)^{\pm}$ & $(u)^{\pm}$ & $(u)^{\pm}$ & $(u)^{\pm}$ & $(u)^{\pm}$ \\
B$_{3g}\otimes$ E$_{1/2g} \to$& E$_{1/2g}$ & & & & & & & $(g)^+$ & $(g)^{\pm}$ & $(g)^{\pm}$ & $(u)^{\pm}$ & $(u)^{\pm}$ &  $(u)^{\pm}$ &  $(u)^{\pm}$ &  $(u)^{\pm}$ &  $(u)^{\pm}$ &  $(u)^{\pm}$ \\
B$_{2g}\otimes$ E$_{1/2g} \to$& E$_{1/2g}$ & & & & & & & & $(g)^+$ & $(g)^{\pm}$ & $(u)^{\pm}$ & $(u)^{\pm}$ & $(u)^{\pm}$ & $(u)^{\pm}$ & $(u)^{\pm}$ & $(u)^{\pm}$ & $(u)^{\pm}$ \\
B$_{1g}\otimes$ E$_{1/2g} \to$& E$_{1/2g}$ & & & & & & & & & $(g)^+$ & $(u)^{\pm}$ & $(u)^{\pm}$ & $(u)^{\pm}$ & $(u)^{\pm}$ & $(u)^{\pm}$ & $(u)^{\pm}$ & $(u)^{\pm}$ \\ \hline
(f) A$_{u}\otimes$ E$_{1/2g} \to$& E$_{1/2u}$ & & & & & & & & & & $(g)^+$ & $(g)^{\pm}$ & $(g)^{\pm}$ & $(g)^{\pm}$ & $(g)^{\pm}$ & $(g)^{\pm}$ & $(g)^{\pm}$ \\
B$_{3u}\otimes$ E$_{1/2g} \to$& E$_{1/2u}$ & & & & & & & & & & & $(g)^+$ &  $(g)^{\pm}$ &  $(g)^{\pm}$ &  $(g)^{\pm}$ &  $(g)^{\pm}$ &  $(g)^{\pm}$ \\
B$_{2u}\otimes$ E$_{1/2g} \to$& E$_{1/2u}$ & & & & & & & & & & & & $(g)^+$ & $(g)^{\pm}$ & $(g)^{\pm}$ & $(g)^{\pm}$ & $(g)^{\pm}$ \\
B$_{1u}\otimes$ E$_{1/2g} \to$& E$_{1/2u}$ & & & & & & & & & & & & & $(g)^+$ & $(g)^{\pm}$ & $(g)^{\pm}$ & $(g)^{\pm}$ \\
B$_{3u}\otimes$ E$_{1/2g} \to$& E$_{1/2u}$ & & & & & & & & & & & & & &  $(g)^+$ &  $(g)^{\pm}$ &  $(g)^{\pm}$ \\
B$_{2u}\otimes$ E$_{1/2g} \to$& E$_{1/2u}$ & & & & & & & & & & & & & & & $(g)^+$ & $(g)^{\pm}$ \\
B$_{1u}\otimes$ E$_{1/2g} \to$& E$_{1/2u}$ & & & & & & & & & & & & & & & & $(g)^+$ 
\end{tabular}
}
\end{table*}

\section{$D_{\rm 6h}$ and its subgroups}
\label{sec:Electric multipoles and CEF parameters}
Here, we summarize various tables for the hexagonal $D_{\rm 6h}$ and its subgroups. 
Tables~\ref{tab_D6h_CEFe} and \ref{tab_D6h_CEFo} represent even- and odd-parity hexagonal harmonics, Tables~\ref{tab_real_D6e} and \ref{tab_real_D6o} represent the operator expressions of even- and odd-parity multipoles in real space, Tables~\ref{tab_k-multipole_D6e} and \ref{tab_k-multipole_D6o} represent the even- and odd-parity multipoles in momentum space, Table~\ref{tab_multipoles_table2} represents the relation of multipoles under the $D_{\rm 6h}$ group and its subgroups, and Tables~\ref{tab_hexa_less} (\ref{tab_hexa_ful}) and \ref{tab_trig_less} (\ref{tab_trig_ful}) represent the active multipoles under the hexagonal and trigonal crystals in the spinless (spinful) basis, respectively. 

\begin{table*}[htb!]
\caption{
Even-parity hexagonal harmonics (E multipoles in unit of $-e$ in the hexagonal $D_{\rm 6h}$ group) up to $l=6$. 
The correspondences to the tesseral harmonics are shown. 
${\rm A}_{1}$, ${\rm A}_{2}$, ${\rm B}_{1}$, ${\rm B}_{2}$, ${\rm E}_{1}$, and ${\rm E}_{2}$ correspond to $\Gamma_{1}$, $\Gamma_{2}$, $\Gamma_{3}$, $\Gamma_{4}$, $\Gamma_{5}$, and $\Gamma_{6}$, respectively, in the Bethe notation.
Note that $Q_{4v}^{\beta1}=Q^{\alpha}_{4z}$, $Q^{\beta2}_{4u}=-Q_{4v}$, $Q_{4v}^{\beta2}=Q_{4z}^{\beta}$, $Q_{6s}=Q_{6z}^{\beta1}$, $Q_{6v}^{\beta1}=Q_{6z}^{\alpha}$, and $Q_{6v}^{\beta2}=Q_{6z}^{\beta2}$ in Table~\ref{tab_Oh_CEFe}.
}
\label{tab_D6h_CEFe}
\centering
\begingroup
\renewcommand{\arraystretch}{1.1}
 \begin{tabular}{ccccc}
 \hline \hline
 rank & irrep. & symbol & definition & correspondence \\ \hline
$0$  & ${\rm A}_{1g}^{+}$ & $Q_0$ & $1$ & $(00)$ \\
\hline
$2$  & ${\rm A}_{1g}^{+}$ & $Q_{u}$ & $\displaystyle \frac{1}{2}(3z^2-r^2)$ & $(20)$ \\
     & ${\rm E}_{1g}^{+}$ & $Q_{zx}$, $Q_{yz}$ & $\displaystyle \sqrt{3}z x$, $\displaystyle \sqrt{3}y z$ & $(21)$, $(21)'$ \\
     & ${\rm E}_{2g}^{+}$ & $Q_{v}$, $Q_{xy}$ & $\displaystyle \frac{\sqrt{3}}{2}(x^2-y^2)$, $\displaystyle \sqrt{3}x y$ & $(22)$, $(22)'$ \\
\hline
$4$ & ${\rm A}_{1g}^{+}$ & $Q_{40}$ & $\displaystyle \frac{1}{8}(35z^4-30 z^2 r^2 + 3 r^4)$ & $(40)$ \\
      & ${\rm B}_{1g}^{+}$ & $Q_{4a}$ & $\displaystyle \frac{\sqrt{70}}{4}yz (3x^2-y^2)$ & $(43)'$ \\
      & ${\rm B}_{2g}^{+}$ & $Q_{4b}$ & $\displaystyle \frac{\sqrt{70}}{4}zx (x^2-3y^2)$ & $(43)$ \\
     & ${\rm E}_{1g}^{+}$ & $Q_{4u}^{\alpha},Q_{4v}^{\alpha}$ & $\displaystyle \frac{\sqrt{10}}{4} zx (7z^2-3r^2)$, $\displaystyle \frac{\sqrt{10}}{4} yz (7z^2-3r^2)$ & $(41)$, $(41)'$ \\
     & ${\rm E}_{2g}^{+}$ & $Q_{4u}^{\beta1},Q_{4v}^{\beta1}$ & $\displaystyle \frac{\sqrt{35}}{8}(x^4-6x^2y^2 +y^4)$, $\displaystyle \frac{\sqrt{35}}{2}xy (x^2-y^2)$ & $(44)$, $(44)'$ \\
     & ${\rm E}_{2g}^{+}$ & $Q_{4u}^{\beta2},Q_{4v}^{\beta2}$ & $\displaystyle \frac{\sqrt{5}}{4} (x^2-y^2)(7z^2-r^2)$, $\displaystyle \frac{\sqrt{5}}{2} xy (7z^2-r^2)$ & $(42)$, $(42)'$ \\
\hline
$6$ & ${\rm A}_{1g}^{+}$  & $Q_{60}$ & $\displaystyle \frac{1}{16}(231z^6-315z^4 r^2 + 105 z^2 r^4 -5 r^6)$ & $(60)$
\\
& ${\rm A}_{1g}^{+}$  & $Q_{6c}$ &  $\displaystyle \frac{\sqrt{462}}{32}[x^6-15x^2y^2 (x^2-y^2)-y^6]$ & $(66)$
\\
& ${\rm A}_{2g}^{+}$  & $Q_{6s}$ &  $\displaystyle \frac{\sqrt{462}}{16}xy (3x^4-10x^2y^2 + 3y^4)$ & $(66)'$
\\
& ${\rm B}_{1g}^{+}$  & $Q_{6a}$ & $\displaystyle \frac{\sqrt{210}}{16}yz(3x^2-y^2)(11z^2-3r^2)$ & $(63)'$
\\
& ${\rm B}_{2g}^{+}$  & $Q_{6b}$ & $\displaystyle \frac{\sqrt{210}}{16}zx(x^2-3y^2)(11z^2-3r^2)$ & $(63)$
\\
& ${\rm E}_{1g}^{+}$  & $Q_{6u}^{\alpha1},Q_{6v}^{\alpha1}$ &  $\displaystyle \frac{3\sqrt{154}}{16}zx (x^4-10x^2y^2+5y^4)$, $\displaystyle \frac{3\sqrt{154}}{16}yz (5x^4-10x^2y^2+y^4)$ & $(65)$, $(65)'$
\\
& ${\rm E}_{1g}^{+}$  & $Q_{6u}^{\alpha2},Q_{6v}^{\alpha2}$ &  $\displaystyle \frac{\sqrt{21}}{8} zx [5r^4+3z^2(11z^2-10r^2)]$, $\displaystyle \frac{\sqrt{21}}{8} yz [5r^4+3z^2(11z^2-10r^2)]$ & $(61)$, $(61)'$
\\
& ${\rm E}_{2g}^{+}$  & $Q_{6u}^{\beta1},Q_{6v}^{\beta1}$ &  $\displaystyle \frac{3\sqrt{7}}{16}(x^4-6x^2 y^2+y^4)(11z^2-r^2)$, $\displaystyle \frac{3\sqrt{7}}{4}xy (x^2-y^2)(11z^2-r^2)$ & $(64)$, $(64)'$
\\
& ${\rm E}_{2g}^{+}$  & $Q_{6u}^{\beta2},Q_{6v}^{\beta2}$ &  $\displaystyle \frac{\sqrt{210}}{32} (x^2-y^2)[r^4+3z^2(11z^2-6r^2)]$, $\displaystyle \frac{\sqrt{210}}{16}xy [r^4 + 3z^2 (11z^2-6r^2)]$ & $(62)$, $(62)'$
\\
\hline\hline
\end{tabular}
\endgroup
\end{table*}
\begin{table*}[htb!]
\caption{
Odd-parity hexagonal harmonics (E multipoles in unit of $-e$ in the hexagonal $D_{\rm 6h}$ group) up to $l=6$.
Note that $Q_{50}=Q^{\alpha1}_{5z}$, $Q_{5u}^{\beta1}=Q_{5z}^{\alpha2}$, $Q_{5v}^{\beta1}=Q_{5u}$, $Q_{5u}^{\beta2}=Q_{5z}^{\beta}$, and $Q_{5v}^{\beta2}=-Q_{5v}$ in Table~\ref{tab_Oh_CEFo}.
}
\label{tab_D6h_CEFo}
\centering
\begingroup
\renewcommand{\arraystretch}{1.1}
 \begin{tabular}{ccccc}
 \hline \hline
rank & irrep. & symbol & definition & correspondence 
\\ \hline
 $1$   & ${\rm A}_{2u}^{+}$ & $Q_z$ & $\displaystyle z$ & $(10)$ \\
   & ${\rm E}_{1u}^{+}$ & $Q_x$, $Q_y$ & $\displaystyle x$, $\displaystyle y$ & $(11)$, $(11)'$ \\ 
 \hline
3  & ${\rm A}_{2u}^{+}$ & $Q_z^\alpha$  & $\displaystyle \frac{1}{2}z (5z^2-3r^2)$ & $(30)$  \\
 & ${\rm B}_{1u}^{+}$ & $Q_{3a}$  & $\displaystyle \frac{\sqrt{10}}{4}x(x^2-3y^2)$ & $(33)$ \\
 & ${\rm B}_{2u}^{+}$ & $Q_{3b}$  & $\displaystyle \frac{\sqrt{10}}{4}y(3x^2-y^2)$ & $(33)'$ \\
 & ${\rm E}_{1u}^{+}$ & $Q_{3u}$, $Q_{3v}$ & $\displaystyle \frac{\sqrt{6}}{4}x(5z^2-r^2)$, $\displaystyle \frac{\sqrt{6}}{4}y(5z^2-r^2)$ & $(31)$, $(31)'$\\
 & ${\rm E}_{2u}^{+}$ & $Q_z^\beta$, $Q_{xyz}$ & $\displaystyle \frac{\sqrt{15}}{2}z(x^2-y^2)$, $\displaystyle \sqrt{15}xyz$ & $(32)$, $(32)'$ \\
\hline
$5$ & ${\rm A}_{2u}^{+}$  & $Q_{50}$ & $\displaystyle \frac{1}{8}z(63 z^4 -70 z^2 r^2 + 15 r^4)$ & $(50)$
\\
& ${\rm B}_{1u}^{+}$ & $Q_{5a}$ & $\displaystyle \frac{\sqrt{70}}{16}x (x^2-3y^2)(9z^2-r^2)$ & $(53)$
\\
& ${\rm B}_{2u}^{+}$ & $Q_{5b}$ & $\displaystyle \frac{\sqrt{70}}{16}y (3x^2-y^2)(9z^2-r^2)$ & $(53)'$
\\
& ${\rm E}_{1u}^{+}$ & $Q_{5u}^{\alpha1},Q_{5v}^{\alpha1}$ & $\displaystyle \frac{3\sqrt{14}}{16}x(x^4-10x^2y^2+5y^4)$, $\displaystyle \frac{3\sqrt{14}}{16}y (5x^4-10x^2y^2+y^4)$ & $(55)$, $(55)'$
\\
& ${\rm E}_{1u}^{+}$ & $Q_{5u}^{\alpha2},Q_{5v}^{\alpha2}$ & $\displaystyle \frac{\sqrt{15}}{8} x [r^4 + 7z^2(3z^2-2r^2)]$, $\displaystyle \frac{\sqrt{15}}{8} y [r^4 + 7z^2(3z^2-2r^2)]$ & $(51)$, $(51)'$
\\
& ${\rm E}_{2u}^{+}$ & $Q_{5u}^{\beta1},Q_{5v}^{\beta1}$ & $\displaystyle \frac{3\sqrt{35}}{8}z (x^4-6 x^2y^2 + y^4)$, $\displaystyle \frac{3\sqrt{35}}{2}xyz (x^2-y^2)$ & $(54)$, $(54)'$
\\
& ${\rm E}_{2u}^{+}$ & $Q_{5u}^{\beta2},Q_{5v}^{\beta2}$ & $\displaystyle \frac{\sqrt{105}}{4}z (x^2-y^2) (3z^2-r^2)$, $\displaystyle \frac{\sqrt{105}}{2}xyz (3z^2-r^2)$ & $(52)$, $(52)'$
\\
\hline\hline
\end{tabular}
\endgroup
\end{table*}
\begin{table*}[htb!]
\begin{center}
\caption{
Operator expressions of even-parity multipoles up to $l=4$ in the hexagonal $D_{\rm 6h}$ group. 
}
\label{tab_real_D6e}
\begingroup
\renewcommand{\arraystretch}{1.2}
\begin{tabular}{ccccc}
\hline\hline
rank & type & irrep. & symbol & definition \\ \hline
$0$ & E  & ${\rm A}_{1g}^{+}$ & $Q_0$ & $1$ \\  
  & MT  & ${\rm A}_{1g}^{-}$ & $T_0'$ & $i$ \\ \hline 
$1$ & M  & ${\rm A}_{2g}^{-}$ & $M_z$ & $m^z_1$ \\
   &  & ${\rm E}_{1g}^{-}$ & $M_x$, $M_y$ & $m^x_1$, $m^y_1$ \\
      & ET &  ${\rm A}_{2g}^{+}$ & $G_z'$ & $z\bm{l}\cdot (\bm{r}\times \bm{\sigma})$ \\ 
    & & ${\rm E}_{1g}^{+}$ & $G_x'$, $G_y'$ & $x\bm{l}\cdot (\bm{r}\times \bm{\sigma})$, $y\bm{l}\cdot (\bm{r}\times \bm{\sigma})$ \\
\hline
$2$ & E  & ${\rm A}_{1g}^{+}$ & $Q_{u}$ & $\displaystyle \frac{1}{2}(3z^2-r^2)$  \\
 &      & ${\rm E}_{1g}^{+}$ & $Q_{zx}$, $Q_{yz}$ & $ \sqrt{3}z x$, $ \sqrt{3}y z$  \\
   &    & ${\rm E}_{2g}^{+}$ & $Q_{v}$, $Q_{xy}$ & $\displaystyle \frac{\sqrt{3}}{2}(x^2-y^2)$, $\sqrt{3}xy$ \\
& MT   & ${\rm A}_{1g}^{-}$ & $T_{u}$ & $3zt_{2}^{z}-\bm{r}\cdot \bm{t}_2$  \\
 &      & ${\rm E}_{1g}^{-}$ & $T_{zx}$, $T_{yz}$ & $\sqrt{3}(zt_{2}^{x}+xt_{2}^{z})$, $\sqrt{3}(yt_{2}^{z}+zt_{2}^{y})$  \\
   &    & ${\rm E}_{2g}^{-}$ & $T_{v}$, $T_{xy}$ & $\sqrt{3} (xt_{2}^{x}-yt_{2}^{y})$, $\sqrt{3}(xt_{2}^{y}+yt_{2}^{x})$  \\
\hline
3  & M  & ${\rm A}_{2g}^{-}$ & $M_z^\alpha$  & $\displaystyle 3\left[\frac{1}{2}(3z^{2}-r^{2})m_{3}^{z}-z(xm_{3}^{x}+ym_{3}^{y})\right]$  \\
 &  & ${\rm B}_{1g}^{-}$ & $M_{3a}$  & $\displaystyle 3\frac{\sqrt{10}}{4}[(x^2-y^2)m^x_3-2 x y m^y_3 ]$  \\
 &  & ${\rm B}_{2g}^{-}$ & $M_{3b}$  & $\displaystyle 3\frac{\sqrt{10}}{4}[2 x y m^x_3 +(x^2-y^2)m^y_3 ]$ \\
 &  & ${\rm E}_{1g}^{-}$ & $M_{3u}$, $M_{3v}$ & $\displaystyle \frac{\sqrt{6}}{4}[(5z^2-r^2)m^x_3+2x (5z m^z_3 - \bm{r} \cdot \bm{m}_3 )]$, $\displaystyle \frac{\sqrt{6}}{4}[(5z^2-r^2)m^y_3+2y (5z m^z_3 -\bm{r}\cdot \bm{m}_3)]$  \\
 &  & ${\rm E}_{2g}^{-}$ & $M_z^\beta$, $M_{xyz}$ & $\displaystyle \sqrt{15}\left[\frac{1}{2}(x^{2}-y^{2})m_{3}^{z}+z(xm_{3}^{x}-ym_{3}^{y})\right]$, $\sqrt{15}(yzm_{3}^{x}+zxm_{3}^{y}+xym_{3}^{z})$ \\
  & ET & ${\rm A}_{2g}^{+}$ & $G_z^\alpha$  & $9zg_{3}^{zz}-6(xg_{3}^{zx}+yg_{3}^{yz})-3z\sum_{\alpha}g_{3}^{\alpha\alpha}$  \\
 &  & ${\rm B}_{1g}^{+}$ & $G_{3a}$  & $\displaystyle 3\sqrt{\frac{5}{2}}[x(g^{xx}_3-g^{yy}_3)-2 y g^{xy}_3]$   \\
 &  & ${\rm B}_{2g}^{+}$ & $G_{3b}$  & $\displaystyle 3\sqrt{\frac{5}{2}}[2 x g^{xy}_3+y(g^{xx}_3-g^{yy}_3)]$ \\
 &  & ${\rm E}_{1g}^{+}$ & $G_{3u}$, $G_{3v}$ & $\displaystyle \sqrt{\frac{3}{2}}[x (5g^{zz}_3 - \sum_{\alpha} g^{\alpha\alpha}_3)+ 2 (5z g^{xz}_3-\sum_{\alpha}\alpha g^{x\alpha})]$, $\displaystyle \sqrt{\frac{3}{2}}[y (5g^{zz}_3 - \sum_{\alpha} g^{\alpha\alpha}_3)+ 2 (5z g^{yz}_3-\sum_{\alpha}\alpha g^{y\alpha})]$   \\
 &  & ${\rm E}_{2g}^{+}$ & $G_z^\beta$, $G_{xyz}$ & $\sqrt{15}[2(xg_{3}^{zx}-yg_{3}^{yz})+z(g_{3}^{xx}-g_{3}^{yy})]$, $2\sqrt{15}(xg_{3}^{yz}+yg_{3}^{zx}+zg_{3}^{xy})$\\
 \hline
$4$ & E  & ${\rm A}_{1g}^{+}$ & $Q_{40}$ & $\displaystyle  \frac{1}{8}(35z^4-30 z^2 r^2 + 3 r^4)$ \\
  &    & ${\rm B}_{1g}^{+}$ & $Q_{4a}$ & $\displaystyle  \frac{\sqrt{70}}{4}yz (3x^2-y^2)$ \\
  &   & ${\rm B}_{2g}^{+}$ & $Q_{4b}$ & $\displaystyle  \frac{\sqrt{70}}{4}zx (x^2-3y^2)$  \\
  &   & ${\rm E}_{1g}^{+}$ & $Q_{4u}^{\alpha},Q_{4v}^{\alpha}$ & $\displaystyle \frac{\sqrt{10}}{4} zx (7z^2-3r^2)$, $\displaystyle \frac{\sqrt{10}}{4} yz (7z^2-3r^2)$  \\
  &   & ${\rm E}_{2g}^{+}$ & $Q_{4u}^{\beta1},Q_{4v}^{\beta1}$ & $\displaystyle  \frac{\sqrt{35}}{8}(x^4-6x^2y^2 +y^4)$, $ \displaystyle \frac{\sqrt{35}}{2}xy (x^2-y^2)$  \\
  &  & ${\rm E}_{2g}^{+}$ & $Q_{4u}^{\beta2},Q_{4v}^{\beta2}$ & $\displaystyle  \frac{\sqrt{5}}{4} (x^2-y^2)(7z^2-r^2)$, $\displaystyle  \frac{\sqrt{5}}{2} xy (7z^2-r^2)$  \\
   & MT & ${\rm A}_{1g}^{-}$ & $T_{40}$ & $\displaystyle \frac{1}{2}[3(5z^2-r^2)(5z t^z_4 - \bm{r}\cdot \bm{t}_4)-40 z t^z_4]$  \\
  &   & ${\rm B}_{1g}^{-}$ & $T_{4a}$ & $\displaystyle \frac{1}{2}\sqrt{\frac{35}{2}}[6 xyz t^x_4 + 3 z (x^2-y^2)t^y_4 + y (3x^2-y^2)t^z_4]$  \\
  &   & ${\rm B}_{2g}^{-}$ & $T_{4b}$ & $\displaystyle \frac{1}{2}\sqrt{\frac{35}{2}}[  3 z (x^2-y^2)t^x_4-6 xyz t^y_4  + x (x^2-3y^2)t^z_4]$    \\
  &  & ${\rm E}_{1g}^{-}$ & $T_{4u}^{\alpha}, T_{4v}^{\alpha}$ & $\bm{t}_4 \cdot \bm{\nabla}Q^{\alpha}_{4u}$, $\bm{t}_4 \cdot \bm{\nabla}Q^{\alpha}_{4v}$  \\
  &  & ${\rm E}_{2g}^{-}$ & $T_{4u}^{\beta1}, T_{4v}^{\beta1}$ & $\displaystyle \frac{\sqrt{35}}{2}[x(x^2-3y^2)t^x_4-y(3x^2-y^2)t^y_4]$, $\displaystyle \frac{\sqrt{35}}{2}[y(3x^2-y^2)t^x_4+ x(x^2-3y^2)t^y_4]$  \\
  & & ${\rm E}_{2g}^{-}$ & $T_{4u}^{\beta2}, T_{4v}^{\beta2}$ & $\displaystyle \sqrt{5}[x(3z^2-x^2)t^x_4+y(y^2-3z^2)t^y_4+3 z(x^2-y^2)t^z_4]$, $\bm{t}_4 \cdot \bm{\nabla}Q^{\beta 2}_{4v}$  \\
  \hline\hline
\end{tabular}
\endgroup
\end{center}
\end{table*}
\begin{table*}[htb!]
\begin{center}
\caption{
Operator expressions of odd-parity multipoles up to $l=4$ in the hexagonal $D_{\rm 6h}$ group. 
}
\label{tab_real_D6o}
\begingroup
\renewcommand{\arraystretch}{1.2}
\begin{tabular}{ccccc}
\hline\hline
rank & type & irrep. & symbol & definition \\ \hline
$0$ & M  & ${\rm A}_{1u}^{-}$ & $M_0'$ & $\bm{r}\cdot \bm{\sigma}$ \\ 
    & ET  & ${\rm A}_{1u}^{+}$ & $G_0'$ & $\bm{l}\cdot (\bm{r}\times \bm{\sigma})$ \\ \hline
$1$ & E  & ${\rm A}_{2u}^{+}$ & $Q_z$ & $z$ \\
   &   & ${\rm E}_{1u}^{+}$ & $Q_x$, $Q_y$ & $x$, $y$ \\
     & MT  & ${\rm A}_{2u}^{-}$ & $T_z$ & $t^z_1$ \\
   &  & ${\rm E}_{1u}^{-}$ & $T_x$, $T_y$ & $t^x_1$, $t^y_1$ \\
\hline
$2$ & M  & ${\rm A}_{1u}^{-}$ & $M_{u}$ & $3zm_{2}^{z}-\bm{r}\cdot \bm{m}_2$  \\
 &     & ${\rm E}_{1u}^{-}$ & $M_{zx}$, $M_{yz}$ & $\sqrt{3}(zm_{2}^{x}+xm_{2}^{z})$, $\sqrt{3}(ym_{2}^{z}+zm_{2}^{y})$  \\
   &   & ${\rm E}_{2u}^{-}$ & $M_{v}$, $M_{xy}$ & $\sqrt{3} (xm_{2}^{x}-ym_{2}^{y})$, $\sqrt{3}(xm_{2}^{y}+ym_{2}^{x})$  \\
& ET  & ${\rm A}_{1u}^{+}$ & $G_{u}$ & $3g_{2}^{zz} -\sum_{\alpha}g_{2}^{\alpha\alpha}$  \\
 &     & ${\rm E}_{1u}^{+}$ & $G_{zx}$, $G_{yz}$ & $2\sqrt{3}g_{2}^{zx}$, $2\sqrt{3}g_{2}^{yz}$  \\
   &  & ${\rm E}_{2u}^{+}$ & $G_{v}$, $G_{xy}$ & $\sqrt{3}(g_{2}^{xx}-g_{2}^{yy})$, $2\sqrt{3}g_{2}^{xy}$  \\
\hline
3  & E & ${\rm A}_{2u}^{+}$ & $Q_z^\alpha$  & $\displaystyle \frac{1}{2}z (5z^2-3r^2)$  \\
 &   & ${\rm B}_{1u}^{+}$ & $Q_{3a}$  & $\displaystyle \frac{\sqrt{10}}{4}x(x^2-3y^2)$ \\
 &   & ${\rm B}_{2u}^{+}$ & $Q_{3b}$  & $\displaystyle \frac{\sqrt{10}}{4}y(3x^2-y^2)$ \\
 &   & ${\rm E}_{1u}^{+}$ & $Q_{3u}$, $Q_{3v}$ & $\displaystyle \frac{\sqrt{6}}{4}x(5z^2-r^2)$, $\displaystyle \frac{\sqrt{6}}{4}y(5z^2-r^2)$ \\
 &   & ${\rm E}_{2u}^{+}$ & $Q_z^\beta$, $Q_{xyz}$ & $\displaystyle \frac{\sqrt{15}}{2}z(x^2-y^2)$, $\sqrt{15}xyz$ \\
 & MT & ${\rm A}_{2u}^{-}$ & $T_z^\alpha$  & $\displaystyle 3\left[\frac{1}{2}(3z^{2}-r^{2})t_{3}^{z}-z(xt_{3}^{x}+yt_{3}^{y})\right]$  \\
 &   & ${\rm B}_{1u}^{-}$ & $T_{3a}$  & $\displaystyle 3\frac{\sqrt{10}}{4}[(x^2-y^2)t^x_3-2 x y t^y_3 ]$  \\
 &   & ${\rm B}_{2u}^{-}$ & $T_{3b}$  & $\displaystyle 3\frac{\sqrt{10}}{4}[2 x y t^x_3 +(x^2-y^2)t^y_3 ]$ \\
 &   & ${\rm E}_{1u}^{-}$ & $T_{3u}$, $T_{3v}$ & $\displaystyle \frac{\sqrt{6}}{4}[(5z^2-r^2)t^x_3+2x (5z t^z_3 - \bm{r}\cdot\bm{t}_3)]$, $\displaystyle \frac{\sqrt{6}}{4}[(5z^2-r^2)t^y_3+2y (5z t^z_3 -\bm{r}\cdot\bm{t}_3 )]$  \\
 &   & ${\rm E}_{2u}^{-}$ & $T_z^\beta$, $T_{xyz}$ & $\displaystyle \sqrt{15}\left[\frac{1}{2}(x^{2}-y^{2})t_{3}^{z}+z(xt_{3}^{x}-yt_{3}^{y})\right]$, $\sqrt{15}(yzt_{3}^{x}+zxt_{3}^{y}+xyt_{3}^{z})$ \\
 \hline
$4$ & M & ${\rm A}_{1u}^{-}$ & $M_{40}$ & $\displaystyle \frac{1}{2}[3(5z^2-r^2)(5z m^z_4- \bm{r}\cdot\bm{m}_4)-40 z m^z_4]$  \\
  &    & ${\rm B}_{1u}^{-}$ & $M_{4a}$ & $\displaystyle \frac{1}{2}\sqrt{\frac{35}{2}}[6 xyz m^x_4 + 3 z (x^2-y^2)m^y_4 + y (3x^2-y^2)m^z_4]$  \\
  &    & ${\rm B}_{2u}^{-}$ & $M_{4b}$ & $\displaystyle \frac{1}{2}\sqrt{\frac{35}{2}}[  3 z (x^2-y^2)m^x_4-6 xyz m^y_4  + x (x^2-3y^2)m^z_4]$    \\
  &   & ${\rm E}_{1u}^{-}$ & $M_{4u}^{\alpha}, M_{4v}^{\alpha}$ & $\bm{m}_4 \cdot \bm{\nabla}Q^{\alpha}_{4u}$, $\bm{m}_4 \cdot \bm{\nabla}Q^{\alpha}_{4v}$  \\
  &   & ${\rm E}_{2u}^{-}$ & $M_{4u}^{\beta1}, M_{4v}^{\beta1}$ & $\displaystyle \frac{\sqrt{35}}{2}[x(x^2-3y^2)m^x_4-y(3x^2-y^2)m^y_4]$, $\displaystyle \frac{\sqrt{35}}{2}[y(3x^2-y^2)m^x_4+ x(x^2-3y^2)m^y_4]$  \\
  &   & ${\rm E}_{2u}^{-}$ & $M_{4u}^{\beta2}, M_{4v}^{\beta2}$ & $\sqrt{5}[x(3z^2-x^2)m^x_4+y(y^2-3z^2)m^y_4+3 z(x^2-y^2)m^z_4]$, $\bm{m}_4 \cdot \bm{\nabla}Q^{\beta 2}_{4v}$  \\
     & ET & ${\rm A}_{1u}^{+}$ & $G_{40}$ & $\displaystyle \frac{3}{2}[ (3x^2+y^2-4z^2)g^{xx}_4+ (x^2+3y^2-4z^2)g^{yy}_4+4(3z^2-r^2)g^{zz}_4+4 xy g^{xy}_4  -16 z(x g^{xz}_4+y g^{yz}_4)]$  \\
  &   & ${\rm B}_{1u}^{+}$ & $G_{4a}$ & $\displaystyle 3\sqrt{\frac{35}{2}}[yz(g^{xx}_4-g^{yy}_4) + 2 x (zg^{xy}_4 + y g^{xz}_4)+(x^2-y^2)g^{yz}_4]$  \\
  &   & ${\rm B}_{2u}^{+}$ & $G_{4b}$ & $\displaystyle 3\sqrt{\frac{35}{2}}[zx(g^{xx}_4-g^{yy}_4) - 2 y (xg^{yz}_4 + z g^{xy}_4)+(x^2-y^2)g^{xz}_4]$    \\
  &  & ${\rm E}_{1u}^{+}$ & $G_{4u}^{\alpha}, G_{4v}^{\alpha}$ & $\sum_{\alpha \beta}g^{\alpha \beta}_4 \nabla_{\alpha}\nabla_{\beta}Q_{4u}^{\alpha}$, $\sum_{\alpha \beta}g^{\alpha \beta}_4 \nabla_{\alpha}\nabla_{\beta}Q_{4v}^{\alpha}$  \\
  &  & ${\rm E}_{2u}^{+}$ & $G_{4u}^{\beta1}, G_{4v}^{\beta1}$ & $\displaystyle 3\frac{\sqrt{35}}{2}[(x^2-y^2)(g^{xx}_4-g^{yy}_4)-4 x y g^{xy}_4]$, $3\sqrt{35}[xy(g^{xx}_4-g^{yy}_4)+(x^2-y^2)g^{xy}_4]$  \\
  &  & ${\rm E}_{2u}^{+}$ & $G_{4u}^{\beta2}, G_{4v}^{\beta2}$ & $3\sqrt{5}[(z^2-x^2)g^{xx}_4+(y^2-z^2)g^{yy}_4+(x^2-y^2)g^{zz}_4+4 z (x g^{zx}_4-y g^{yz}_4)]$, $\sum_{\alpha \beta}g^{\alpha \beta}_4 \nabla_{\alpha}\nabla_{\beta}Q_{4v}^{\beta 2}$  \\
  \hline\hline
\end{tabular}
\endgroup
\end{center}
\end{table*}
\begin{table*}[htb!]
\caption{
Even-parity multipoles in momentum space up to $l=4$ in the hexagonal $D_{6{\rm h}}$ group.
The higher-order representation with respect to $\bm{k}$ is also shown for $Q_0$ and $(M_{x}, M_{y}, M_{z})$ in the bracket. 
}
\label{tab_k-multipole_D6e}
\centering
\begingroup
\renewcommand{\arraystretch}{1.05}
 \begin{tabular}{ccccc}
 \hline \hline
rank & type & irrep. & symbol & definition \\ \hline
$0$ & E & ${\rm A}_{1g}^{+}$ & $Q_0$ & $\sigma_{0}$ [$(k_x^2+k_y^2+k_z^2)\sigma_0$]  \\  
 & MT & ${\rm A}_{1g}^{-}$ & $T'_0$ & $i\sigma_{0}$ \\  
\hline
$1$ & M & ${\rm A}_{2g}^{-}$ & $M_z$ & $\sigma_z$ $\displaystyle  \left[(\bm{k}\cdot \bm{\sigma})k_z-\frac{1}{3}k^2\sigma_z \right]$ \\
   &  & ${\rm E}_{1g}^{-}$ & $M_x$, $M_y$ & $ \sigma_{x}$, $\sigma_{y}$$\displaystyle \left[(\bm{k}\cdot \bm{\sigma})k_x-\frac{1}{3}k^2\sigma_x, (\bm{k}\cdot \bm{\sigma})k_y-\frac{1}{3}k^2\sigma_y \right]$ \\
 & ET & ${\rm A}_{2g}^{+}$ & $G'_z$ & $i\sigma_z$  \\
   &  & ${\rm E}_{1g}^{+}$ & $G'_x$, $G'_y$ & $ i\sigma_{x}$, $i\sigma_{y}$ \\
\hline
$2$ & E & ${\rm A}_{1g}^{+}$ & $Q_{u}$ & $\displaystyle \frac{1}{2}(3k_{z}^2-k^2)\sigma_{0}$  \\
 &   & ${\rm E}_{1g}^{+}$ & $Q_{zx}$, $Q_{yz}$ & $\sqrt{3}k_{z}k_{x}\sigma_{0}$, $\sqrt{3}k_{y}k_{z}\sigma_{0}$  \\
   &   & ${\rm E}_{2g}^{+}$ & $Q_{v}$, $Q_{xy}$ & $\displaystyle \frac{\sqrt{3}}{2}(k_{x}^2-k_{y}^2)\sigma_{0}$, $\sqrt{3}k_{x}k_{y}\sigma_{0}$ \\
& MT & ${\rm A}_{1g}^{-}$ & $T_{u}$ & $3k_{z}Q_{z}-\bm{k}\cdot\bm{Q}$ \\
 &    & ${\rm E}_{1g}^{-}$ & $T_{zx}$, $T_{yz}$ & $\sqrt{3}(k_{z}Q_{x}+k_{x}Q_{z})$, $\sqrt{3}(k_{y}Q_{z}+k_{z}Q_{y})$ \\
   & & ${\rm E}_{2g}^{-}$ & $T_{v}$, $T_{xy}$ & $\sqrt{3}(k_{x}Q_{x}-k_{y}Q_{y})$, $\sqrt{3}(k_{x}Q_{y}+k_{y}Q_{x})$ \\
\hline
3  & M & ${\rm A}_{2g}^{-}$ & $M_z^\alpha$  & $\displaystyle  3\left[\frac{1}{2}(3k_{z}^{2}-k^{2})\sigma_{z}-k_{z}(k_{x}\sigma_{x}+k_{y}\sigma_{y})\right]$  \\
 &  & ${\rm B}_{1g}^{-}$ & $M_{3a}$  & $\displaystyle  3\frac{\sqrt{10}}{4}[(k_x^2-k_y^2)\sigma_x-2 k_x k_y \sigma_y ]$ \\
 &  & ${\rm B}_{2g}^{-}$ & $M_{3b}$  & $\displaystyle  3\frac{\sqrt{10}}{4}[2 k_x k_y \sigma_x +(k_x^2-k_y^2) \sigma_y ]$ \\
 &  & ${\rm E}_{1g}^{-}$ & $M_{3u}$, $M_{3v}$ & $\displaystyle  \frac{\sqrt{6}}{4}[(5k_z^2-k^2)\sigma_x+2k_x (5 k_z \sigma_z - \bm{k}\cdot\bm{\sigma})]$, $\displaystyle  \frac{\sqrt{6}}{4}[(5 k_z^2-k^2)\sigma_y+2 k_y (5 k_z \sigma_z - \bm{k}\cdot \bm{\sigma})]$  \\
 &  & ${\rm E}_{2g}^{-}$ & $M_z^\beta$, $M_{xyz}$ & $\displaystyle  \sqrt{15}\left[\frac{1}{2}(k_{x}^{2}-k_{y}^{2})\sigma_{z}+k_{z}(k_{x}\sigma_{x}-k_{y}\sigma_{y})\right]$, $\displaystyle  \sqrt{15}(k_{y}k_{z}\sigma_{x}+k_{z}k_{x}\sigma_{y}+k_{x}k_{y}\sigma_{z})$ \\
  & ET & ${\rm A}_{2g}^{+}$ & $G_z^{\alpha'}$  & $\displaystyle  3i\left[\frac{1}{2}(3k_{z}^{2}-k^{2})\sigma_{z}-k_{z}(k_{x}\sigma_{x}+k_{y}\sigma_{y})\right]$  \\
 &  & ${\rm B}_{1g}^{+}$ & $G'_{3a}$  & $\displaystyle  3i\frac{\sqrt{10}}{4}[(k_x^2-k_y^2)\sigma_x-2 k_x k_y \sigma_y ]$ \\
 &  & ${\rm B}_{2g}^{+}$ & $G'_{3b}$  & $\displaystyle  3i\frac{\sqrt{10}}{4}[2 k_x k_y \sigma_x +(k_x^2-k_y^2) \sigma_y ]$ \\
 &  & ${\rm E}_{1g}^{+}$ & $G'_{3u}$, $G'_{3v}$ & $\displaystyle  \frac{\sqrt{6}}{4}i[(5k_z^2-k^2)\sigma_x+2k_x (5 k_z \sigma_z - \bm{k}\cdot \bm{\sigma})]$, $\displaystyle  \frac{\sqrt{6}}{4}i[(5 k_z^2-k^2)\sigma_y+2 k_y (5 k_z\sigma_z - \bm{k}\cdot\bm{\sigma})]$  \\
 &  & ${\rm E}_{2g}^{+}$ & $G_z^{\beta'}$, $G'_{xyz}$ & $\displaystyle  \sqrt{15}i\left[\frac{1}{2}(k_{x}^{2}-k_{y}^{2})\sigma_{z}+k_{z}(k_{x}\sigma_{x}-k_{y}\sigma_{y})\right]$, $\displaystyle  \sqrt{15}i(k_{y}k_{z}\sigma_{x}+k_{z}k_{x}\sigma_{y}+k_{x}k_{y}\sigma_{z})$ \\
 \hline
$4$ & E & ${\rm A}_{1g}^{+}$ & $Q_{40}$ & $\displaystyle  \frac{1}{8}(35k_z^4-30 k_z^2 k^2 + 3 k^4) \sigma_0$ \\
  &   & ${\rm B}_{1g}^{+}$ & $Q_{4a}$ & $\displaystyle  \frac{\sqrt{70}}{4}k_y k_z (3k_x^2-k_y^2)\sigma_0$ \\
  &   & ${\rm B}_{2g}^{+}$ & $Q_{4b}$ & $\displaystyle  \frac{\sqrt{70}}{4}k_z k_x (k_x^2-3k_y^2)\sigma_0$  \\
  &  & ${\rm E}_{1g}^{+}$ & $Q_{4u}^{\alpha},Q_{4v}^{\alpha}$ & $\displaystyle \frac{\sqrt{10}}{4} k_z k_x (7 k_z^2-3k^2)\sigma_0$, $\displaystyle \frac{\sqrt{10}}{4} k_y k_z (7k_z^2-3k^2)\sigma_0$  \\
  &  & ${\rm E}_{2g}^{+}$ & $Q_{4u}^{\beta1},Q_{4v}^{\beta1}$ & $\displaystyle  \frac{\sqrt{35}}{8}(k_x^4-6k_x^2k_y^2 +k_y^4)\sigma_0$, $\displaystyle  \frac{\sqrt{35}}{2}k_x k_y (k_x^2-k_y^2)\sigma_0$  \\
  &  & ${\rm E}_{2g}^{+}$ & $Q_{4u}^{\beta2},Q_{4v}^{\beta2}$ & $\displaystyle  \frac{\sqrt{5}}{4} (k_x^2-k_y^2)(7k_z^2-k^2)\sigma_0$, $\displaystyle  \frac{\sqrt{5}}{2} k_xk_y (7k_z^2-k^2)\sigma_0$  \\
   & MT & ${\rm A}_{1g}^{-}$ & $T_{40}$ & $\displaystyle \frac{1}{2}[3(5 k_z^2-k^2)(5 k_z Q_z  - \bm{k}\cdot \bm{Q})-40 k_z Q_z ]$  \\
  &   & ${\rm B}_{1g}^{-}$ & $T_{4a}$ &  $\displaystyle \frac{1}{2}\sqrt{\frac{35}{2}}[6 k_x k_y k_z Q_x + 3 k_z (k_x^2-k_y^2)Q_y + k_y (3k_x^2-k_y^2)Q_z]$  \\
  &   & ${\rm B}_{2g}^{-}$ & $T_{4b}$ &  $\displaystyle \frac{1}{2}\sqrt{\frac{35}{2}}[  3 k_z (k_x^2-k_y^2)Q_x-6 k_x k_y k_z Q_y  + k_x (k_x^2-3k_y^2)Q_z]$   \\
  &  & ${\rm E}_{1g}^{-}$ & $T_{4u}^{\alpha}, T_{4v}^{\alpha}$ & $\bm{Q} \cdot \bm{\nabla}Q^{\alpha}_{4u}$, $\bm{Q} \cdot \bm{\nabla}Q^{\alpha}_{4v}$  \\
  &  & ${\rm E}_{2g}^{-}$ & $T_{4u}^{\beta1}, T_{4v}^{\beta1}$ & $\displaystyle \frac{\sqrt{35}}{2}[k_x(k_x^2-3k_y^2)Q_x-k_y(3k_x^2-k_y^2)Q_y]$, $\displaystyle \frac{\sqrt{35}}{2}[k_y(3k_x^2-k_y^2)Q_x+ k_x(k_x^2-3k_y^2)Q_y]$  \\
  &  & ${\rm E}_{2g}^{-}$ & $T_{4u}^{\beta2}, T_{4v}^{\beta2}$ & $\sqrt{5}[k_x(3k_z^2-k_x^2)Q_x+k_y(k_y^2-3k_z^2)Q_y+3 k_z(k_x^2-k_y^2)Q_z]$, $\bm{Q} \cdot \bm{\nabla}Q^{\beta 2}_{4v}$  \\
\hline\hline
\end{tabular}
\endgroup
\end{table*}
\begin{table*}[htb!]
\caption{
Odd-parity multipoles in momentum space up to $l=4$ in the hexagonal $D_{\rm 6h}$ group.
The higher-order representation with respect to $\bm{k}$ is also shown for $Q_0$ and $(M_{x}, M_{y}, M_{z})$ in the bracket. 
}
\label{tab_k-multipole_D6o}
\centering
\begingroup
\renewcommand{\arraystretch}{1.05}
 \begin{tabular}{ccccc}
 \hline \hline
rank & type & irrep. & symbol & definition \\ \hline
$0$ & M  & ${\rm A}_{1u}^{-}$ & $M'_0$ & $i\bm{k}\cdot\bm{\sigma}$ \\ 
 & ET  & ${\rm A}_{1u}^{+}$ & $G_0$ & $\bm{k}\cdot\bm{\sigma}$ \\ 
\hline
$1$ & E  & ${\rm A}_{2u}^{+}$ & $Q_z$ & $k_{x}\sigma_{y}-k_{y}\sigma_{x}$ \\
   &   & ${\rm E}_{1u}^{+}$ & $Q_x$, $Q_y$ & $k_{y}\sigma_{z}-k_{z}\sigma_{y}$, $k_{z}\sigma_{x}-k_{x}\sigma_{z}$ \\
     & MT  & ${\rm A}_{2u}^{-}$ & $T_z$ & $k_{z}\sigma_{0}$ \\
   &   & ${\rm E}_{1u}^{-}$ & $T_x$, $T_y$ & $k_{x}\sigma_{0}$, $k_{y}\sigma_{0}$ \\
\hline
$2$ & M  & ${\rm A}_{1u}^{-}$ & $M'_{u}$ & $3ik_{z}\sigma_{z}-i\bm{k}\cdot\bm{\sigma}$\\
 &     & ${\rm E}_{1u}^{-}$ & $M'_{zx}$, $M'_{yz}$ & $\sqrt{3}i(k_{z}\sigma_{x}+k_{x}\sigma_{z})$, $\sqrt{3}i(k_{y}\sigma_{z}+k_{z}\sigma_{y})$ \\
   &  & ${\rm E}_{2u}^{-}$ & $M'_{v}$, $M'_{xy}$ & $\sqrt{3}i(k_{x}\sigma_{x}-k_{y}\sigma_{y})$, $\sqrt{3}i(k_{x}\sigma_{y}+k_{y}\sigma_{x})$ \\
 & ET  & ${\rm A}_{1u}^{+}$ & $G_{u}$ & $3k_{z}\sigma_{z}-\bm{k}\cdot\bm{\sigma}$\\
 &     & ${\rm E}_{1u}^{+}$ & $G_{zx}$, $G_{yz}$ & $\sqrt{3}(k_{z}\sigma_{x}+k_{x}\sigma_{z})$, $\sqrt{3}(k_{y}\sigma_{z}+k_{z}\sigma_{y})$ \\
   &  & ${\rm E}_{2u}^{+}$ & $G_{v}$, $G_{xy}$ & $\sqrt{3}(k_{x}\sigma_{x}-k_{y}\sigma_{y})$, $\sqrt{3}(k_{x}\sigma_{y}+k_{y}\sigma_{x})$ \\
\hline
3  & E & ${\rm A}_{2u}^{+}$ & $Q_z^\alpha$  & $\displaystyle 3\left[\frac{1}{2}(3k_{z}^{2}-k^{2})Q_{z}-k_{z}(k_{x}Q_{x}+k_{y}Q_{y})\right]$ \\
 &  & ${\rm B}_{1u}^{+}$ & $Q_{3a}$  & $\displaystyle  3\frac{\sqrt{10}}{4}\left[(k_x^2-k_y^2)Q_x-2 k_x k_y Q_y \right]$ \\
 &  & ${\rm B}_{2u}^{+}$ & $Q_{3b}$  & $\displaystyle  3\frac{\sqrt{10}}{4}\left[2 k_x k_y Q_x +(k_x^2-k_y^2)Q_y \right]$ \\
 &  & ${\rm E}_{1u}^{+}$ & $Q_{3u}$, $Q_{3v}$ & $\displaystyle  \frac{\sqrt{6}}{4}[(5k_z^2-k^2)Q_x+2k_x (5 k_z Q_z  - \bm{k} \cdot \bm{Q})]$, $\displaystyle  \frac{\sqrt{6}}{4}[(5 k_z^2-k^2)Q_y+2 k_y (5 k_z Q_z - \bm{k}\cdot\bm{Q})]$\\
 &  & ${\rm E}_{2u}^{+}$ & $Q_z^\beta$, $Q_{xyz}$ &$\displaystyle  \sqrt{15}\left[\frac{1}{2}(k_{x}^{2}-k_{y}^{2})Q_{z}+k_{z}(k_{x}Q_{x}-k_{y}Q_{y})\right]$, $\sqrt{15}(k_{y}k_{z}Q_{x}+k_{z}k_{x}Q_{y}+k_{x}k_{y}Q_{z})$  \\
 & MT & ${\rm A}_{2u}^{-}$ & $T_z^\alpha$  &  $\displaystyle  \frac{1}{2}k_{z}(5k_{z}^{2}-3k^{2})\sigma_{0}$ \\
 &  & ${\rm B}_{1u}^{-}$ & $T_{3a}$  & $\displaystyle  \frac{\sqrt{10}}{4}k_x(k_x^2-3k_y^2)\sigma_0$  \\
 &  & ${\rm B}_{2u}^{-}$ & $T_{3b}$  & $\displaystyle  \frac{\sqrt{10}}{4}k_y(3k_x^2-k_y^2)\sigma_0$ \\
 &  & ${\rm E}_{1u}^{-}$ & $T_{3u}$, $T_{3v}$ & $\displaystyle  \frac{\sqrt{6}}{4}k_x(5k_z^2-k^2)\sigma_0$, $\displaystyle  \frac{\sqrt{6}}{4}k_y(5k_z^2-k^2)\sigma_0$  \\
 &  & ${\rm E}_{2u}^{-}$ & $T_z^\beta$, $T_{xyz}$ & $\displaystyle  \frac{\sqrt{15}}{2}k_{z}(k_{x}^{2}-k_{y}^{2})\sigma_{0}$, $\sqrt{15}k_{x}k_{y}k_{z}\sigma_{0}$ \\
 \hline
$4$ & M & ${\rm A}_{1u}^{-}$ & $M'_{40}$ & $\displaystyle  \frac{1}{2}i[3(5 k_z^2-k^2)(5k_z \sigma_z -\bm{k}\cdot\bm{\sigma})-40k_z \sigma_z]$  \\
  &   & ${\rm B}_{1u}^{-}$ & $M'_{4a}$ &  $\displaystyle  \frac{1}{2}\sqrt{\frac{35}{2}}i[6 k_x k_y k_z \sigma_x + 3 k_z (k_x^2-k_y^2)\sigma_y + k_y (3k_x^2-k_y^2)\sigma_z]$ \\
  &  & ${\rm B}_{2u}^{-}$ & $M'_{4b}$ & $\displaystyle  \frac{1}{2}\sqrt{\frac{35}{2}}i[  3 k_z (k_x^2-k_y^2)\sigma_x-6 k_x k_y k_z \sigma_y  + k_x (k_x^2-3k_y^2)\sigma_z]$    \\
  & & ${\rm E}_{1u}^{-}$ & $M_{4u}^{\alpha'}, M_{4v}^{\alpha'}$ & $i\bm{\sigma} \cdot \bm{\nabla}Q^{\alpha}_{4u}$, $i\bm{\sigma} \cdot \bm{\nabla}Q^{\alpha}_{4v}$  \\
  &  & ${\rm E}_{2u}^{-}$ & $M_{4u}^{\beta1'}, M_{4v}^{\beta1'}$ &  $\displaystyle  \frac{\sqrt{35}}{2}i[k_x(k_x^2-3k_y^2)\sigma_x-k_x(3k_x^2-k_y^2)\sigma_y]$, $\displaystyle  \frac{\sqrt{35}}{2}i[k_y(3k_x^2-k_y^2)\sigma_x+ k_x(k_x^2-3k_y^2)\sigma_y]$ \\
  & & ${\rm E}_{2u}^{-}$ & $M_{4u}^{\beta2'}, M_{4v}^{\beta2'}$ &  $\sqrt{5}i[k_x(3k_z^2-k_x^2)\sigma_x+k_y(k_y^2-3k_z^2)\sigma_y+3 k_z(k_x^2-k_y^2)\sigma_z]$, $i\bm{\sigma} \cdot \bm{\nabla}Q^{\beta 2}_{4v}$ \\
 & ET & ${\rm A}_{1u}^{+}$ & $G_{40}$ & $\displaystyle  \frac{1}{2}[3(5 k_z^2-k^2)(5k_z \sigma_z - \bm{k}\cdot \bm{\sigma} )-40 k_z \sigma_z]$  \\
  &   & ${\rm B}_{1u}^{+}$ & $G_{4a}$ &  $\displaystyle  \frac{1}{2}\sqrt{\frac{35}{2}}[6 k_x k_y k_z \sigma_x + 3 k_z (k_x^2-k_y^2)\sigma_y + k_y (3k_x^2-k_y^2)\sigma_z]$ \\
  &  & ${\rm B}_{2u}^{+}$ & $G_{4b}$ & $\displaystyle  \frac{1}{2}\sqrt{\frac{35}{2}}[  3 k_z (k_x^2-k_y^2)\sigma_x-6 k_x k_y k_z \sigma_y  + k_x (k_x^2-3k_y^2)\sigma_z]$    \\
  & & ${\rm E}_{1u}^{+}$ & $G_{4u}^{\alpha}, G_{4v}^{\alpha}$ & $\bm{\sigma} \cdot \bm{\nabla}Q^{\alpha}_{4u}$, $\bm{\sigma} \cdot \bm{\nabla}Q^{\alpha}_{4v}$  \\
  &  & ${\rm E}_{2u}^{+}$ & $G_{4u}^{\beta1}, G_{4v}^{\beta1}$ &  $\displaystyle  \frac{\sqrt{35}}{2}[k_x(k_x^2-3k_y^2)\sigma_x-k_y(3k_x^2-k_y^2)\sigma_y]$, $\displaystyle  \frac{\sqrt{35}}{2}[k_y(3k_x^2-k_y^2)\sigma_x+ k_x(k_x^2-3k_y^2)\sigma_y]$ \\
  & & ${\rm E}_{2u}^{+}$ & $G_{4u}^{\beta2}, G_{4v}^{\beta2}$ &  $\sqrt{5}[k_x(3k_z^2-k_x^2)\sigma_x+k_y(k_y^2-3k_z^2)\sigma_y+3 k_z(k_x^2-k_y^2)\sigma_z]$, $\bm{\sigma} \cdot \bm{\nabla}Q^{\beta 2}_{4v}$ \\
\hline\hline
\end{tabular}
\endgroup
\end{table*}
\begin{table*}
\caption{
Multipoles under hexagonal and trigonal crystals. 
The upper and lower columns represent even-parity ($g$) and odd-parity ($u$) multipoles. 
We take the $x$ axis as the $C_{2}'$ rotation axis.
For $C_{3 {\rm v}}$, we take the $zx$ plane as the $\sigma_v$ mirror plane.
$C_{\rm 3i}=S_{6}$. 
The basis functions are taken as $(x,y)\to x\pm iy$ for $C_{\rm 6h}$, $C_{6}$, $C_{\rm 3h}$, and $C_{\rm 3i}$. 
}
\label{tab_multipoles_table2}
\centering
\begin{tabular}{llll|ccccccc|ccccc} \hline\hline
E & ET & M & MT &
$D_{\rm 6h}$ & $D_{6}$ & $C_{\rm 6h}$ & $C_{\rm 6v}$ & $C_{6}$ & $D_{\rm 3h}$ & $C_{\rm 3h}$ &
$D_{\rm 3d}$ & $D_{3}$ & $C_{\rm 3v}$ & $C_{\rm 3i}$ & $C_{3}$ \\ \hline
$Q_{0}$, $Q_{u}$, $Q_{40}$ & --- & --- & $T_{0}$, $T_{u}$, $T_{40}$ &
A$_{1g}$ & A$_{1}$ & A$_{g}$ & A$_{1}$ & A & A$_{1}'$ & A$'$ &
A$_{1g}$ & A$_{1}$ & A$_{1}$ & A$_{g}$ & A
\\
--- &$G_{z}$, $G_{z}^{\alpha}$ & $M_{z}$, $M_{z}^{\alpha}$ & --- &
A$_{2g}$ & A$_{2}$ & A$_{g}$ & A$_{2}$ & A & A$_{2}'$ & A$'$ &
A$_{2g}$ & A$_{2}$ & A$_{2}$ & A$_{g}$ & A
\\
$Q_{4a}$ & $G_{3a}$ & $M_{3a}$ & $T_{4a}$ &
B$_{1g}$ & B$_{1}$ & B$_{g}$ & B$_{2}$ & B & A$_{1}''$ & A$''$ &
A$_{1g}$ & A$_{1}$ & A$_{2}$ & A$_{g}$ & A
\\
$Q_{4b}$ & $G_{3b}$ & $M_{3b}$ & $T_{4b}$ &
B$_{2g}$ & B$_{2}$ & B$_{g}$ & B$_{1}$ & B & A$_{2}''$ & A$''$ &
A$_{2g}$ & A$_{2}$ & A$_{1}$ & A$_{g}$ & A
\\
$Q_{zx}$, $Q_{4u}^\alpha$ & $G_{x}$, $G_{3u}$ & $M_{x}$, $M_{3u}$ & $T_{zx}$, $T_{4u}^\alpha$ &
E$_{1g}$ & E$_{1}$ & E$_{1g}$ & E$_{1}$ & E$_{1}$ & E$''$ & E$''$ &
E$_{g}$ & E & E & E$_{g}$ & E
\\
$Q_{yz}$, $Q_{4v}^\alpha$ & $G_{y}$, $G_{3v}$ & $M_{y}$, $M_{3v}$ & $T_{yz}$, $T_{4v}^\alpha$ &
& & & & & & &
& & & &
\\
$Q_{v}$, $Q_{4u}^{\beta 1}$, $Q_{4u}^{\beta 2}$ & $G_{xyz}$ & $M_{xyz}$ & $T_{v}$, $T_{4u}^{\beta 1}$, $T_{4u}^{\beta 2}$ &
E$_{2g}$ & E$_{2}$ & E$_{2g}$ & E$_{2}$ & E$_{2}$ & E$'$ & E$'$ &
E$_{g}$ & E & E & E$_{g}$ & E
\\
$Q_{xy}$, $Q^{\beta 1}_{4v}$, $Q^{\beta 2}_{4v}$ & $G_{z}^{\beta}$ & $M_{z}^{\beta}$ & $T_{xy}$, $T^{\beta 2}_{4v}$, $T^{\beta 2}_{4v}$ &
& & & & & & &
& & & &
\\ \hline
--- & $G_{0}$, $G_{u}$, $G_{40}$ & $M_{0}$, $M_{u}$, $M_{40}$ & --- &
A$_{1u}$ & A$_{1}$ & A$_{u}$ & A$_{2}$ & A & A$_{1}''$ & A$''$ &
A$_{1u}$ & A$_{1}$ & A$_{2}$ & A$_{u}$ & A
\\
$Q_{z}$, $Q_{z}^{\alpha}$ & --- & --- & $T_{z}$, $T_{z}^{\alpha}$ &
A$_{2u}$ & A$_{2}$ & A$_{u}$ & A$_{1}$ & A & A$_{2}''$ & A$''$ &
A$_{2u}$ & A$_{2}$ & A$_{1}$ & A$_{u}$ & A
\\
$Q_{3a}$ & $G_{4a}$ & $M_{4a}$ & $T_{3a}$ &
B$_{1u}$ & B$_{1}$ & B$_{u}$ & B$_{1}$ & B & A$_{1}'$ & A$'$ &
A$_{1u}$ & A$_{1}$ & A$_{1}$ & A$_{u}$ & A
\\
$Q_{3b}$ & $G_{4b}$ & $M_{4b}$ & $T_{3b}$ &
B$_{2u}$ & B$_{2}$ & B$_{u}$ & B$_{2}$ & B & A$_{2}'$ & A$'$ &
A$_{2u}$ & A$_{2}$ & A$_{2}$ & A$_{u}$ & A
\\
$Q_{x}$, $Q_{3u}$ & $G_{zx}$, $G_{4u}^\alpha$ & $M_{zx}$, $M_{4u}^\alpha$ & $T_{x}$, $T_{3u}$ &
E$_{1u}$ & E$_{1}$ & E$_{1u}$ & E$_{1}$ & E$_{1}$ & E$'$ & E$'$ &
E$_{u}$ & E & E & E$_{u}$ & E
\\
$Q_{y}$, $Q_{3v}$ & $G_{yz}$, $G_{4v}^\alpha$ & $M_{yz}$, $M_{4v}^\alpha$ & $T_{y}$, $T_{3v}$ &
& & & & & & &
& & & &
\\
$Q_{xyz}$  & $G_{v}$, $G^{\beta1}_{4u}$, $G^{\beta2}_{4u}$ & $M_{v}$, $M_{4u}^{\beta 1}$, $M_{4u}^{\beta 2}$ & $T_{xyz}$ &
E$_{2u}$ & E$_{2}$ & E$_{2u}$ & E$_{2}$ & E$_{2}$ & E$''$ & E$''$ &
E$_{u}$ & E & E & E$_{u}$ & E
\\
$Q_{z}^{\beta}$ & $G_{xy}$, $G^{\beta1}_{4v}$, $G^{\beta2}_{4v}$ & $M_{xy}$, $M^{\beta 1}_{4v}$, $M^{\beta 2}_{4v}$ & $T_{z}^{\beta}$ &
& & & & & & &
& & & &
\\
\hline\hline
\end{tabular}
\end{table*}
\begin{table*}[h]
\caption{
Multipoles under the hexagonal crystal system ($D_{6\rm h}$) in the spinless basis. 
$(g,u)^{\pm}={\rm A}^{\pm}_{1g,u} \oplus {\rm A}^{\mp}_{2g,u} \oplus {\rm E}^{\pm}_{2g,u}$ and $\braket{g,u}^{\pm}={\rm B}^{\pm}_{1g,u} \oplus {\rm B}^{\pm}_{2g,u} \oplus {\rm E}^{\pm}_{1g,u}$. 
The subscripts $g,u$ are omitted for the $D_{6}$ group. 
The subscripts $1,2$ in A and B are omitted for the $C_{6{\rm h}}$ group. 
A$_{1u}$ $\leftrightarrow$ A$_{2u}$ and B$_{1g}$ $\leftrightarrow$ B$_{2g}$, and then, the subscripts $g,u$ are omitted for the $C_{6{\rm v}}$ group. 
The subscripts $g,u$ and $1,2$ in A and B are omitted for the $C_{6}$ group. 
A$_{1g}$, B$_{1u}$ $\to$ A$'_{1}$, A$_{2g}$, B$_{2u}$ $\to$ A$'_{2}$, A$_{1u}$, B$_{1g}$ $\to$ A$''_{1}$, A$_{2u}$, B$_{2g}$ $\to$ A$''_{2}$, E$_{2u}$, E$_{1g}$ $\to$ E$''$, E$_{2g}$, and E$_{1u}$ $\to$ E' for the $D_{3{\rm h}}$ group. 
A$_{1g}$, A$_{2g}$, B$_{1u}$, B$_{2u}$ $\to$ A$'$, A$_{1u}$, A$_{2u}$, B$_{1g}$, B$_{2g}$ $\to$ A$''$, E$_{2u}$, E$_{1g}$ $\to$ E$''$, E$_{2g}$, and E$_{1u}$ $\to$ E$'$ for the $C_{3{\rm h}}$ group. 
}
\label{tab_hexa_less}
\centering
\begin{tabular}{c|c|cc|ccc|ccccc}
& (s) A$_{1g}$ & (p) A$_{2u}$ & E$_{1u}$ & (d) A$_{1g}$ & E$_{1g}$ & E$_{2g}$ & (f) A$_{2u}$ & B$_{1u}$ & B$_{2u}$ & E$_{1u}$ & E$_{2u}$ \\ \hline
(s) A$_{1g}$ & A$^+_{1g}$  & A$^{\pm}_{2u}$ & E$^{\pm}_{1u}$ & A$^{\pm}_{1g}$ & E$^{\pm}_{1g}$ & E$^{\pm}_{2g}$ & A$^{\pm}_{2u}$ & B$^{\pm}_{1u}$ & B$^{\pm}_{2u}$ & E$^{\pm}_{1u}$ & E$^{\pm}_{2u}$ \\ \hline
(p) A$_{2u}$ & & A$^+_{1g}$ & E$^{\pm}_{1g}$ & A$^{\pm}_{2u}$ & E$^{\pm}_{1u}$ & E$^{\pm}_{2u}$ & A$^{\pm}_{1g}$ & B$^{\pm}_{2g}$ & B$^{\pm}_{1g}$ & E$^{\pm}_{1g}$ & E$^{\pm}_{2g}$ \\
E$_{1u}$ & & & $(g)^+$ & E$^{\pm}_{1u}$ & $(u)^{\pm}$ & $\braket{u}^{\pm}$ & E$^{\pm}_{1g}$ & E$^{\pm}_{2g}$ & E$^{\pm}_{2g}$ & $(g)^{\pm}$ & $\braket{g}^{\pm}$\\ \hline
(d) A$_{1g}$ & & & & A$^+_{1g}$ & E$^{\pm}_{1g}$ & E$^{\pm}_{2g}$ & A$^{\pm}_{2u}$ & B$^{\pm}_{1u}$ & B$^{\pm}_{2u}$ & E$^{\pm}_{1u}$ & E$^{\pm}_{2u}$ \\
E$_{1g}$ & & & & & $(g)^+$ & $\braket{g}^{\pm}$ & E$^{\pm}_{1u}$ & E$^{\pm}_{2u}$ & E$^{\pm}_{2u}$ & $(u)^{\pm}$ & $\braket{u}^{\pm}$ \\
E$_{2g}$ & & & & & & $(g)^+$ & E$^{\pm}_{2u}$ & E$^{\pm}_{1u}$ & E$^{\pm}_{1u}$ & $\braket{u}^{\pm}$ & $(u)^{\pm}$ \\ \hline
(f) A$_{2u}$ & & & & & & & A$^+_{1g}$ & B$^{\pm}_{2g}$ & B$^{\pm}_{1g}$ & E$^{\pm}_{1g}$ & E$^{\pm}_{2g}$ \\
B$_{1u}$ & & & & & & & & A$^+_{1g}$ & A$^{\pm}_{2g}$ & E$^{\pm}_{2g}$ & E$^{\pm}_{1g}$ \\
B$_{2u}$ & & & & & & & & & A$^+_{1g}$ & E$^{\pm}_{2g}$ & E$^{\pm}_{1g}$ \\
E$_{1u}$ & & & & & & & & & & $(g)^+$ & $\braket{g}^{\pm}$ \\
E$_{2u}$ & & & & & & & & & & & $(g)^+$ 
\end{tabular}
\end{table*}
\begin{table*}[h]
\caption{
Multipoles under the hexagonal crystal system ($D_{6\rm h}$) in the spinful basis. 
$(g,u)^{\pm}={\rm A}^{\pm}_{1g,u} \oplus {\rm A}^{\mp}_{2g,u} \oplus {\rm E}^{\mp}_{1g,u}$, $(g',u')^{\pm}={\rm A}^{\pm}_{1g,u} \oplus {\rm A}^{\mp}_{2g,u}\oplus {\rm B}^{\mp}_{1g,u} \oplus {\rm B}^{\mp}_{2g,u}$, and $\braket{g,u}^{\pm}={\rm B}^{\pm}_{1g,u} \oplus {\rm B}^{\pm}_{2g,u} \oplus {\rm E}^{\pm}_{2g,u}$. 
The subscripts $g,u$ are omitted for the $D_{6}$ group. 
The subscripts $1,2$ in A and B are omitted for the $C_{6{\rm h}}$ group. 
${\rm A}_{1u} \leftrightarrow {\rm A}_{2u}$ and ${\rm B}_{1g} \leftrightarrow {\rm B}_{2g}$, and then, the subscripts $g,u$ are omitted for the $C_{6{\rm v}}$ group. 
The subscripts $g,u$ and $1,2$ in A and B are omitted for the $C_{6}$ group. 
${\rm A}_{1g}, {\rm B}_{1u} \to {\rm A}'_{1}$, ${\rm A}_{2g}, {\rm B}_{2u} \to {\rm A}'_{2}$, ${\rm A}_{1u}, {\rm B}_{1g} \to {\rm A}''_{1}$, ${\rm A}_{2u}, {\rm B}_{2g} \to {\rm A}''_{2}$, ${\rm E}_{2u}, {\rm E}_{1g} \to {\rm E}''$, ${\rm E}_{2g}, {\rm E}_{1u} \to {\rm E}'$, and ${\rm E}_{1/2u} \leftrightarrow {\rm E}_{3/2u}$ for the $D_{3{\rm h}}$ group. 
${\rm A}_{1g}, {\rm A}_{2g}, {\rm B}_{1u}, {\rm B}_{2u} \to {\rm A}'$, ${\rm A}_{1u}, {\rm A}_{2u}, {\rm B}_{1g}, {\rm B}_{2g} \to {\rm A}''$, ${\rm E}_{2u}, {\rm E}_{1g} \to {\rm E}''$, ${\rm E}_{2g}, {\rm E}_{1u} \to{\rm E}'$, and ${\rm E}_{1/2u} \leftrightarrow {\rm E}_{3/2u}$ for the $C_{3{\rm h}}$ group. 
}
\label{tab_hexa_ful}
\centering
\scalebox{0.7}{
\begin{tabular}{cc|c|ccc|ccccc|ccccccc}
& & (s) E$_{1/2g}$ & (p) E$_{1/2u}$ & E$_{1/2u}$ & E$_{5/2u}$ & (d) E$_{1/2g}$ & E$_{1/2g}$ & E$_{5/2g}$& E$_{3/2g}$& E$_{5/2g}$ & (f) E$_{1/2u}$ & E$_{3/2u}$ & E$_{3/2u}$ & E$_{1/2u}$ & E$_{5/2u}$& E$_{3/2u}$& E$_{5/2u}$ \\ \hline
(s) A$_{1g}\otimes$ E$_{1/2g} \to$ & E$_{1/2g}$ & $(g)^+$ & $(u)^{\pm}$ & $(u)^{\pm}$ & E$^{\pm}_{1u}$, E$^{\pm}_{2u}$ & $(g)^{\pm}$ & $(g)^{\pm}$ & E$^{\pm}_{1g}$, E$^{\pm}_{2g}$ & $\braket{g}^{\pm}$ & E$^{\pm}_{1g}$, E$^{\pm}_{2g}$ & $(u)^{\pm}$ & $\braket{u}^{\pm}$ & $\braket{u}^{\pm}$ & $(u)^{\pm}$ & E$^{\pm}_{1u}$, E$^{\pm}_{2u}$ & $\braket{u}^{\pm}$ & E$^{\pm}_{1u}$, E$^{\pm}_{2u}$ \\ \hline
(p) A$_{2u}\otimes$ E$_{1/2g} \to$ & E$_{1/2u}$ & & $(g)^+$ & $(g)^{\pm}$ & E$^{\pm}_{1g}$, E$^{\pm}_{2g}$ & $(u)^{\pm}$ & $(u)^{\pm}$ & E$^{\pm}_{1u}$, E$^{\pm}_{2u}$ & $\braket{u}^{\pm}$ & E$^{\pm}_{1u}$, E$^{\pm}_{2u}$ & $(g)^{\pm}$ & $\braket{g}^{\pm}$ & $\braket{g}^{\pm}$ & $(g)^{\pm}$ & E$^{\pm}_{1g}$, E$^{\pm}_{2g}$ & $\braket{g}^{\pm}$ & E$^{\pm}_{1g}$, E$^{\pm}_{2g}$ \\
E$_{1u}\otimes$ E$_{1/2g} \to$ & E$_{1/2u}$ & & & $(g)^+$ & E$^{\pm}_{1g}$, E$^{\pm}_{2g}$ & $(u)^{\pm}$ & $(u)^{\pm}$ & E$^{\pm}_{1u}$, E$^{\pm}_{2u}$ & $\braket{u}^{\pm}$ & E$^{\pm}_{1u}$, E$^{\pm}_{2u}$ & $(g)^{\pm}$ & $\braket{g}^{\pm}$ & $\braket{g}^{\pm}$ & $(g)^{\pm}$ & E$^{\pm}_{1g}$, E$^{\pm}_{2g}$ & $\braket{g}^{\pm}$ & E$^{\pm}_{1g}$, E$^{\pm}_{2g}$\\ 
 & E$_{5/2u}$ & & & & $(g')^+$ & E$^{\pm}_{1u}$, E$^{\pm}_{2u}$ & E$^{\pm}_{1u}$, E$^{\pm}_{2u}$ & $(u')^{\pm}$ & E$^{\pm}_{1u}$, E$^{\pm}_{2u}$ & $(u')^{\pm}$ & E$^{\pm}_{1g}$, E$^{\pm}_{2g}$ & E$^{\pm}_{1g}$, E$^{\pm}_{2g}$ &  E$^{\pm}_{1g}$, E$^{\pm}_{2g}$  & E$^{\pm}_{1g}$, E$^{\pm}_{2g}$ & $(g')^{\pm}$ &  E$^{\pm}_{1g}$, E$^{\pm}_{2g}$  & $(g')^{\pm}$\\ \hline
(d) A$_{1g}\otimes$ E$_{1/2g} \to$ & E$_{1/2g}$ & & & & & $(g)^+$ & $(g)^{\pm}$ & E$^{\pm}_{1g}$, E$^{\pm}_{2g}$ & $\braket{g}^{\pm}$ & E$^{\pm}_{1g}$, E$^{\pm}_{2g}$ & $(u)^{\pm}$ & $\braket{u}^{\pm}$ & $\braket{u}^{\pm}$ & $(u)^{\pm}$ & E$^{\pm}_{1u}$, E$^{\pm}_{2u}$ & $\braket{u}^{\pm}$ & E$^{\pm}_{1u}$, E$^{\pm}_{2u}$ \\
E$_{1g}\otimes$ E$_{1/2g} \to$ & E$_{1/2g}$  & & & & & & $(g)^+$ & E$^{\pm}_{1g}$, E$^{\pm}_{2g}$ & $\braket{g}^{\pm}$ & E$^{\pm}_{1g}$, E$^{\pm}_{2g}$ & $(u)^{\pm}$ & $\braket{u}^{\pm}$ & $\braket{u}^{\pm}$ & $(u)^{\pm}$ & E$^{\pm}_{1u}$, E$^{\pm}_{2u}$ & $\braket{u}^{\pm}$ & E$^{\pm}_{1u}$, E$^{\pm}_{2u}$ \\
& E$_{5/2g}$  & & & & & & & $(g')^+$ & E$^{\pm}_{1g}$, E$^{\pm}_{2g}$  & $(g')^{\pm}$ & E$^{\pm}_{1u}$, E$^{\pm}_{2u}$ & E$^{\pm}_{1u}$, E$^{\pm}_{2u}$  &  E$^{\pm}_{1u}$, E$^{\pm}_{2u}$ & E$^{\pm}_{1u}$, E$^{\pm}_{2u}$ & $(u')^{\pm}$ &  E$^{\pm}_{1u}$, E$^{\pm}_{2u}$  & $(u')^{\pm}$ \\
E$_{2g}\otimes$ E$_{1/2g} \to$ & E$_{3/2g}$ & & & & & & & & $(g)^+$ & E$^{\pm}_{1g}$, E$^{\pm}_{2g}$ & $\braket{u}^{\pm}$ & $(u)^{\pm}$ & $(u)^{\pm}$ & $\braket{u}^{\pm}$ & E$^{\pm}_{1u}$, E$^{\pm}_{2u}$ & $(u)^{\pm}$ & E$^{\pm}_{1u}$, E$^{\pm}_{2u}$ \\
& E$_{5/2g}$ & & & & & & & & & $(g')^+$ & E$^{\pm}_{1u}$, E$^{\pm}_{2u}$ &  E$^{\pm}_{1u}$, E$^{\pm}_{2u}$  &  E$^{\pm}_{1u}$, E$^{\pm}_{2u}$  & E$^{\pm}_{1u}$, E$^{\pm}_{2u}$ & $(u')^{\pm}$ & E$^{\pm}_{1u}$, E$^{\pm}_{2u}$  & $(u')^{\pm}$ \\ \hline
(f) A$_{2u}\otimes$ E$_{1/2g} \to$ & E$_{1/2u}$ & & & & & & & & & & $(g)^+$ & $\braket{g}^{\pm}$ & $\braket{g}^{\pm}$ & $(g)^{\pm}$ & E$^{\pm}_{1g}$, E$^{\pm}_{2g}$ & $\braket{g}^{\pm}$ & E$^{\pm}_{1g}$, E$^{\pm}_{2g}$ \\
B$_{1u}\otimes$ E$_{1/2g} \to$ & E$_{3/2u}$ & & & & & & & & & & & $(g)^+$ & $(g)^{\pm}$ & $\braket{g}^{\pm}$ & E$^{\pm}_{1g}$, E$^{\pm}_{2g}$ & $(g)^{\pm}$ & E$^{\pm}_{1g}$, E$^{\pm}_{2g}$ \\
B$_{2u}\otimes$ E$_{1/2g} \to$ & E$_{3/2u}$ & & & & & & & & & & & & $(g)^+$ & $\braket{g}^{\pm}$ & E$^{\pm}_{1g}$, E$^{\pm}_{2g}$ & $(g)^{\pm}$ & E$^{\pm}_{1g}$, E$^{\pm}_{2g}$ \\
E$_{1u}\otimes$ E$_{1/2g} \to$ & E$_{1/2u}$ & & & & & & & & & & & & & $(g)^+$ & E$^{\pm}_{1g}$, E$^{\pm}_{2g}$ & $\braket{g}^{\pm}$ & E$^{\pm}_{1g}$, E$^{\pm}_{2g}$ \\
& E$_{5/2u}$ & & & & & & & & & & & & & & $(g')^+$ &  E$^{\pm}_{1g}$, E$^{\pm}_{2g}$  & $(g')^{\pm}$ \\
E$_{2u}\otimes$ E$_{1/2g} \to$& E$_{3/2u}$ & & & & & & & & & & & & & & & $(g)^+$ & E$^{\pm}_{1g}$, E$^{\pm}_{2g}$ \\
& E$_{5/2u}$ & & & & & & & & & & & & & & & & $(g')^+$
\end{tabular}
}
\end{table*}

\begin{table*}[h]
\caption{
Multipoles under the trigonal crystal system ($D_{3\rm d}$) in the spinless basis. 
$(g,u)^{\pm}={\rm A}^{\pm}_{1g,u} \oplus {\rm A}^{\mp}_{2g,u} \oplus {\rm E}^{\pm}_{g,u}$. 
The subscripts $g,u$ are omitted for the $D_{3}$ group. 
The subscripts $g,u$ are omitted, and $Q_{3a} \leftrightarrow Q_{3b}$ for the $C_{3{\rm v}}$ group. 
The subscripts $1,2$ are omitted for the $C_{3{\rm i}}$ group. 
The subscripts $g,u$ and $1,2$ are omitted for the $C_{3}$ group. 
}
\label{tab_trig_less}
\centering
\begin{tabular}{c|c|cc|ccc|ccccc}
& (s) A$_{1g}$ & (p) A$_{2u}$ & E$_{u}$ & (d) A$_{1g}$ & E$_{g}$ & E$_{g}$ & (f) A$_{2u}$ & A$_{1u}$ & A$_{2u}$ & E$_{u}$ & E$_{u}$ \\ \hline
(s) A$_{1g}$ & A$^+_{1g}$  & A$^{\pm}_{2u}$ & E$^{\pm}_{u}$ & A$^{\pm}_{1g}$ & E$^{\pm}_{g}$ & E$^{\pm}_{g}$ & A$^{\pm}_{2u}$ & A$^{\pm}_{1u}$ & A$^{\pm}_{2u}$ & E$^{\pm}_{u}$ & E$^{\pm}_{u}$ \\ \hline
(p) A$_{2u}$ & & A$^+_{1g}$ & E$^{\pm}_{g}$ & A$^{\pm}_{2u}$ & E$^{\pm}_{u}$ & E$^{\pm}_{u}$ & A$^{\pm}_{1g}$ & A$^{\pm}_{2g}$ & A$^{\pm}_{1g}$ & E$^{\pm}_{g}$ & E$^{\pm}_{g}$ \\
E$_{u}$ & & & $(g)^+$ & E$^{\pm}_{u}$ & $(u)^{\pm}$ & $(u)^{\pm}$ & E$^{\pm}_{g}$ & E$^{\pm}_{g}$ & E$^{\pm}_{g}$ & $(g)^{\pm}$ & $(g)^{\pm}$\\ \hline
(d) A$_{1g}$ & & & & A$^+_{1g}$ & E$^{\pm}_{g}$ & E$^{\pm}_{g}$ & A$^{\pm}_{2u}$ & A$^{\pm}_{1u}$ & A$^{\pm}_{2u}$ & E$^{\pm}_{u}$ & E$^{\pm}_{u}$ \\
E$_{g}$ & & & & & $(g)^+$ & $(g)^{\pm}$ & E$^{\pm}_{u}$ & E$^{\pm}_{u}$ & E$^{\pm}_{u}$ & $(u)^{\pm}$ & $(u)^{\pm}$ \\
E$_{g}$ & & & & & & $(g)^+$ & E$^{\pm}_{u}$ & E$^{\pm}_{u}$ & E$^{\pm}_{u}$ & $(u)^{\pm}$ & $(u)^{\pm}$ \\ \hline
(f) A$_{2u}$ & & & & & & & A$^+_{1g}$ & A$^{\pm}_{2g}$ & A$^{\pm}_{1g}$ & E$^{\pm}_{g}$ & E$^{\pm}_{g}$ \\
A$_{1u}$ & & & & & & & & A$^+_{1g}$ & A$^{\pm}_{2g}$ & E$^{\pm}_{g}$ & E$^{\pm}_{g}$ \\
A$_{2u}$ & & & & & & & & & A$^+_{1g}$ & E$^{\pm}_{g}$ & E$^{\pm}_{g}$ \\
E$_{u}$ & & & & & & & & & & $(g)^+$ & $(g)^{\pm}$ \\
E$_{u}$ & & & & & & & & & & & $(g)^+$ 
\end{tabular}
\end{table*}

\begin{table*}[h]
\caption{
Multipoles under the trigonal crystal system ($D_{3\rm d}$) in the spinful basis. 
$(g,u)^{\pm}={\rm A}^{\pm}_{1g,u} \oplus {\rm A}^{\mp}_{2g,u} \oplus {\rm E}^{\mp}_{g,u}$ and $(g',u')^{\mp}={\rm A}^{\mp}_{1g,u} \oplus {\rm A}^{\pm}_{1g,u} \oplus 2{\rm A}^{\pm}_{2g,u}$. 
The subscripts $g,u$ are omitted for the $D_{3}$ group. 
The subscripts $g,u$ are omitted, and $Q_{3a} \leftrightarrow Q_{3b}$ for the $C_{3{\rm v}}$ group. 
The subscripts $1,2$ are omitted for the $C_{3{\rm i}}$ group. 
The subscripts $g,u$ and $1,2$ are omitted for the $C_{3}$ group. 
}
\label{tab_trig_ful}
\centering
\scalebox{0.85}{
\begin{tabular}{cc|c|ccc|ccccc|ccccccc}
& & (s) E$_{1/2g}$ & (p) E$_{1/2u}$ & E$_{1/2u}$ & E$_{3/2u}$ & (d) E$_{1/2g}$ & E$_{1/2g}$ & E$_{3/2g}$& E$_{1/2g}$& E$_{3/2g}$ & (f) E$_{1/2u}$ & E$_{1/2u}$ & E$_{1/2u}$ & E$_{1/2u}$ & E$_{3/2u}$& E$_{1/2u}$& E$_{3/2u}$ \\ \hline
(s) A$_{1g}\otimes$ E$_{1/2g} \to$ & E$_{1/2g}$ & $(g)^+$ & $(u)^{\pm}$ & $(u)^{\pm}$ & 2E$^{\pm}_{u}$ & $(g)^{\pm}$ & $(g)^{\pm}$ & 2E$^{\pm}_{g}$ & $(g)^{\pm}$ & 2E$^{\pm}_{g}$ & $(u)^{\pm}$ & $(u)^{\pm}$ & $(u)^{\pm}$ & $(u)^{\pm}$ & 2E$^{\pm}_{u}$ & $(u)^{\pm}$ & 2E$^{\pm}_{u}$ \\ \hline
(p) A$_{2u}\otimes$ E$_{1/2g} \to$ & E$_{1/2u}$ & & $(g)^+$ & $(g)^{\pm}$ & 2E$^{\pm}_{g}$ & $(u)^{\pm}$ & $(u)^{\pm}$ & 2E$^{\pm}_{u}$ & $(u)^{\pm}$ & 2E$^{\pm}_{u}$ & $(g)^{\pm}$ & $(g)^{\pm}$ & $(g)^{\pm}$ & $(g)^{\pm}$ & 2E$^{\pm}_{g}$ & $(g)^{\pm}$ & 2E$^{\pm}_{g}$ \\
E$_{u}\otimes$ E$_{1/2g} \to$ & E$_{1/2u}$ & & & $(g)^+$ & 2E$^{\pm}_{g}$ & $(u)^{\pm}$ & $(u)^{\pm}$ & 2E$^{\pm}_{u}$ & $(u)^{\pm}$ & 2E$^{\pm}_{u}$ & $(g)^{\pm}$ & $(g)^{\pm}$ & $(g)^{\pm}$ & $(g)^{\pm}$ & 2E$^{\pm}_{g}$ & $(g)^{\pm}$ & 2E$^{\pm}_{g}$\\ 
 & E$_{3/2u}$ & & & & $(g')^+$ & 2E$^{\pm}_{u}$ & 2E$^{\pm}_{u}$ & $(u')^{\pm}$ & 2E$^{\pm}_{u}$ & $(u')^{\pm}$ & 2E$^{\pm}_{g}$ & 2E$^{\pm}_{g}$ &  2E$^{\pm}_{g}$ & 2E$^{\pm}_{g}$ & $(g')^{\pm}$ &  2E$^{\pm}_{g}$ & $(g')^{\pm}$\\ \hline
(d) A$_{1g}\otimes$ E$_{1/2g} \to$ & E$_{1/2g}$ & & & & & $(g)^+$ & $(g)^{\pm}$ & 2E$^{\pm}_{g}$ & $(g)^{\pm}$ & 2E$^{\pm}_{g}$ & $(u)^{\pm}$ & $(u)^{\pm}$ & $(u)^{\pm}$ & $(u)^{\pm}$ & 2E$^{\pm}_{u}$ & $(u)^{\pm}$ & 2E$^{\pm}_{u}$ \\
E$_{g}\otimes$ E$_{1/2g} \to$ & E$_{1/2g}$  & & & & & & $(g)^+$ & 2E$^{\pm}_{g}$ & $(g)^{\pm}$ & 2E$^{\pm}_{g}$ & $(u)^{\pm}$ & $(u)^{\pm}$ & $(u)^{\pm}$ & $(u)^{\pm}$ & 2E$^{\pm}_{u}$ & $(u)^{\pm}$ & 2E$^{\pm}_{u}$ \\
& E$_{3/2g}$  & & & & & & & $(g')^+$ & 2E$^{\pm}_{g}$ & $(g'^{\pm})$ & 2E$^{\pm}_{u}$ & 2E$^{\pm}_{u}$ & 2E$^{\pm}_{u}$ & 2E$^{\pm}_{u}$ & $(u')^{\pm}$ & 2E$^{\pm}_{u}$ & $(u')^{\pm}$ \\
E$_{g}\otimes$ E$_{1/2g} \to$ & E$_{1/2g}$ & & & & & & & & $(g)^+$ & 2E$^{\pm}_{g}$ & $(u)^{\pm}$ & $(u)^{\pm}$ & $(u)^{\pm}$ & $(u)^{\pm}$ & 2E$^{\pm}_{u}$ & $(u)^{\pm}$ & 2E$^{\pm}_{u}$ \\
& E$_{3/2g}$ & & & & & & & & & $(g')^+$ & 2E$^{\pm}_{u}$ & 2E$^{\pm}_{u}$ & 2E$^{\pm}_{u}$ & 2E$^{\pm}_{u}$ & $(u')^{\pm}$ & 2E$^{\pm}_{u}$ & $(u')^{\pm}$ \\ \hline
(f) A$_{2u}\otimes$ E$_{1/2g} \to$ & E$_{1/2u}$ & & & & & & & & & & $(g)^+$ & $(g)^{\pm}$ & $(g)^{\pm}$ & $(g)^{\pm}$ & 2E$^{\pm}_{g}$ & $(g)^{\pm}$ & 2E$^{\pm}_{g}$ \\
A$_{1u}\otimes$ E$_{1/2g} \to$ & E$_{1/2u}$ & & & & & & & & & & & $(g)^+$ & $(g)^{\pm}$ & $(g)^{\pm}$ & 2E$^{\pm}_{g}$ & $(g)^{\pm}$ & 2E$^{\pm}_{g}$ \\
A$_{2u}\otimes$ E$_{1/2g} \to$ & E$_{1/2u}$ & & & & & & & & & & & & $(g)^+$ & $(g)^{\pm}$ & 2E$^{\pm}_{g}$ & $(g)^{\pm}$ & 2E$^{\pm}_{g}$ \\
E$_{u}\otimes$ E$_{1/2g} \to$ & E$_{1/2u}$ & & & & & & & & & & & & & $(g)^+$ & 2E$^{\pm}_{g}$ & $(g)^{\pm}$ & 2E$^{\pm}_{g}$ \\
& E$_{3/2u}$ & & & & & & & & & & & & & & $(g')^+$ & 2E$^{\pm}_{g}$ & $(g')^{\pm}$ \\
E$_{u}\otimes$ E$_{1/2g} \to$& E$_{1/2u}$ & & & & & & & & & & & & & & & $(g)^+$ & 2E$^{\pm}_{g}$ \\
& E$_{3/2u}$ & & & & & & & & & & & & & & & & $(g')^+$
\end{tabular}
}
\end{table*}

\bibliographystyle{apsrev}
\bibliography{ref}

\end{document}